\documentclass[mnsc,nonblindrev]{arxiv}

\DoubleSpacedXI 

\usepackage[utf8]{inputenc}
\usepackage{algorithm}
\usepackage{algpseudocode}
\usepackage{tikz}
\usepackage{amsmath, amsfonts, amssymb}
\usepackage{booktabs} 
\usepackage{url}            
\usepackage{bbm}
\usepackage{bm}
\usepackage{epstopdf}
\usepackage{layout, color}
\usepackage{soul}
\usepackage{array, makecell}
\usepackage{multirow}
\usepackage{threeparttable}
\usepackage{silence}
\usepackage{subcaption}
\usepackage{caption}

\newcommand*\mystrut[1]{\vrule width0pt height0pt depth#1\relax}

\usepackage{endnotes}
\let\footnote=\endnote

\usepackage{natbib}
 \bibpunct[, ]{(}{)}{,}{a}{}{,}%
 \def\bibfont{\small}%
\TheoremsNumberedThrough     
\ECRepeatTheorems

\EquationsNumberedThrough    

\begin{document}


\RUNAUTHOR{Zhu et al.}

\RUNTITLE{Quantifying Grid Resilience Against Extreme Weather Using Large-Scale Customer Power Outage Data}

\TITLE{Quantifying Grid Resilience\\Against Extreme Weather Using\\Large-Scale Customer Power Outage Data}

\ARTICLEAUTHORS{%
\AUTHOR{Shixiang Zhu, Rui Yao, Yao Xie, Feng Qiu, Yueming (Lucy) Qiu, Xuan Wu}
} 

\ABSTRACT{%
In recent years, extreme weather events frequently cause large-scale power outages. Resilience, the capability of withstanding, adapting to, and recovering from a large-scale disruption, has become a top priority for the power sector. However, a system-level understanding of power grid resilience remains limited, with most studies yielding conceptual insights or focusing on isolated technical issues. Using a spatio-temporal model, this study adopts a data-driven approach and analyzes quarter-hourly, customer-level power outage data and corresponding weather records from three major service territories on the U.S. East Coast. Our findings reveal that excessive weather stress and planning vulnerabilities at specific grid nodes are key drivers of prolonged local outages, which propagate system-wide. Simulations show that targeted interventions, such as isolating critical nodes and protecting vulnerable nodes from transient faults, can reduce customer outages by 45.5\% and 49.5\%, respectively. These insights inform actionable strategies for decision-makers to enhance grid resilience and mitigate future disruptions.
}%
\newcommand*{\eg}{\emph{e.g.}{}}
\newcommand*{\ie}{\emph{i.e.}{}}
\newcommand*{\iid}{\emph{i.i.d.}{}}
\newcommand*{\etc}{\emph{etc}{}}
\newcommand{\indep}{\perp \!\!\! \perp}
\newcommand{\norm}[1]{\left\lVert#1\right\rVert}



\maketitle

%


\section{Introduction}

Extreme weather events, such as hurricanes, winter storms, and tornadoes, have become a major cause of large-scale electric power outages in recent years \citep{Handmer2012, Kenward2014}. 
For example, in March 2018, the Northeastern United States was hit by three winter storms in just 14 days, and power failures occurred across the New England region, affecting more than 2,755,000 customers, causing total economic losses of \$4 billion, including \$2.9 billion in insured losses \citep{Captive2018}. Such extreme weather events often left millions of people without electricity for days, caused substantial economic losses \citep{Executive2013, Smith2018}, and even human lives in some cases \citep{Shapiro2017}.
In light of the extensive losses from extreme weather since the early 2000s, regulatory entities of the United States at different levels have requested the industry to investigate the power grid resilience and adopt hardening measures against the extreme weather \citep{AbiSamra2013, Executive2013}.
Accurate assessment of power grid resilience is important in estimating extreme weather damage, carrying out preventive measures to reduce losses, and following up on energy policy-making. 

However, the study of grid resilience has been plagued by two major challenges \citep{Jufri2019, Wang2015}.
First, there is a profound disconnect between the real-world power failure data researchers need and what is available. 
Utility companies and system operators may collect fine-grained data on failure and recovery mainly for reporting purposes, but they are generally not shared beyond service territories \citep{Bryan2012}. 
For example, US federal and state governments require only aggregated information such as the total customer service interruption duration from investor-owned service regions during major storms \citep{Campbell2012}.
These data are often aggregated into daily statistics over the entire region, too crude to study the resilience of the infrastructure and services \citep{Executive2013}. 
Apart from data scarcity, such studies are also hampered by a lack of attendant mathematical models that can capture intricate dynamics between disruption and restoration processes under weather impacts.
As the power grid is a complex and chaotic system \citep{Fairley2004},
comprehensive and continuous influence from extreme weather incidents on one part of the system can accumulate and magnify to cause large effects on the system. 
As a result, identifying the key factors that contribute to the massive blackouts has long been a very complicated problem. 


\begin{figure}[!t]
\centering
\FIGURE{\begin{subfigure}{0.45\linewidth}
\includegraphics[width=\linewidth]{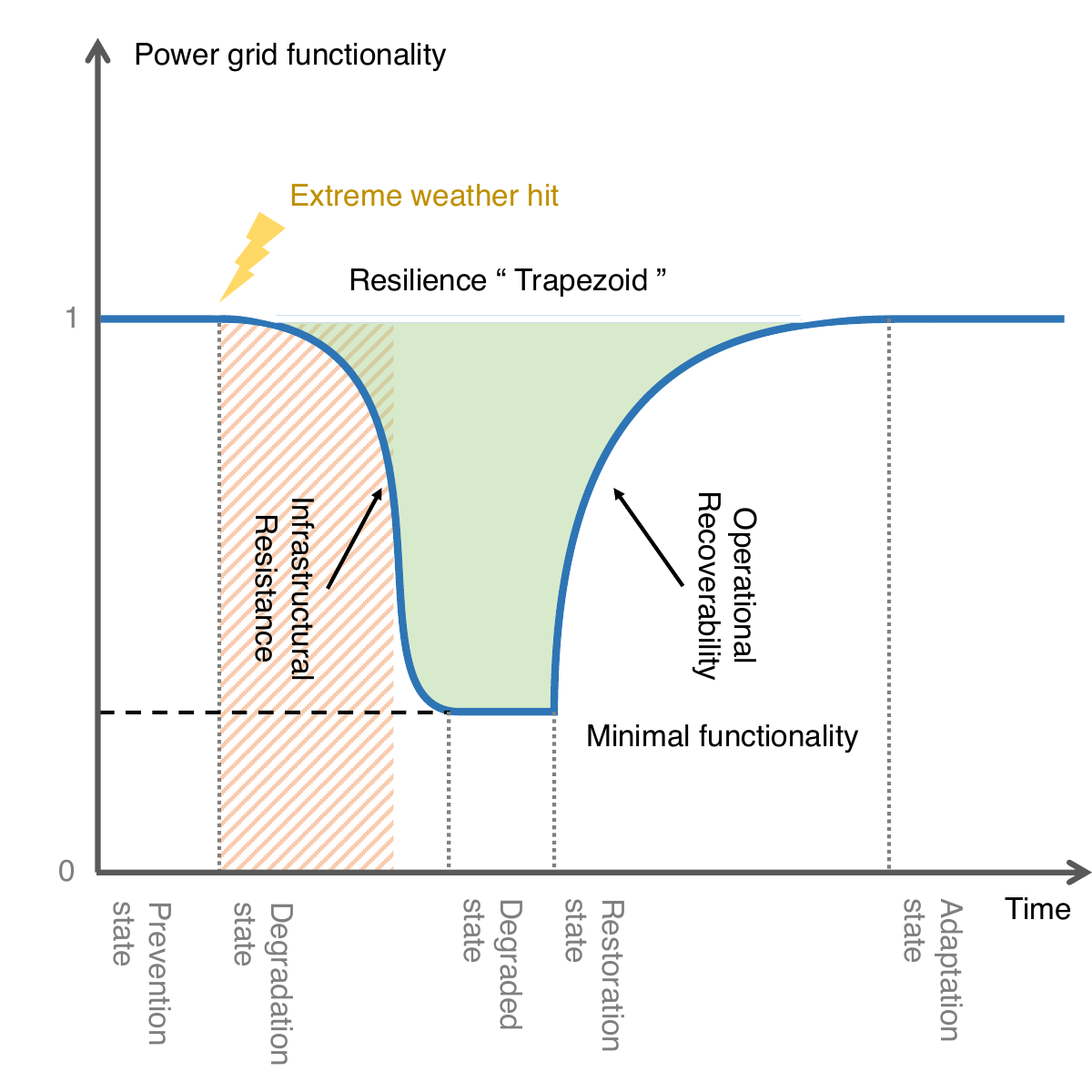}
\caption{}
\end{subfigure}
\hfill
\begin{subfigure}{0.45\linewidth}
\includegraphics[width=\linewidth]{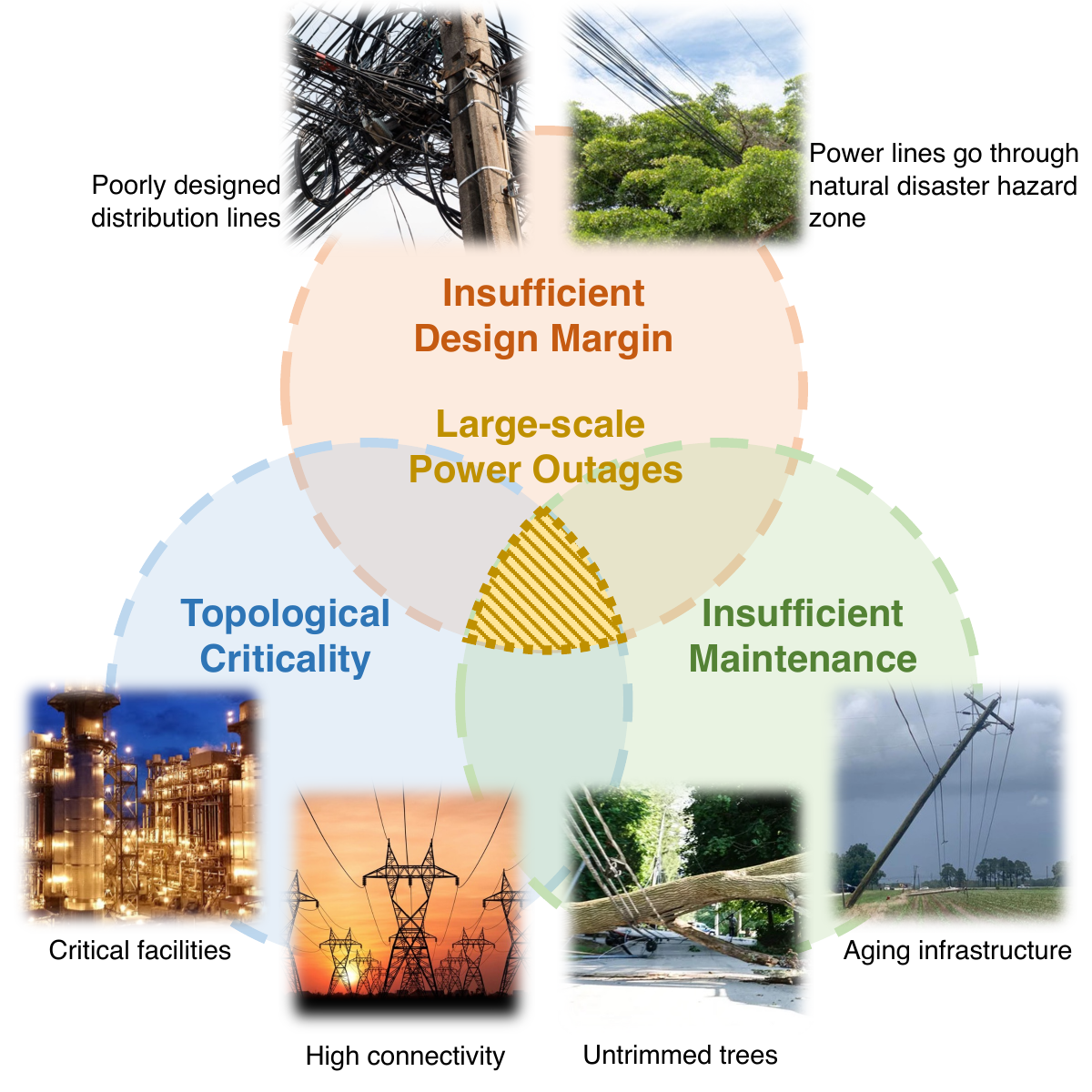}
\caption{}
\end{subfigure}}
{Resilience trapezoid. \label{fig:resilience-trapezoid}}
{(a) An illustration of the concept of resilience trapezoid (green area). The horizontal and vertical coordinates represent time and power grid functionality, respectively. The red-shaded area represents the period affected by extreme weather. The blue line represents the power grid functionality during an extreme weather incident, where the timeline can be divided into five states based on the status of the power grid: prevention, degradation, restoration, and adaptation. Power grid resilience in this study consists of two key characteristics, i.e., infrastructural resistance to extreme weather and operational recoverability from such damages. (b) The proposed model captures three key factors of infrastructural resistance that are closely tied to large-scale power outages: planning vulnerability, maintenance sufficiency, and criticality.}
\end{figure}

To circumvent these challenges and bridge this research gap, this paper quantitatively studies power grid resilience to extreme weather by taking advantage of a unique set of fine-grained customer-level power outage records and comprehensive weather data collected from three weather events across different service regions (covering four U.S. states).
Unlike detailed power grid facility outage records, which are normally inaccessible to the public, customer-level power outage records reported by local utilities \citep{MEMA2020, dobson2016obtaining, Duke2020, Georgia2020} are an important volume data source that is easily accessible but often overlooked. 
The data record the number of customers without power supplies in each \emph{geographical unit} (town, county, or zip code) every 15 minutes, portraying the entire evolution processes of multiple large-scale power outages. 
High-Resolution Rapid Refresh (HRRR) Archive \citep{Blaylock2017, Blaylock2015} is another publicly available data set that provides comprehensive and fine-grained near real-time weather information estimated by the National Centers for Environmental Prediction's (NCEP) HRRR data assimilation model.
Specifically, we developed a statistical model for the number of customer outages without knowing detailed disruption and restoration records. 
The proposed model aims to capture intricate spatio-temporal dynamics of customer outages between geographical units and over time based on a multivariate random process.
Using the fitted model, we first present a descriptive analysis that demonstrates the existence of grid resilience from real data and shows that the \emph{accumulation of weather effect} plays a pivotal role in creating large-scale power outages. 
The proposed model also leads to a general definition of power grid resilience as shown in Figure~\ref{fig:resilience-trapezoid}, which can be quantitatively measured by the fitted model's parameters from four general aspects: planning vulnerability, maintenance sufficiency, criticality, recoverability.
Our model suggests that local power outages directly induced by extreme weather is a non-linear response to the accumulation of weather effect and causes subsequent large-scale and long-duration blackout by propagating failures through some critical nodes in power networks.
According to the simulation, large-scale power outages can be effectively mitigated by isolating or de-escalating a small set of nodes in the power network.

\subsection{Related work}
The study of grid resilience has gained popularity in recent years \citep{Bhusal2020, Bie2017, Jufri2019, Wang2015}.
The existing works on resilience evaluation usually fall into two groups: qualitative methods and quantitative methods.

There is a large body of qualitative resilience evaluation methods. For example, \cite{carlson2012resilience, mcmanus2007resilience, Panteli2015grid} provide frameworks for system- and regional-level resilience overview using investigation, questionnaires, and individual ratings to address personal, business, governmental, and infrastructure aspects of resilience. 
A scoring matrix is formulated in \cite{Roege2014} to evaluate the system function from different perspectives; analytic methods such as the analytic hierarchy process (AHP) can be conveniently carried out to turn subjective opinions into comparable quantities, which is easy to use in decision making \citep{orencio2013localized}. 
These qualitative frameworks can guide long-term energy policy-making, providing a generally thorough picture of the system.
However, there are recent signs that qualitative methods are becoming increasingly expensive and difficult to carry out with the surge in large-scale blackouts over the past decade due to climate change. 
In addition, these methods are also restricted for lacking enough scientific evidence with limited justification of the methods adopted in some scenarios. 

On the other hand, quantitative methods are often based on quantifying system performances. 
Due to the very limited and insufficient access to real-world data \citep{EPRI2020}, the quantitative approaches to investigating and modeling the resilience of actual power grids to weather events are still nascent and poorly established.
Previous studies, such as \cite{Panteli2015modeling, Panteli2017, vugrin2017resilience}, attempt to quantify grid resilience by introducing operational and infrastructure resilience metrics based on different indicators, which usually require detailed disruption and restoration records at each phase. 
Some other works attempt to tackle this issue through a statistical approach without sufficient failure and recovery data. The Monte Carlo simulation is used in \cite{Baranski2003, Panteli2015modeling, Wang2016} to assess the capability of the grid and to determine the amount of impact caused by extreme weather.
Recently, a few previous studies \citep{dobson2016obtaining, Hines2017, Ji2016} have made significant progress in analyzing some specific aspects of power grid resilience or a particular extreme weather event by taking advantage of the limited amount of aggregated power failure data.
However, a resilience study that systematically explores the customer power outage data, quantitatively analyzes different weather effects on power systems, and evaluates the power grid resilience across multiple service territories has not hitherto existed. 

Two relevant studies \citep{wei2012non, zapata2008modeling} use similar nonstationary random processes to model the number of failed nodes in a power network, assuming that the expected number of nodes in a failure state equals the difference between expected failures and recoveries. This approach characterizes the transition of a distribution network from the present to the near future, encompassing the entire lifecycle of large-scale failures and recoveries.
Our study differs from these works in several key ways:
($i$) While their studies focus on a relatively small region (only ten counties) at the component level (including poles, circuits, and substations), our research aims to investigate system-level dynamics;
($ii$) Their models analyze a single large-scale failure event without considering weather conditions. In contrast, our study seeks to generalize grid resilience across different regions by examining the relationship between observed outage processes and various weather factors;
($iii$) They model the failure and restoration processes separately using detailed records about the components, including failure time and duration. By comparison, we model the observed number of customer outages as a single stochastic process.

There is another line of literature focused on power system resilience to extreme weather events, particularly through fragility modeling and probabilistic risk assessment. For instance, \cite{dobson2023long} introduced Poisson process models for outage and restore processes using North American transmission data. \cite{espinoza2016multi} developed a multi-phase resilience framework for power systems, incorporating fragility curves and Monte Carlo simulations. \cite{hou2022review} reviewed failure risk and outage prediction under wind hazards, emphasizing machine learning techniques. \cite{mukherjee2018multi} proposed a multi-hazard risk assessment model using hybrid SVM-RF techniques for state-level analysis in the U.S. \cite{panteli2015influence} presented a resilience assessment methodology for transmission systems, focusing on fragility curves and adaptation measures. 
\cite{panteli2016power} further provided a detailed resilience framework for extreme weather impacts, integrating fragility modeling, probabilistic assessment, and adaptation strategies for the Great Britain transmission network. \cite{teoh2019probabilistic} developed a risk-informed framework for power distribution poles under wind events, using finite element analysis and fragility functions. 
In contrast to these studies, our approach emphasizes the analysis of large-scale customer outage data to derive resilience metrics, using statistical and machine learning models tailored to spatio-temporal dynamics. While the reviewed literature often focuses on component-level analysis or specific natural hazards, our study addresses a broader range of extreme weather events and their impacts on the grid. 

Other recent theoretical studies, such as \cite{Dobson2007}, have pointed out that the power grid could show vulnerability to massive blackouts, and such vulnerability is mainly attributed to a very small portion of the facilities and external factors \citep{yang2017small}. 
Such characteristics would be very useful for the analysis and enhancement of resilience to weather events, but related quantitative studies on real-world power grids are extremely scarce.

\section{Motivating example of grid resilience}

\begin{figure}[!t]
\centering
\FIGURE{\includegraphics[width=.9\linewidth]{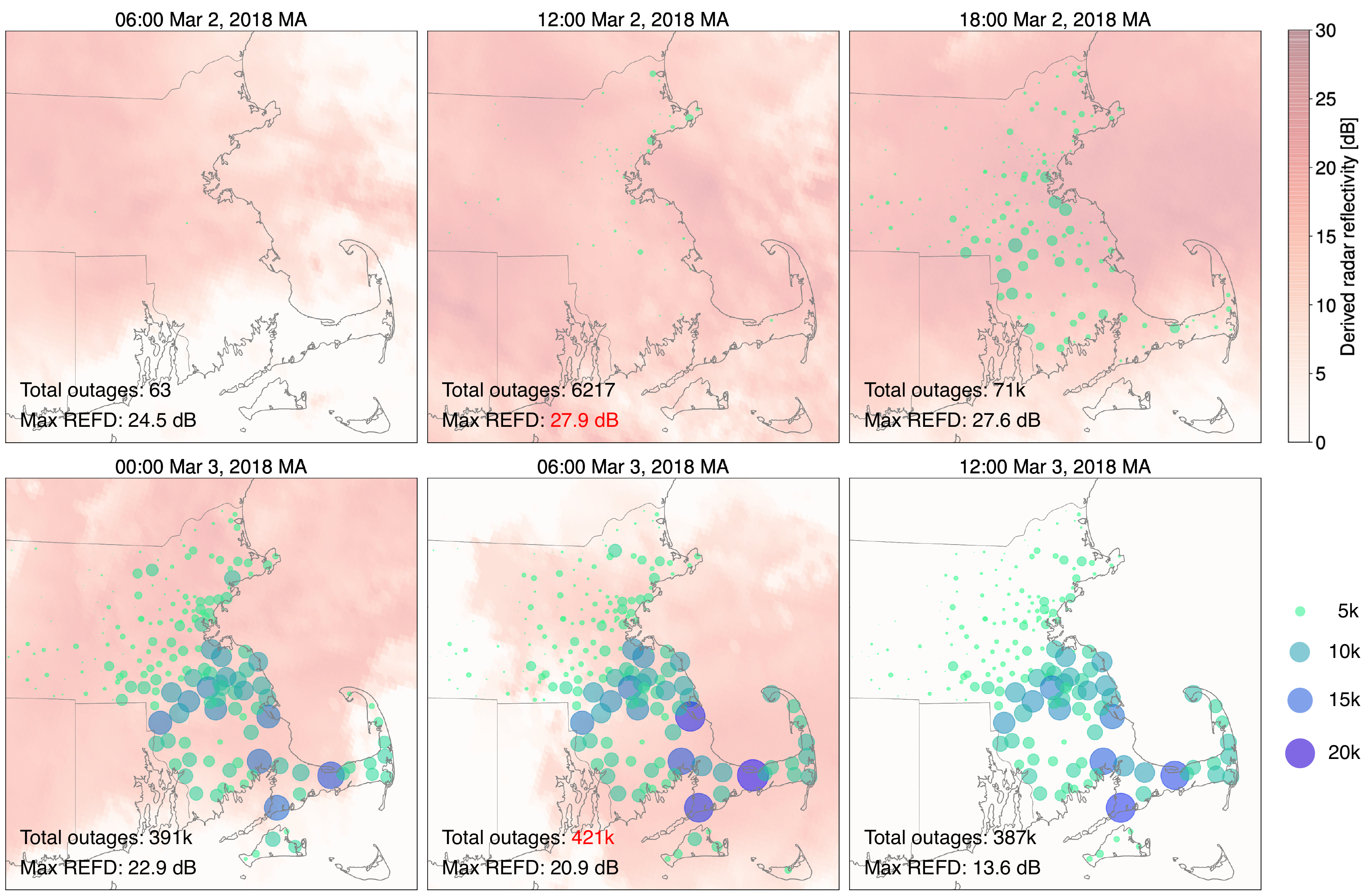}}
{Weather intensity and power outage during the first Nor'easter on March 2-3, 2018, in Massachusetts. \label{fig:motivating-exp-ma}}
{Snapshots of outage and weather maps in every six hours during the first Nor'easter in March 2018, Massachusetts. 
Each bubble corresponds to a geographical unit. 
The size and color depth of the bubbles represent the number of reported customer power outages per 15 minutes. 
The depth of the red cloud over the map represents the level of derived radar reflectivity (REFD) in the corresponding region, which reflects the intensity of the storm. 
The total number of customer power outages and the maximum REFD in the region of the map for each snapshot is shown in the lower left of the figure. The largest total number of outages and the maximum REFD over this period are highlighted in red.}
\end{figure}

\begin{figure}[!t]
\centering
\FIGURE{\begin{subfigure}[h]{0.3\linewidth}
\includegraphics[width=\linewidth]{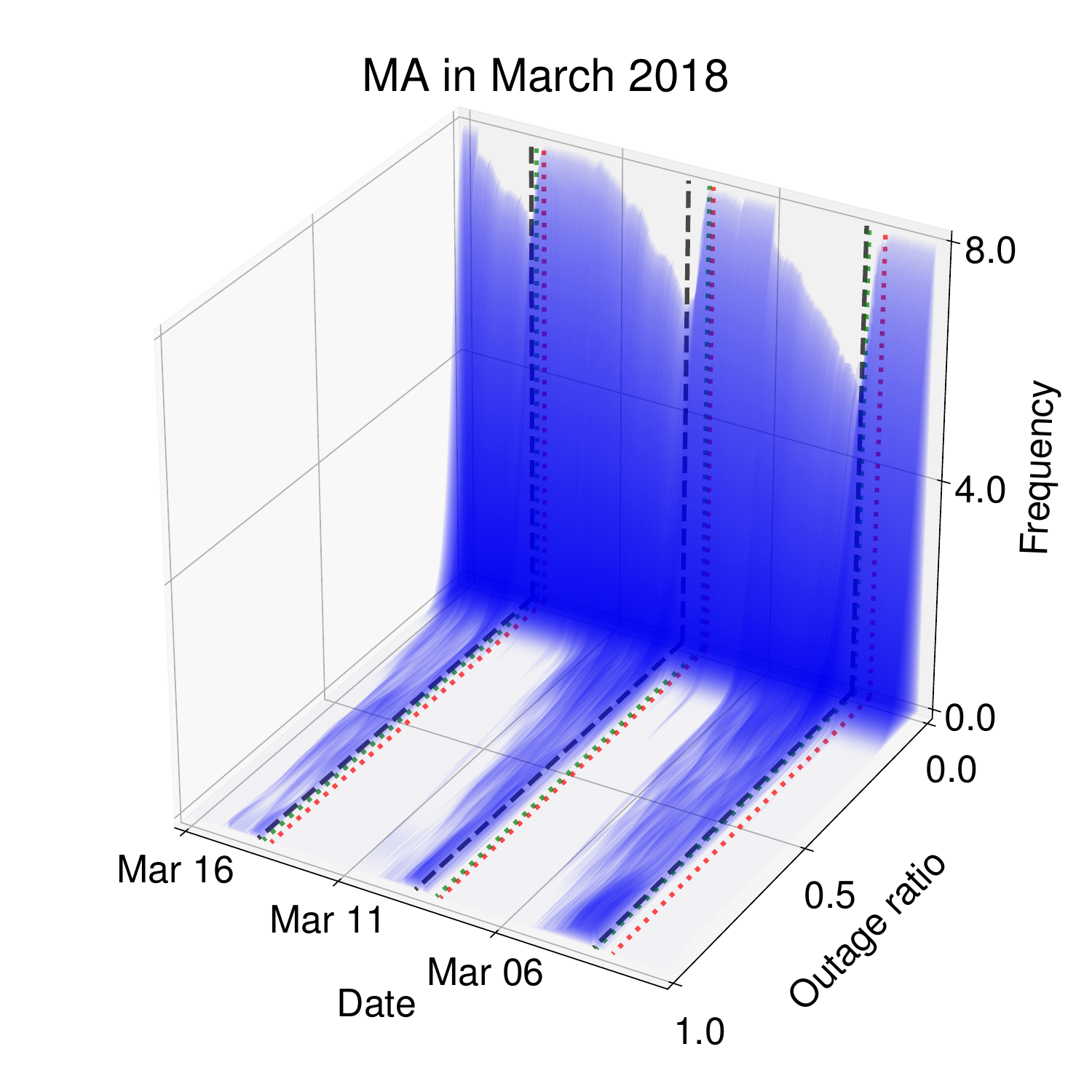}
\end{subfigure}
\begin{subfigure}[h]{0.3\linewidth}
\includegraphics[width=\linewidth]{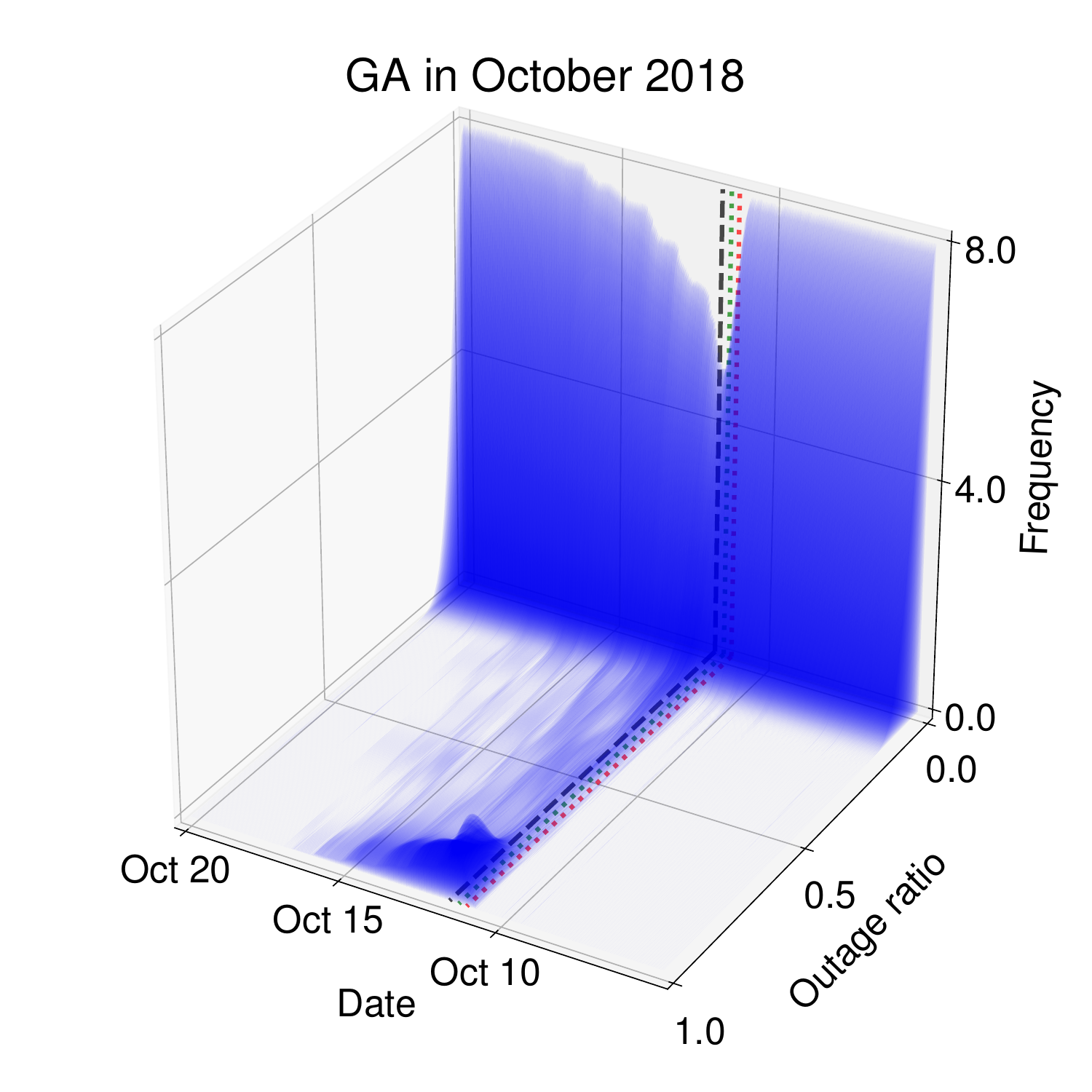}
\end{subfigure}
\begin{subfigure}[h]{0.3\linewidth}
\includegraphics[width=\linewidth]{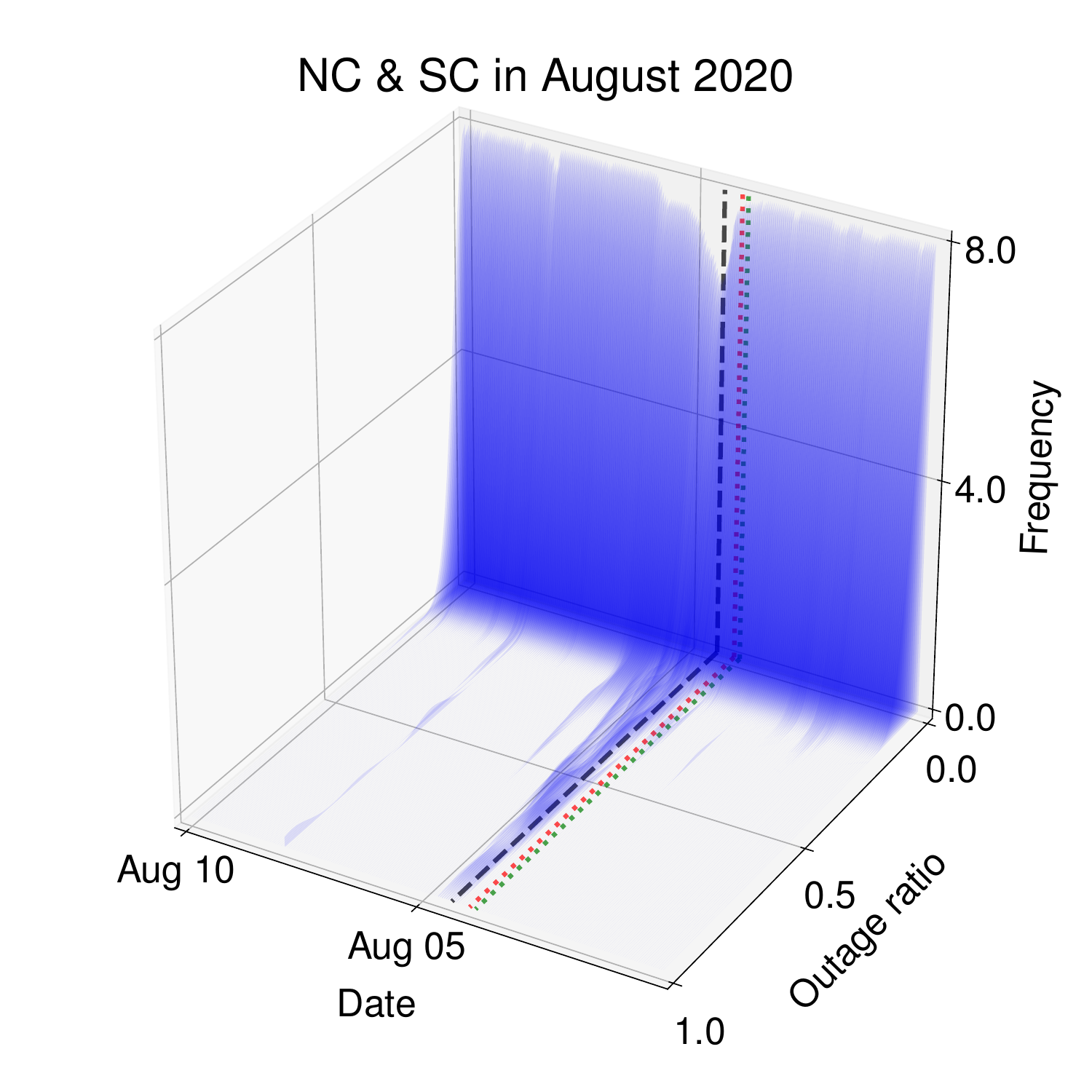}
\end{subfigure}}
{Histogram of outage ratio across all cities in the service territory over time. \label{fig:outage-rate-distribution}}
{The outage ratio refers to the ratio of the number of outages to the total number of customers in a given city.
Black dash lines indicate the moment when the number of customer power outages in the system reaches the peak during the incident. Red and green dotted lines indicate the moment when the weather intensities (REFD and WIND, respectively) reach their peak during extreme weather incidents.}
\end{figure}

We first demonstrate the existence of power grid resilience by presenting a real example as shown in Figure~\ref{fig:motivating-exp-ma}.
It demonstrates a typical large-scale blackout during the period of the first Nor'easter (winter storm) that mainly affected eastern Massachusetts in early March 2018. 
The storm began impacting Massachusetts in the early morning of March 2nd and peaked around noon. The peak wind gust reached hurricane level (about 97 miles per hour) and the storm brought strong precipitation (maximum REFD is 27.9 dB). 
However, large-scale power failures appeared in this area almost 18 hours after the storm peak, which eventually knocked out power to 4.21 million customers in Massachusetts, and resulted in 53 out of 351 cities having an outage ratio larger than 50\% (Figure~\ref{fig:outage-rate-distribution}), according to the outage records.
By that time, the storm had greatly weakened and had been moving out from Massachusetts.

We find a similar behavior from other extreme weather events as illustrated in Figure~\ref{fig:outage-rate-distribution}, where the occurrence of large-scale power outages is normally 12-36 hours behind the peak of the extreme weather.
The time lag between extreme weather and large-scale power outages is a result by the existing power grid resilience and can be mainly explained by the following four aspects:
\begin{enumerate}
    \item Severe power disruptions are the result of cumulative damage in the power system due to various effects caused by extreme weather, which usually build up through time. 
    According to recent outage reports \citep{Campbell2012, FERC2020, Giuliano2018, Kenward2014}, these customer outages were majorly caused by damages to distribution systems. 
    Much of the transmission and distribution networks, across the United States, particularly in the Midwest and Eastern regions, are still above ground, leaving them vulnerable to the effects of extreme weather \citep{Kenward2014}.
    Continuous rainfalls and constant strong winds exert pressure on trees and power poles over time.
    Eventually, electric line poles may fall, downed trees could bring down power lines, or continuous precipitation could raise waterlines and lead to flooding and damage to transformers.
    
    \item Rather than being directly affected/damaged by extreme weather, some areas may see outages hours or even days later, caused by power failures from other connected areas in the system. 
    Such outage interdependence or propagation phenomenon can be observed often due to damages to transmission network infrastructures, which typically consist of a meshed transmission network and carry a significant amount of power transfer (typically more than 100 MVA per transmission line). Therefore, a loss of a transmission line will congest other adjacent lines and what's worse, cause load shedding on a remote load center due to severe congestion on the remaining lines. 
    
    \item Utility companies and system operators are obliged to plan timely restoration for critical power facilities and impose preemptive measures in response to potential damages. These operational restoration measures conducted throughout the downtime can slow the process of large-scale power outages.
    
    \item Power grid has a certain \emph{disruption tolerance capacity} (DTC) within which the power grid can withstand external impacts and does not cause power outages. 
    Such principles are quite common in power grids. The transmission system is operated under N-1 requirements \citep{Ovaere2016}, which means the system is guaranteed secure after losing any one component. 
    The power grid facilities are also designed with some safety margins to endure external hazards up to a certain extent.
    The accumulation of weather impacts, including rainfall, wind, or snow, normally takes hours or even days to exceed the DTC of a power system, after which large-scale outages begin to occur.
\end{enumerate}

Throughout the downtime, as we see from the example, power grid resilience plays a crucial role in withstanding impacts from the extreme weather, slowing the propagation of blackouts, and bringing the blackout areas back to normal.
Motivated by this, we first develop a model with the above presumptions made, aiming to capture such an intricate relationship between the number of outages and the extreme weather effect. 
We then present an analysis based on the fitted model, which quantitatively describes the damages caused by extreme weather and intuitively explains how the system reacts to and recovers from these damages. 

\section{Grid resilience modeling}

This section presents a spatio-temporal model for the number of customer power outages that jointly capture the spatio-temporal dynamics of customer power outages across the region in a data-driven manner. 
We first investigate the impact of extreme weather and define the accumulation of weather effect; 
Then we propose a non-homogeneous multivariate Poisson process for the number of customer power outages across the service territory. 
Assume there are $K$ geographical units in the service territory and $T$ time slots in the time horizon that span the entire event. 
We also consider $M$ different weather variables. 
Let $i, j \in \{1,\dots,K\}$ denote the index of units, $t \in \{1, \dots, T\}$ denote the index of time slots, and $m \in \{1, \dots, M\}$ denote the index of weather variables.
Let $N_{it} \in \mathbb{Z}_+$ be the number of customer power outages we observed in unit $i$ at time $t$ and let $x_{i,t,m} \in \mathbb{R}$ be the value of weather variable $m$ in unit $i$ at time $t$ recorded by weather data.

\begin{figure}[!t]
\centering
\FIGURE{
\centering
  \begin{tabular}[c]{cccc}
    \begin{subfigure}[c]{0.22\linewidth}
    \includegraphics[width=\linewidth]{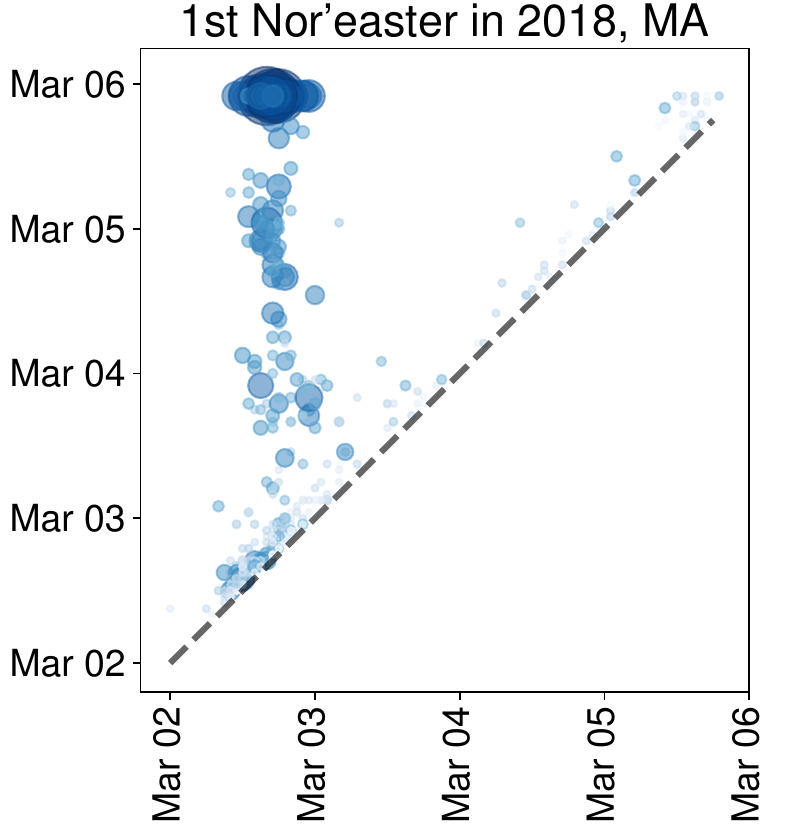}
    \end{subfigure} &
    \begin{subfigure}[c]{0.22\linewidth}
    \includegraphics[width=\linewidth]{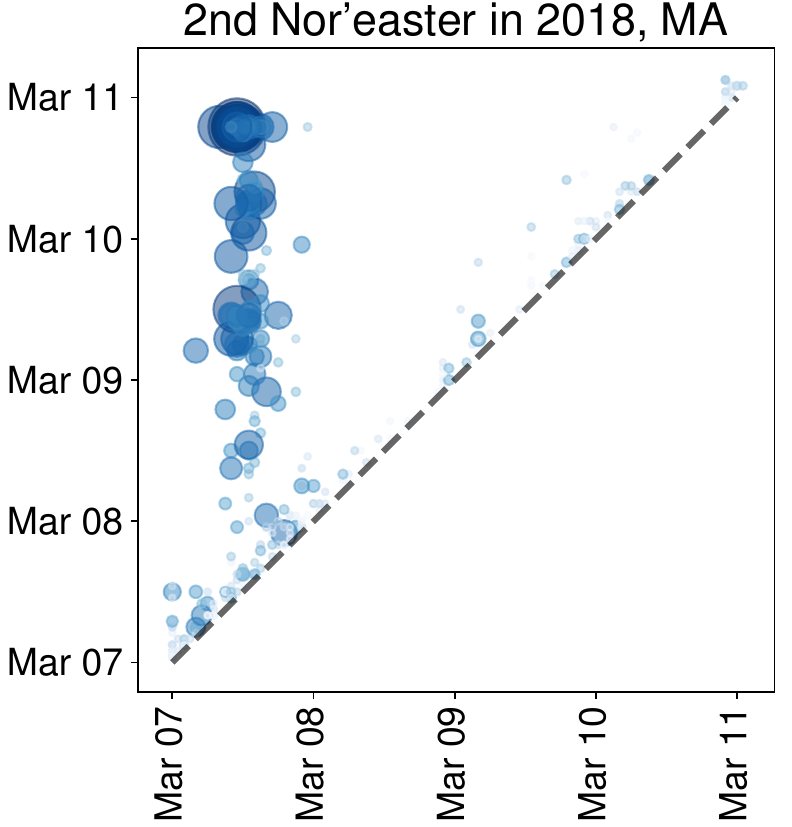}
    \end{subfigure} &
    \begin{subfigure}[c]{0.22\linewidth}
    \includegraphics[width=\linewidth]{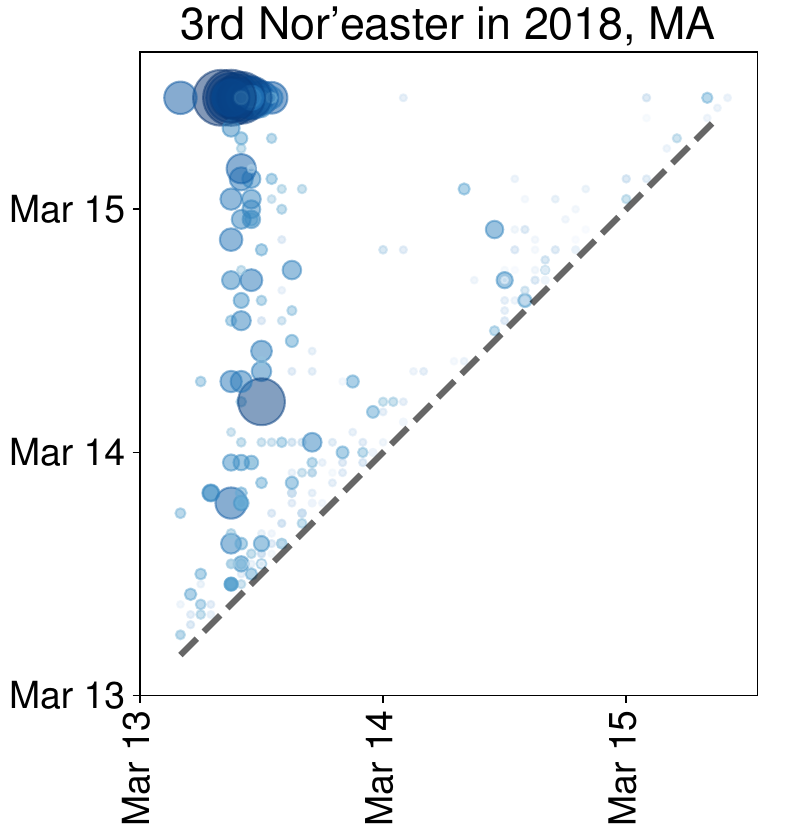}
    \end{subfigure} &
    \begin{subfigure}[c]{0.22\linewidth}
    \includegraphics[width=\linewidth]{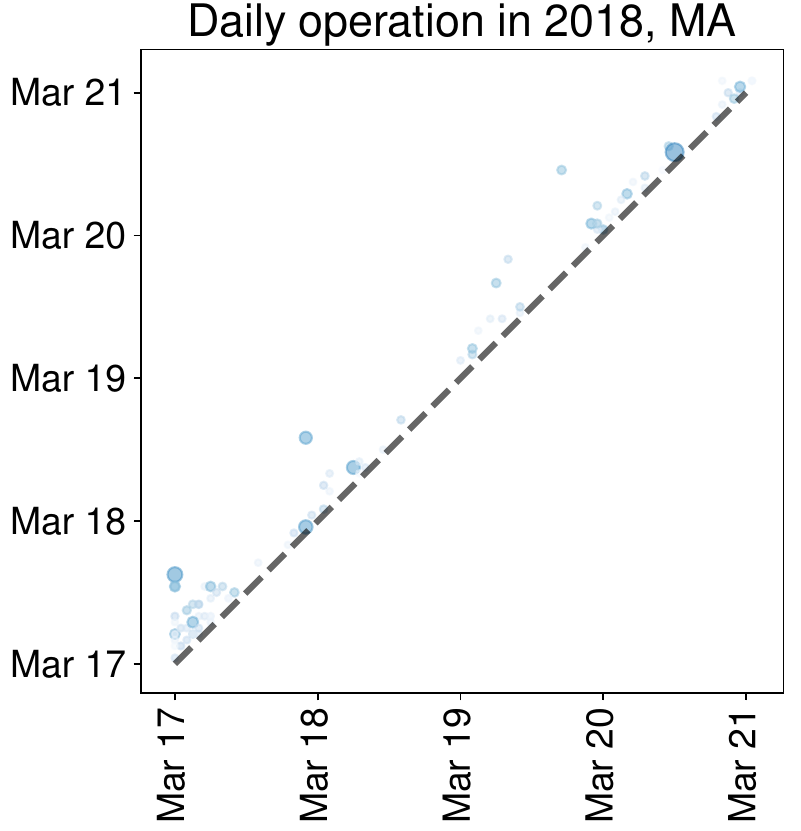}
    \end{subfigure}
    \\
    \begin{subfigure}[c]{0.22\linewidth}
    \includegraphics[width=\linewidth]{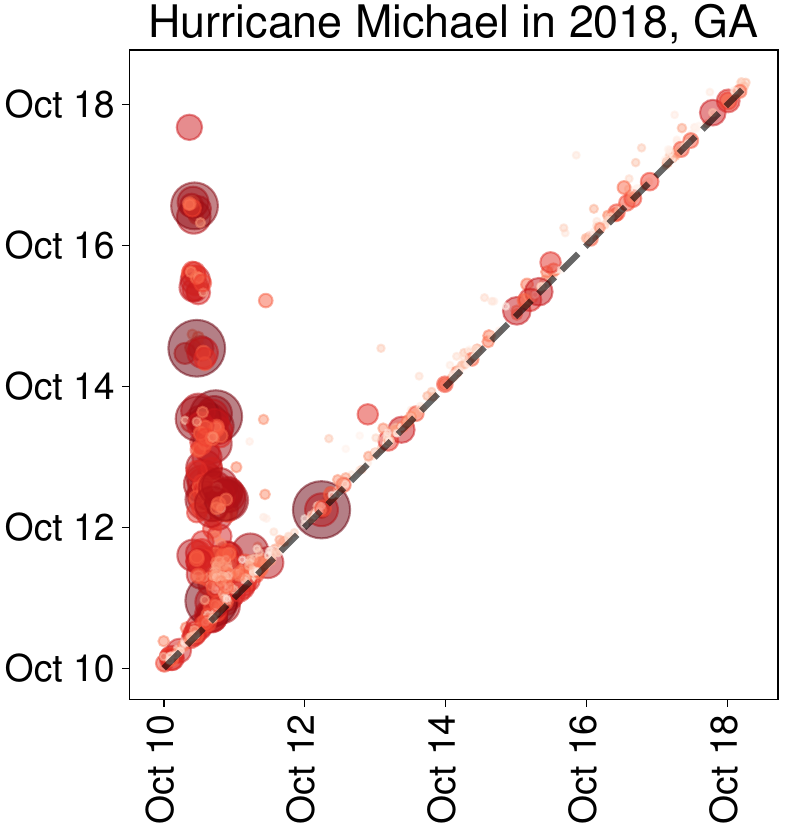}
    \end{subfigure} &
    \begin{subfigure}[c]{0.22\linewidth}
    \includegraphics[width=\linewidth]{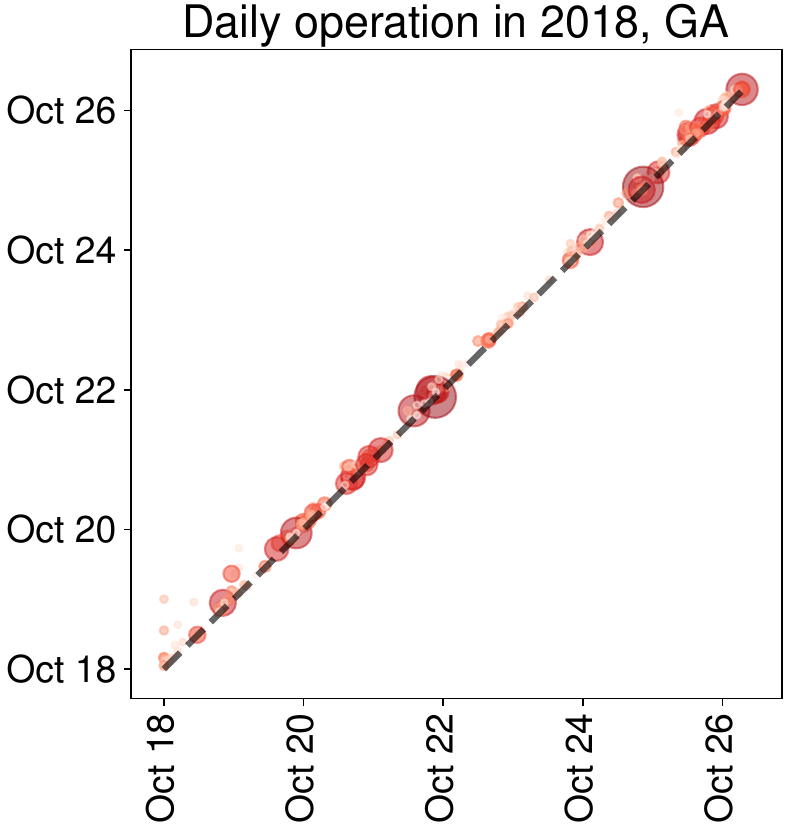}
    \end{subfigure} &
    \begin{subfigure}[c]{0.22\linewidth}
    \includegraphics[width=\linewidth]{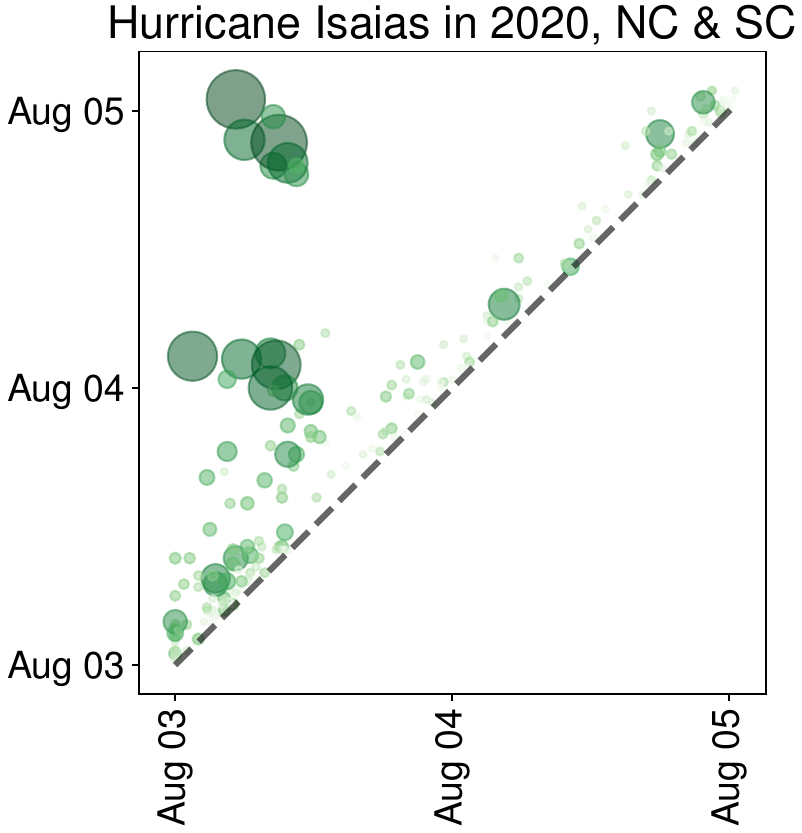}
    \end{subfigure} &
    \begin{subfigure}[c]{0.22\linewidth}
    \includegraphics[width=\linewidth]{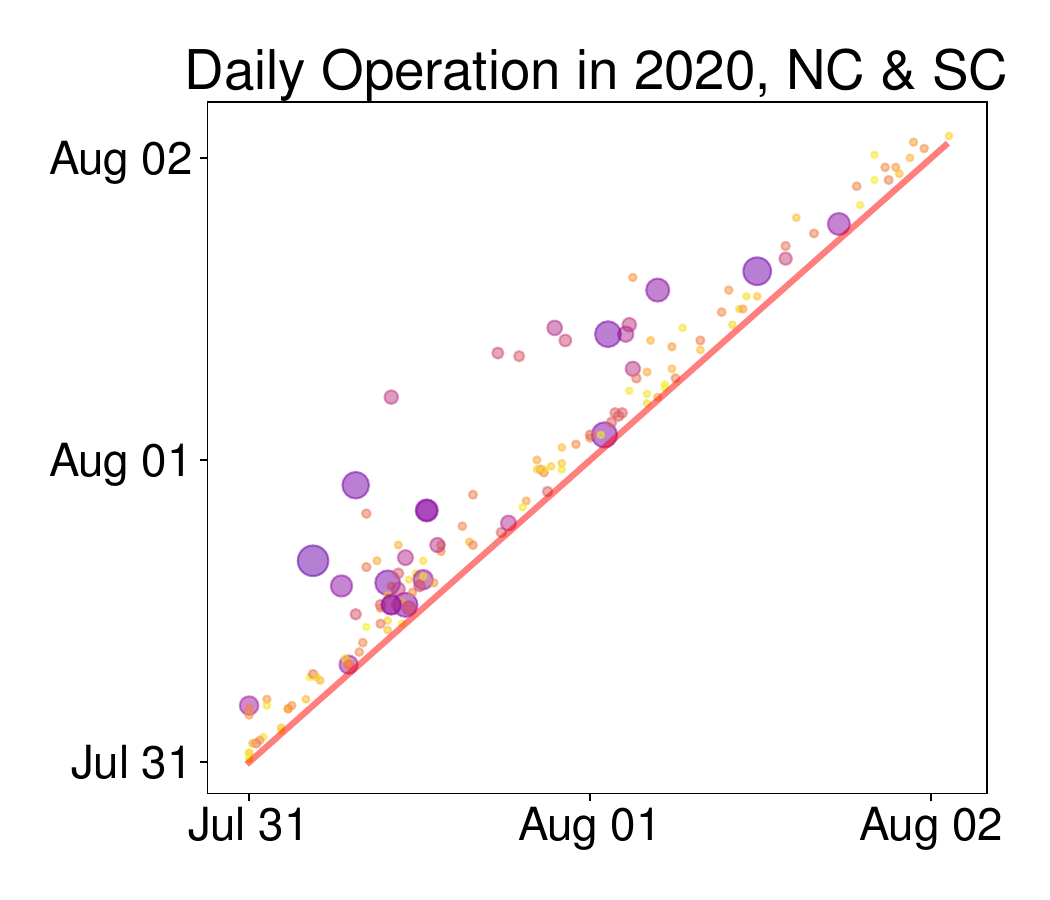}
    \end{subfigure}
  \end{tabular}}
{Scatter plots of the duration of restoration state during five extreme weather events and daily operations. \label{fig:recovery-period}}
{Each bubble corresponds to the duration of the restoration state of a geographical unit. 
The duration is approximated by the time period during which the outage ratio in a city exceeded 5\%. 
The horizontal and vertical coordinates correspond to the start and end time of the restoration, respectively. The distance of the bubbles above the diagonal line indicates the length of the restoration state. The diagonal dashed line represents the length of the restoration state is zero. Bubble sizes represent the maximum number of customer outages that occurred in the corresponding unit. The colors represent different service regions.}
\end{figure}

\subsection{Accumulation of weather effect}

We find that the accumulation of weather effects plays a pivotal role in large-scale power disruptions. 
Regional and temporal weather variations rarely cause long-lasting and severe damages to the power system in daily operations, whereas sustained customer power outages can build up as damages to tree limbs or critical power facilities are excessively accumulated under extreme weather conditions (Figure~\ref{fig:recovery-period}). 
This implies that the impact of immediate weather effect on the system fades away quickly over time. 
We therefore define the accumulation of weather effects as a discounted sum of recent weather influences. 
Here we focus on 34 weather variables defined in weather data (Table~\ref{tab:weather} in Appendix~\ref{append:data}) that characterize the near-surface atmospheric activities, which are directly linked to the weather impacts on power grids. 
Formally, for a weather variable $m$, the accumulation of weather effect $v_{i,t,m}$ in unit $i$ at time $t$ is written as:
\begin{equation}
v_{i,t,m} = \sum_{\tau=t-d+1}^t x_{i,\tau,m} \exp\{ - \omega_m (t - \tau)\}, 
\end{equation}
where $d$ denotes the time window we consider, $x_{i,\tau,m}$ is the value of weather variable $m$ in unit $i$ at time $\tau$, and $\omega_m \ge 0$ is a learnable discount rate for weather variable $m$. The larger the $\omega_m$ is, the faster the weather effect decays; The weather effect does not fade away if $\omega_m = 0$. 

\subsection{Spatio-temporal non-homogeneous Poisson process}

We now model the number of outages without knowing detailed disruption and restoration records. 
It should be noted that outage occurrence and system restoration may occur concurrently and the two opposite dynamics constitute the overall power outage evolution process.
Thus, separately modeling these two dynamics is not practical.
Rather than modeling the physical process, we directly build a spatio-temporal model from customer power outages and weather data, where the number of outages in each geographical unit is regarded as a non-homogeneous Poisson process, and these processes interact with each other in an underlying topological space determined by distance and possible grid connectivity. 

The distribution of customer power outages in a unit at any given time is specified by an \emph{occurrence rate of customer power outage} (we simply refer to it by outage occurrence rate in the rest of the paper).
To be specific, we model the outage occurrence rate in each unit and at arbitrary time as a Poisson process, i.e., $N_{it} \sim \text{Poisson}(\lambda_{it})$, where the outage occurrence rate in unit $i$ at time $t$ can be characterized by $\lambda_{it}$. 
To simulate the customer outages in a power system under hazardous weather conditions, 
we assume that 
($i$) The outage occurrence rate of a unit affected by extreme weather will grow as the accumulation of weather effect builds up until it reaches the system's minimal functionality;
($ii$) A unit with a larger number of customer power outages has a higher chance of raising the outage occurrence rate in its neighboring units (in the topological space) due to their connectivity;
($iii$) The outage occurrence rate of a unit will also decay exponentially over time as restoration plans are being conducted throughout the region. 

\begin{figure}[!t]
\centering
\FIGURE{\includegraphics[width=\linewidth]{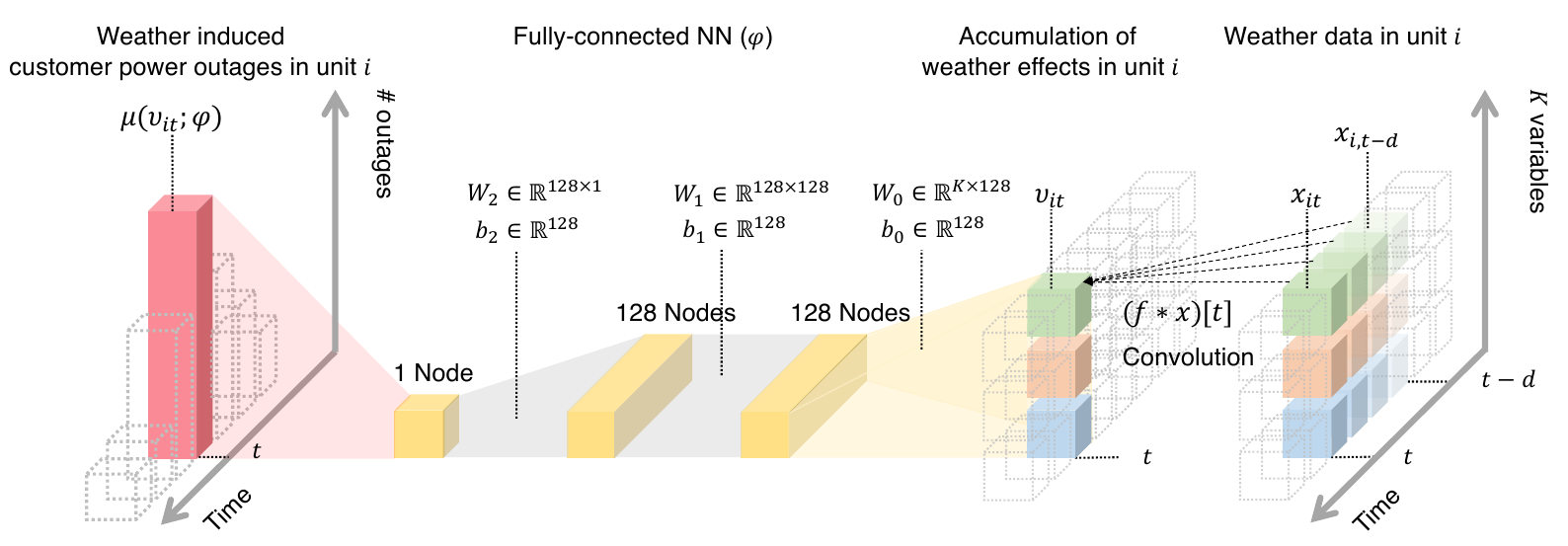}}
{Architecture of the neural network. \label{fig:nn-architecture}}
{}
\end{figure}

We denote the accumulation of all weather variables in unit $i$ at time $t$ as a vector $\boldsymbol{v}_{it} = [v_{i,t,1}, v_{i,t,2}, \dots, v_{1,t,M}]^\top \in \mathbb{R}^M$.
We define the power network in the service territory as a directed graph $\mathcal{G} = (\mathcal{V}, \mathcal{E})$, where $\mathcal{V} = \{i: 1 \le i \le K \}$ is a set of vertices representing the units, and $\mathcal{E} \subseteq \{(i, j): i, j \in \mathcal{V}^2\} $ is a set of directed edges (ordered pairs of vertices), which represents underlying connections between units.
As we assumed that the occurrence of customer power outages can be possibly induced by either the direct impact of cumulative weather effect in the past or indirect impact from previous outages that occurred in the connected units, we define the outage occurrence rate as:
\begin{equation}
    \lambda_{it} = \underbrace{\mystrut{3.4ex} 
        \gamma_i \mu(\boldsymbol{v}_{it};\varphi)
    }_{\substack{\text{direct impact}\\\text{induced by weather}}} + \underbrace{\mystrut{2.ex}
        \sum_{t'<t} \sum_{(i,j) \in \mathcal{E}} g(i, j, t, t')
    }_{\substack{\text{indirect impact}\\\text{from connected units}}},
\end{equation}
where $\gamma_i \ge 0$ is the planning vulnerability of unit $i$ 
and $\mu \ge 0$ is a function that returns the weather-induced outage occurrence rate.
Function $g(i, j, t, t') \ge 0$ is the triggering effects on unit $i$ at time $t$ from previous outages that occurred in unit $j$ at time $t'$.
There are many possible forms of functions $g$. Here we adopt one of the most commonly used forms, which assumes the triggering effects decay exponentially over time \citep{Reinhart2018}. 
For $(i, j) \in \mathcal{E}$ and $t > t'$, we define:
\begin{equation}
    g(i, j, t, t') = \alpha_{ij} N_{jt'} \beta_j e^{-\beta_j (t-t')},
\end{equation}
where $\beta_j \ge 0$ captures the discount rate of the influence; note that the kernel function integrates to one over $t$. 
We assume each edge $(i, j) \in \mathcal{E}$ is associated with a non-negative weight $\alpha_{ij} \ge 0$ indicating the correlation between unit $i$ and $j$. The larger the weight $\alpha_{ij}$ is, the more likely the unit $i$ will be influenced by unit $j$. 
We also assume $\alpha_{ii} = 1,~i \in \mathcal{V}$; thus, $\beta_j$ can be regarded as the recovery rate of unit $j$.
In particular, the term $g(i, i, \cdot, \cdot),~i\in\mathcal{V}$ captures the dynamics of the restoration process of unit $i$ and the term $g(i, j, \cdot, \cdot),~i,j \in \mathcal{V},~i \neq j$ captures the influence of unit $j$ to unit $j$. 
We also disallow the presence of loops between two arbitrary units, i.e., $\alpha_{ij} = 0$ if $\alpha_{ji} \neq 0$.

Here we use a deep neural network (DNN) to approximate $\mu(\cdot; \varphi)$ parametrized by $\varphi$. The neural network's input consists of accumulated weather effects from the past. 
The DNN is introduced to capture complex, non-linear relationships between the input variables (weather values) and the output (outage occurrences) that simpler models may fail to identify. 
The architecture of the neural network is described in Figure~\ref{fig:nn-architecture}. 
The reason of choosing this architecture is two-fold: 
($i$) The first layer is designed to capture the cumulative weather effects by summarizing historical weather conditions through a convolutional structure.
($ii$) The subsequent three layers are developed to model the complex interactions between the accumulated weather effects and the resulting direct damage, measured by the number of customer power outages. 
More details on the DNN's specifications are provided in Appendix~\ref{append:dnn}.

The model can be estimated by maximizing the log-likelihood of the parameters, which can be solved by stochastic gradient descent \citep{Kingma2014}. 
Denote the set of all the parameters $\{\alpha_{ij}\}_{(i,j)\in\mathcal{E},i \neq j}$, $\{\beta_i\}_{i \in \mathcal{V}}$, $\{\gamma_i\}_{i \in \mathcal{V}}$, $\{\omega_i\}_{i \in \mathcal{V}}$, and $\varphi$ as $\theta$.
Denote the corresponding parameter space as $\Theta = \mathbb{R}_+^{K\times K} \times \mathbb{R}_+^K \times \mathbb{R}_+^K \times \mathbb{R}_+^K \times \Phi$, where $\Phi$ is the parameter space of the neural network defined in $\mu$.
Denote all the observed customer power outages and weather data in the studied service territory and time horizon as $N = \{N_{it}\}_{1\le i \le K, 1\le t \le T}$ and $X = \{x_{i,t,m}\}_{1\le i \le K, 1\le t \le T, 1 \le m \le M}$.
Our objective is to find the optimal $\hat \theta$ by maximizing the log-likelihood function:
\begin{equation}
\begin{aligned}
\underset{\theta \in \Theta}{\mbox{maximize}} &\quad - \sum_{t=1}^T \sum_{i=1}^K \lambda_{it}(\theta) + N_{it} \log\left(\lambda_{it}(\theta)\right)\\
\mbox{subject to} 
&\quad \alpha_{ii} = 1,~\quad\forall i,\\
&\quad \alpha_{ij} \alpha_{ji} = 0,~\quad i \neq j.
\end{aligned}
\end{equation}

\section{Numerical results}

This section discusses comprehensive numerical studies based on the fitted model using real data. The detailed data description can be found in Appendix~\ref{append:data}. We first present the predictive performance of our model. Then, we perform a descriptive analysis for extreme weather effects. Lastly, we explain the grid resilience by interpreting the fitted model's parameters. 

\subsection{Predictive performance}


\begin{figure}[!t]
\centering
\FIGURE{\begin{subfigure}[b]{.45\textwidth}
        \includegraphics[width=\linewidth]{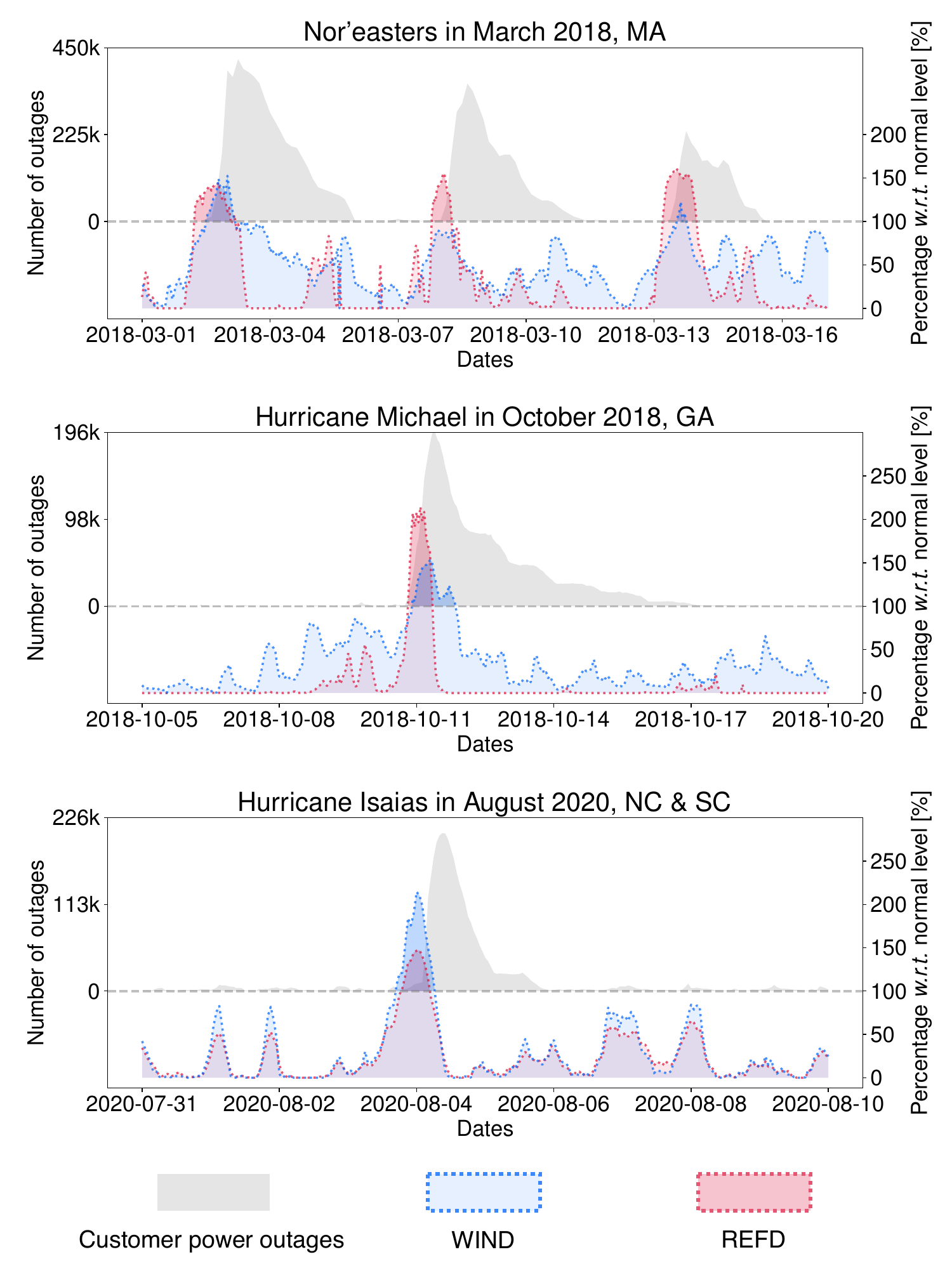}
        \caption{}
    \end{subfigure}
    \begin{subfigure}[b]{.45\textwidth}
        \includegraphics[width=\linewidth]{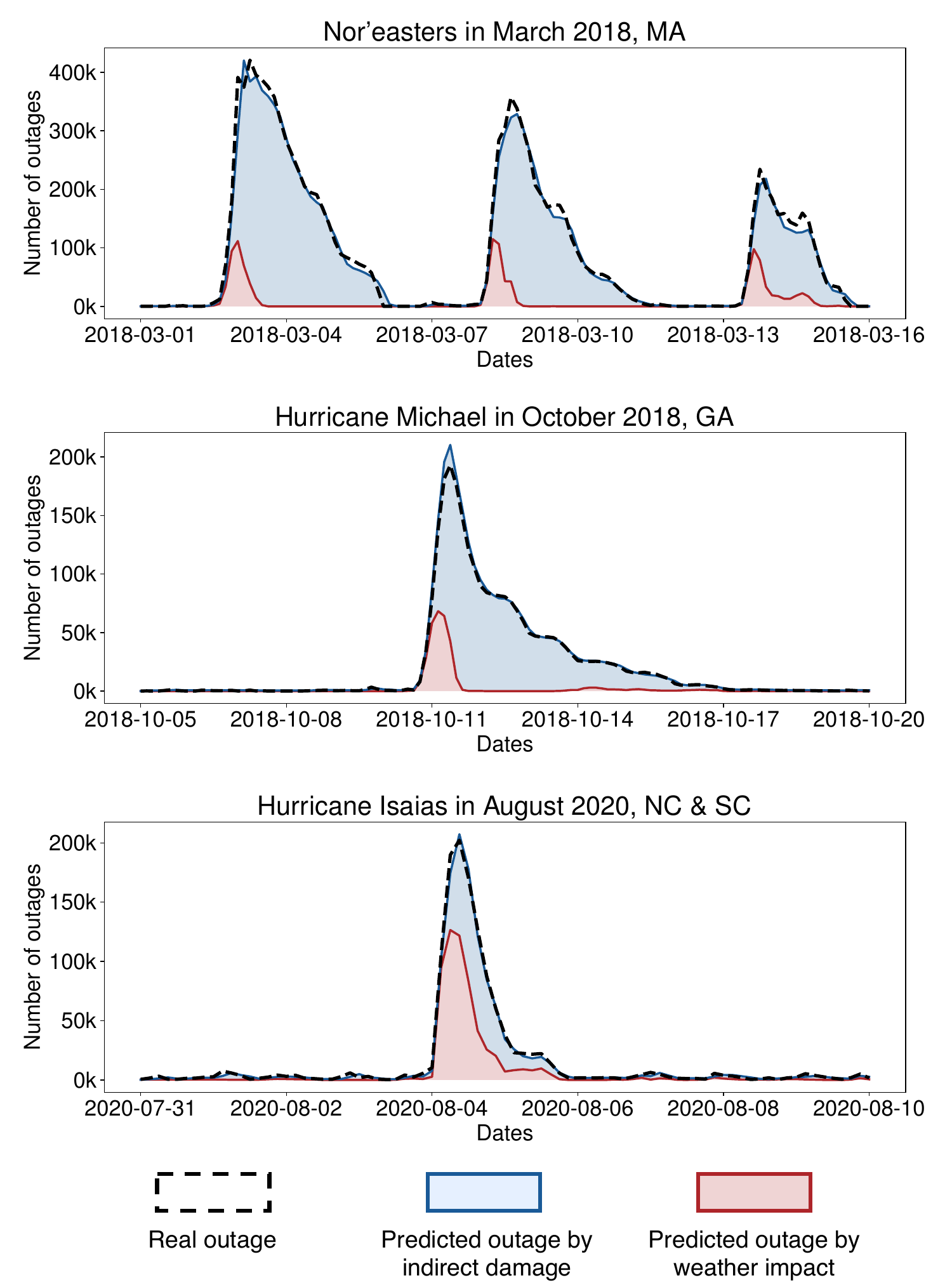}
        \caption{}
    \end{subfigure}}
{Spatially aggregated power outage and weather effect. \label{fig:aggregated-power-outage-vs-weather}}
{(a) Spatially aggregated the total number of customer outages per minute and corresponding two leading weather factors with respect to their maximum normal levels. The left vertical coordinate corresponds to the total number of customer outages per minute reported in the entire service territory, and the right vertical coordinate corresponds to the relative percentage of the weather data with respect to its Three-Sigma limits in the daily operation. (b) Total number of customer outages and corresponding decoupling according to our model estimation. Blue and red regions represent the estimated number of outages induced by indirect damage and the number of outages directly induced by extreme weather, respectively. Black dash lines represent the true total number of customer outages.}
\end{figure}

We fit the model using real data and evaluate its predictive performance. 
We first compare the unit-level in-sample estimation on the number of outages. The in-sample estimation is a process of evaluating the model’s explanatory capabilities using observed data to see how effective the model is in reproducing the data. The process can be carried out as follows: We first fit the model using the entire dataset of a service territory. The in-sample estimation can then be obtained by feeding the same data into the fitted model and recovering the estimation for the number of outages.
The result shows that our model achieves promising accuracy in unit-level prediction for the number of customer power outages three hours ahead.
In addition to the in-sample estimation, we assess the model’s predictive power by performing the one-step-ahead (out-of-sample) prediction. More details on out-of-sample prediction can be found in Appendix~\ref{append:results}.

Figure~\ref{fig:aggregated-power-outage-vs-weather} gives an example of our prediction result (indicated by blue and red region combined) using our model for the three service territories under different weather events, which have achieved promising results compared to the ground truth indicated by black dash lines.
Our result reveals that, in some events (e.g., Nor'easters in March 2018, MA), the direct impact of cumulative extreme weather only causes a small number of power outages in localized areas. However, as indirect damages spread, large-scale outages emerge. In contrast, events like Hurricane Isaias show a more substantial direct impact.
One hypothesis for this discrepancy is that hurricanes typically combine strong winds, heavy rainfall, and storm surges, causing significant immediate damage to power infrastructure. The intense, sustained winds lead to widespread tree falls and damage to power lines and poles, resulting in a higher number of direct outages. Additionally, flooding from heavy rainfall and storm surges can severely impact substations and other critical infrastructure.
In comparison, winter storms primarily cause outages through the accumulation of snow and ice on power lines and trees, leading to line breaks and falls. While this can cause substantial outages, the impact is often less immediate and more gradual as the weight of snow and ice builds up over time.

\begin{figure}[!t]
\centering
\FIGURE{
\centering
  \begin{tabular}[c]{ccc}
    \begin{subfigure}[c]{0.28\linewidth}
    \includegraphics[width=\linewidth]{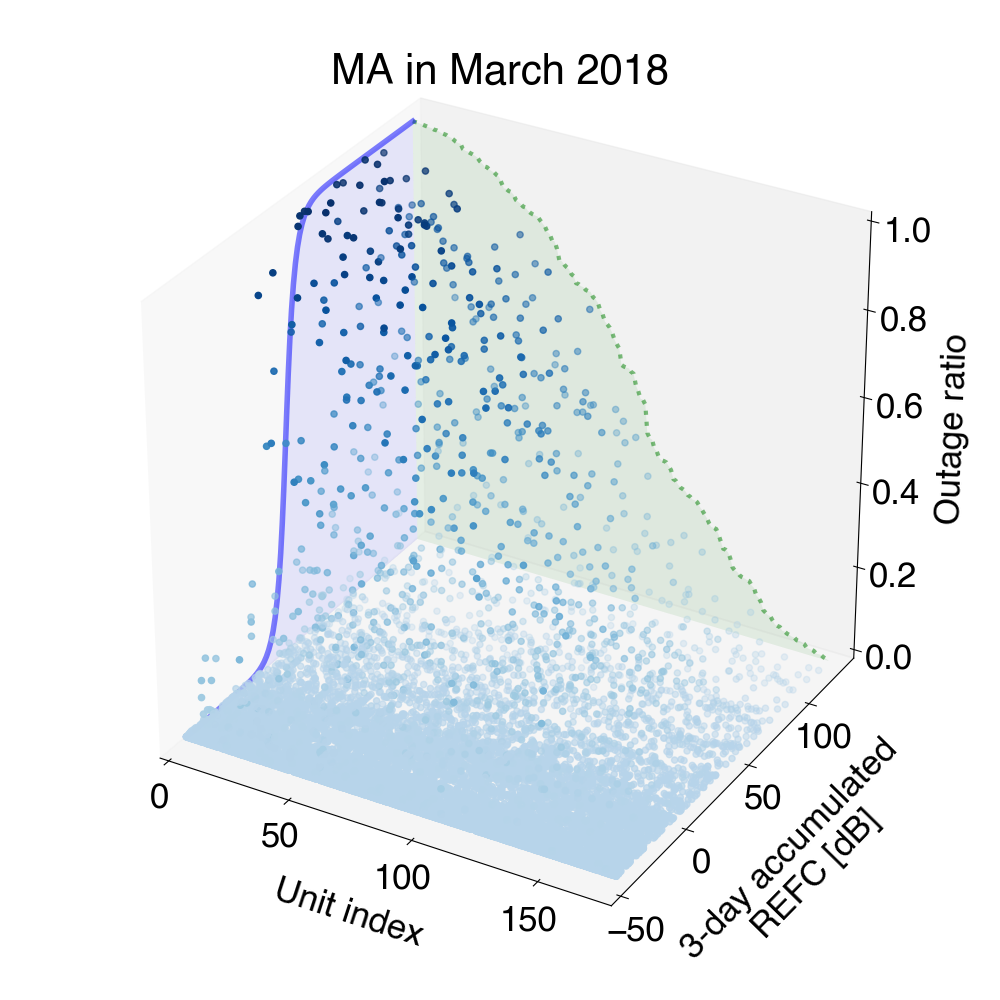}
    \end{subfigure} & 
    \begin{subfigure}[c]{0.28\linewidth}
    \includegraphics[width=\linewidth]{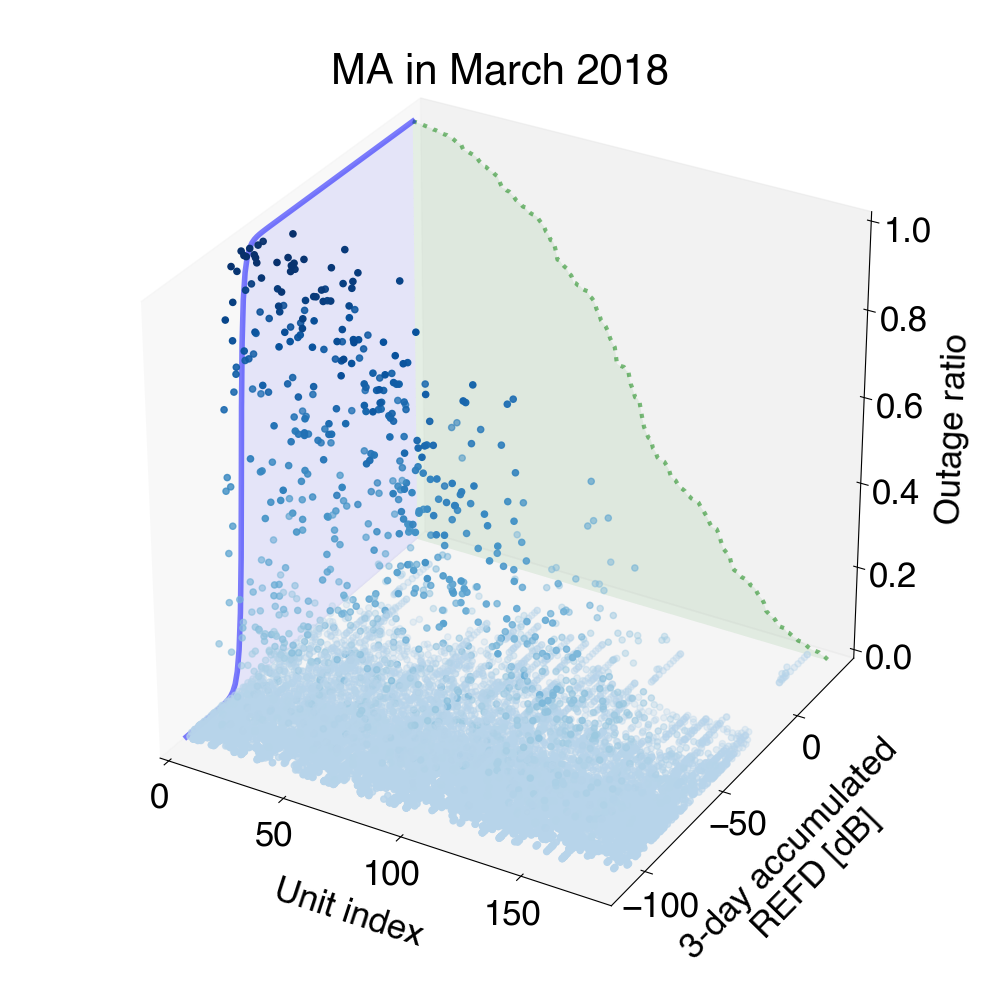}
    \end{subfigure} & 
    \begin{subfigure}[c]{0.28\linewidth}
    \includegraphics[width=\linewidth]{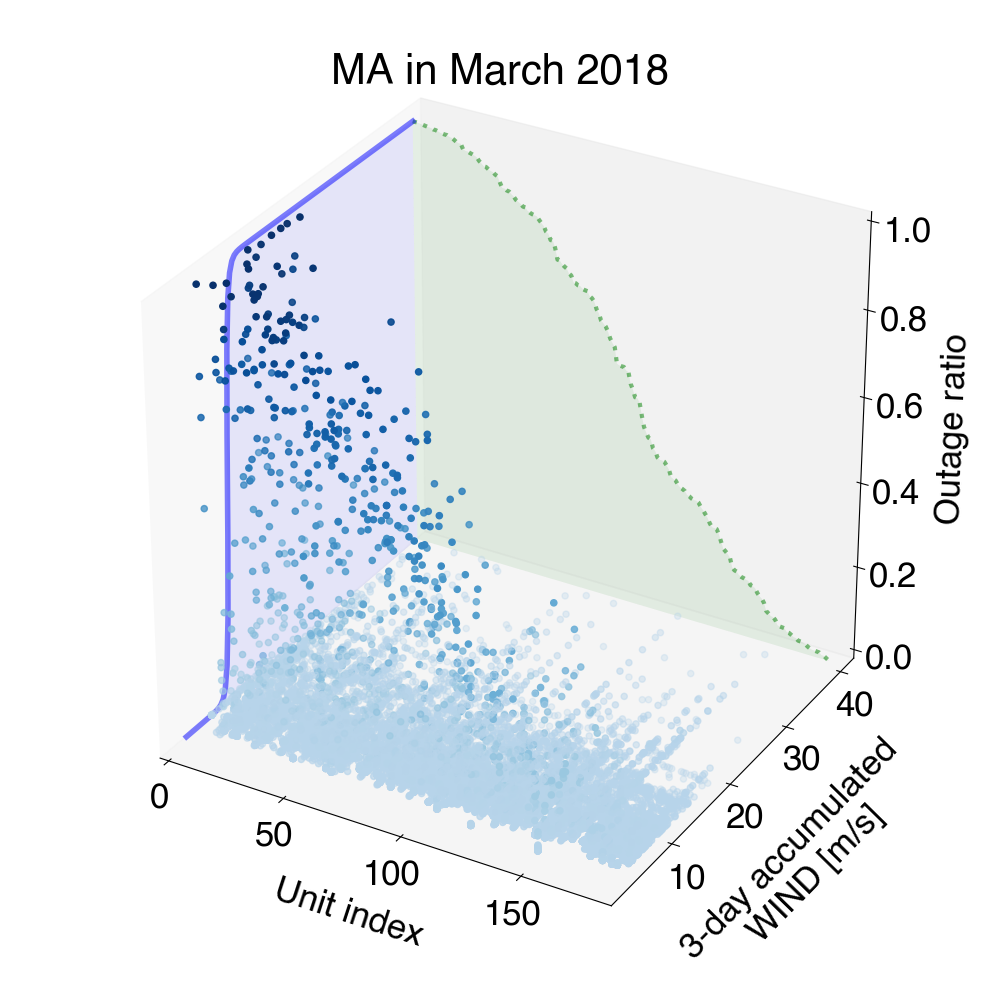}
    \end{subfigure}
    \\
    \begin{subfigure}[c]{0.28\linewidth}
    \includegraphics[width=\linewidth]{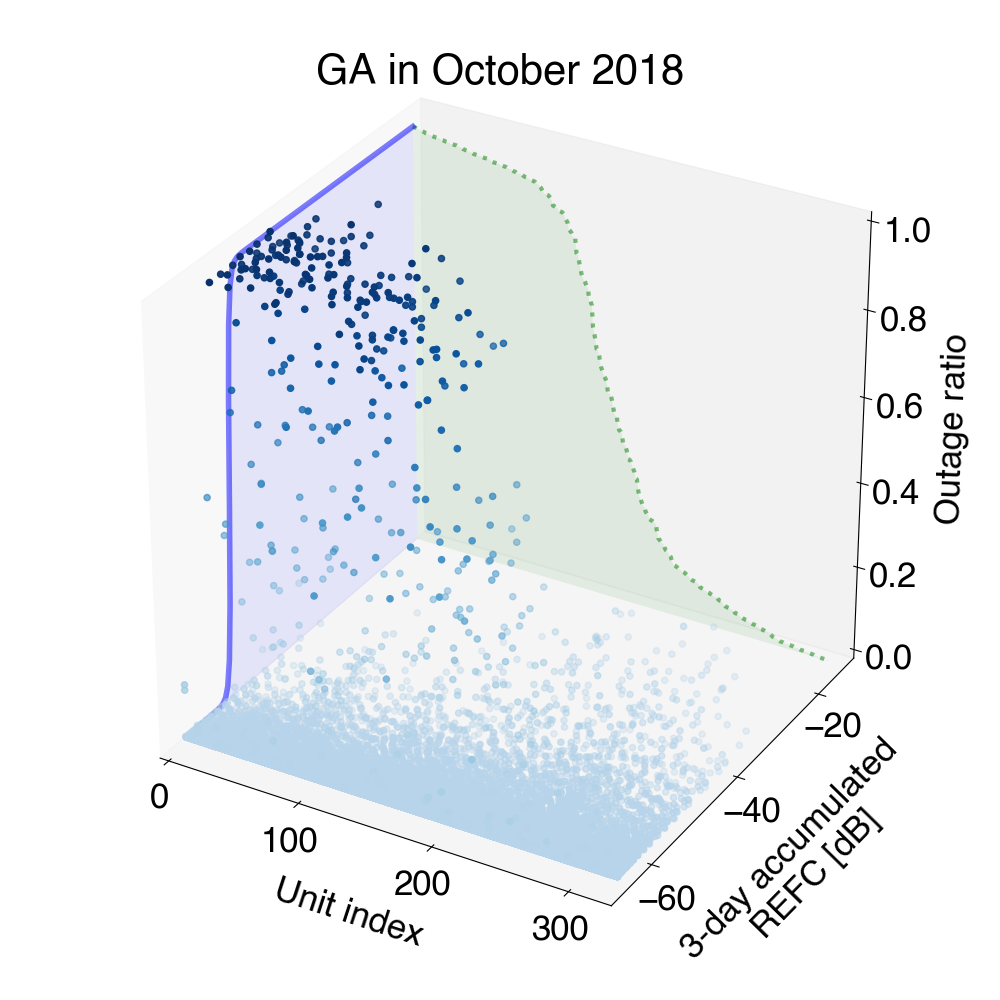}
    \end{subfigure} & 
    \begin{subfigure}[c]{0.28\linewidth}
    \includegraphics[width=\linewidth]{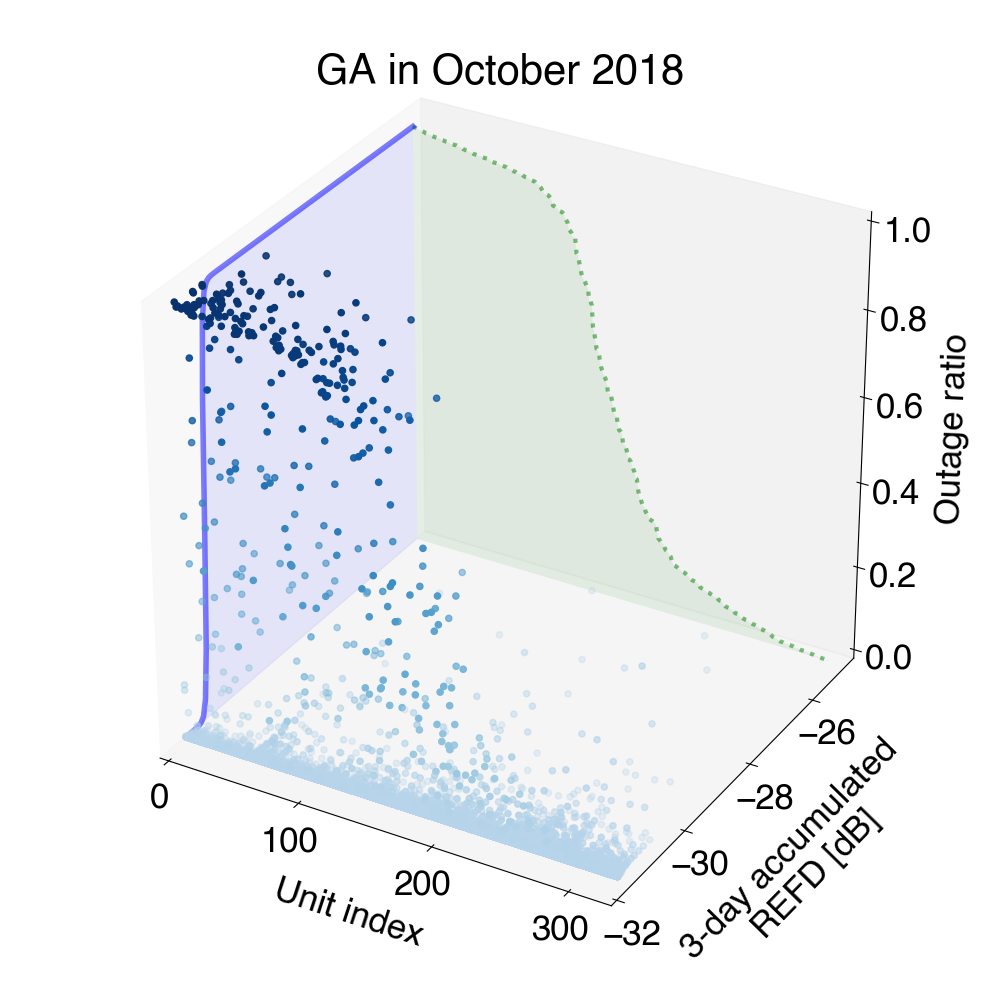}
    \end{subfigure} & 
    \begin{subfigure}[c]{0.28\linewidth}
    \includegraphics[width=\linewidth]{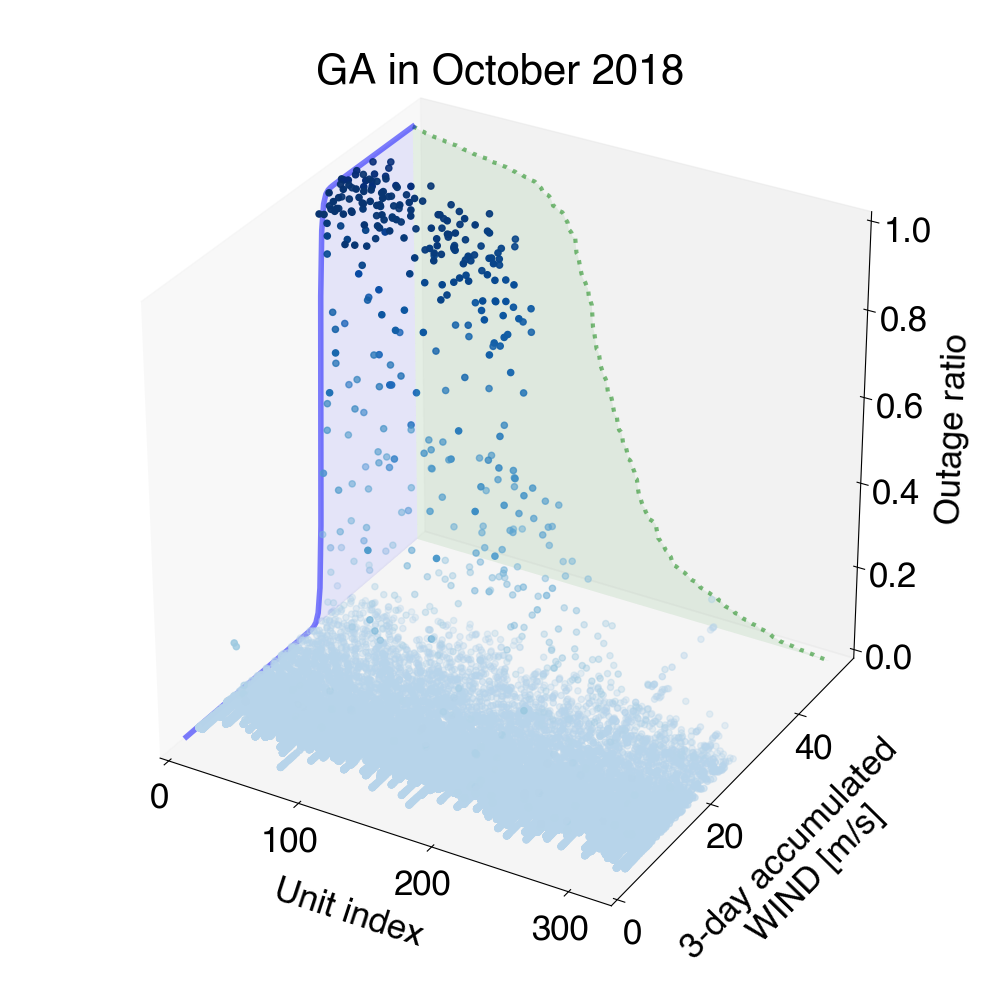}
    \end{subfigure}
    \\
    \begin{subfigure}[c]{0.28\linewidth}
    \includegraphics[width=\linewidth]{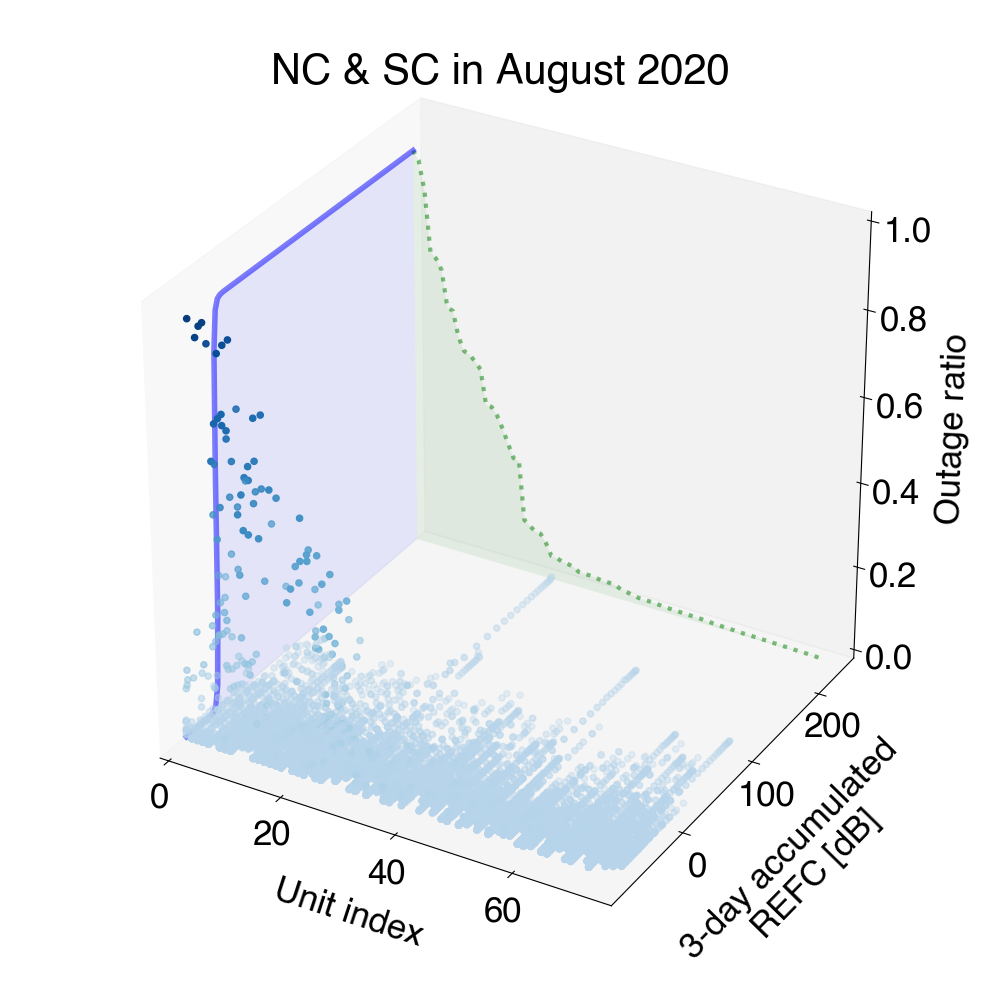}
    \end{subfigure} & 
    \begin{subfigure}[c]{0.28\linewidth}
    \includegraphics[width=\linewidth]{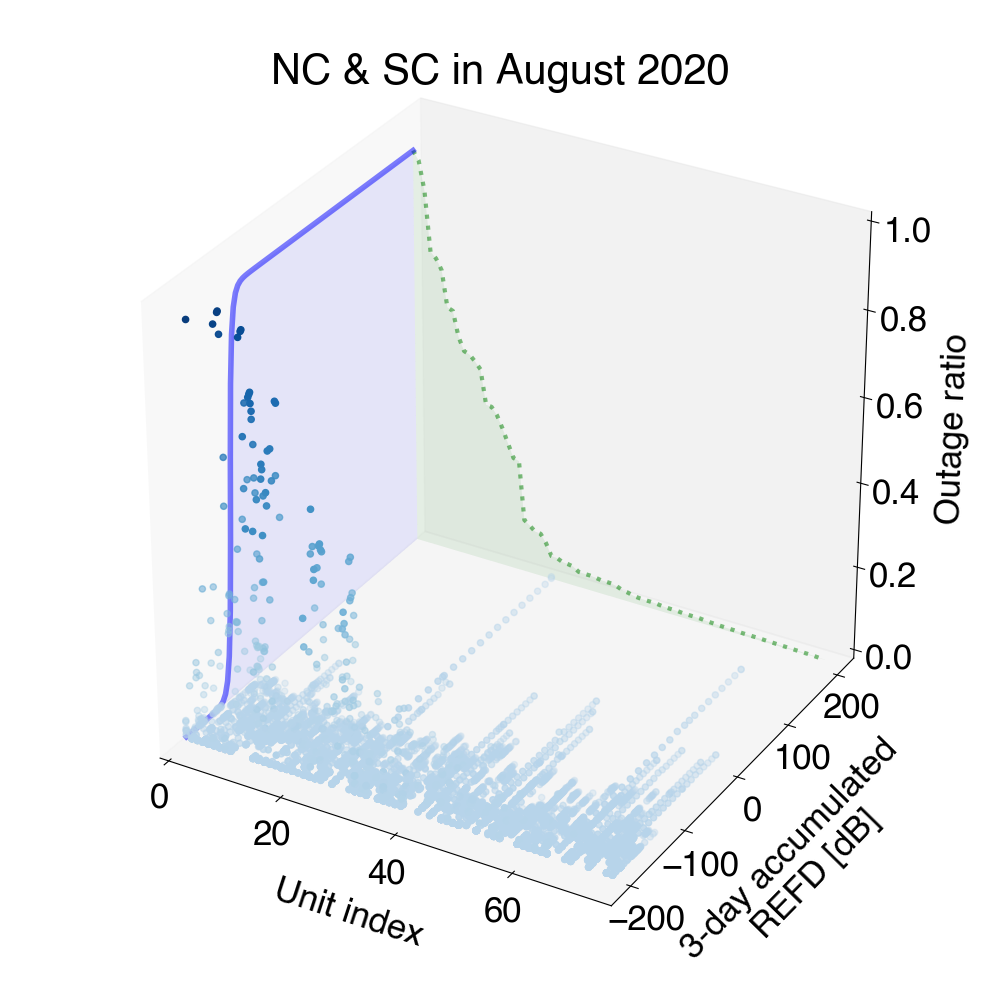}
    \end{subfigure} & 
    \begin{subfigure}[c]{0.28\linewidth}
    \includegraphics[width=\linewidth]{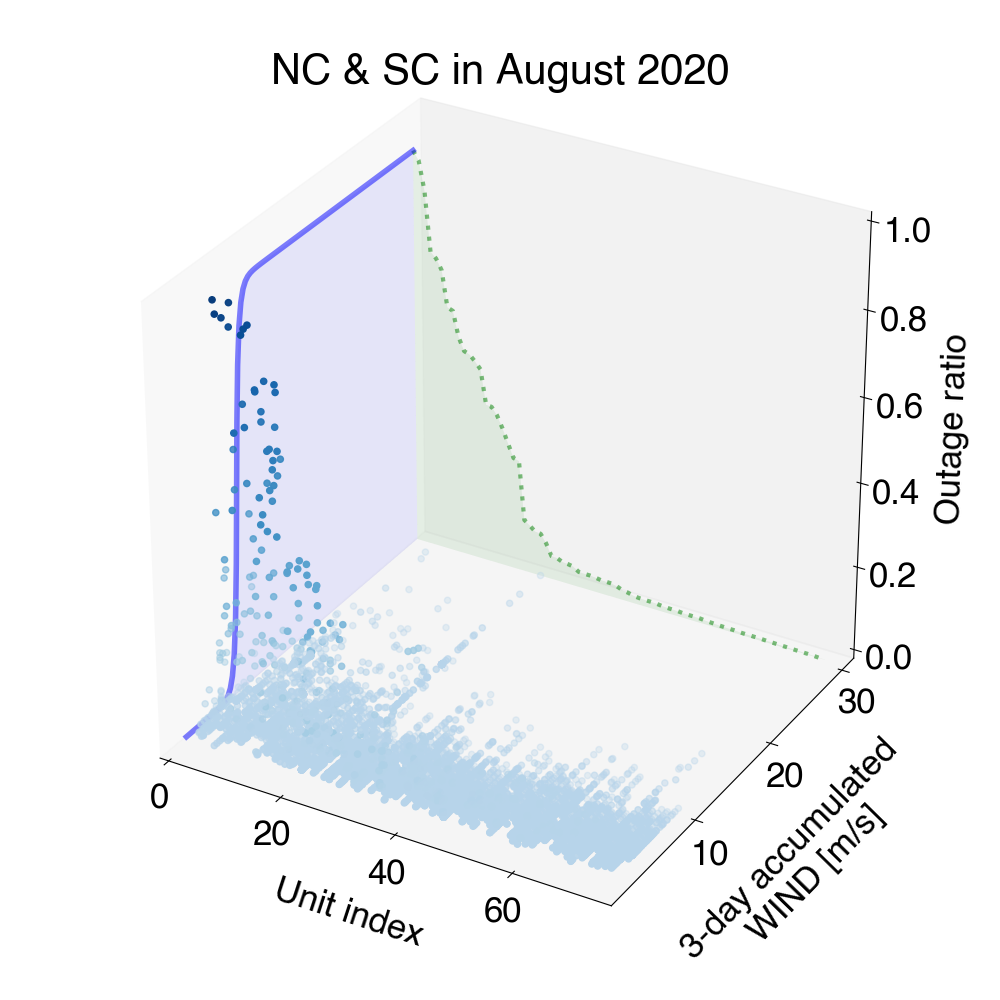}
    \end{subfigure}
  \end{tabular}}
{Accumulation of weather effect versus outage ratio during degradation stage. \label{fig:acc-weather-vs-outage}}
{Scatter plots of outage ratio during the degradation stage given the accumulation of three different types of weather effects.
The vertical and left horizontal coordinates are the outage ratio and unit ID, respectively. Each data point corresponds to the outage ratio (total number of customer outages/total number of customers) of a certain unit at a certain time.
The color depth of data points also indicates the outage ratio.
The unit IDs have been sorted by their historical maximal outage ratios in descending order. 
The right horizontal coordinate is the corresponding accumulation of weather effects in the past three days. 
The blue line on the left vertical plane is a sigmoid curve fitted by all the data points with an outage ratio larger than 0. The green dotted line on the right vertical plane is the maximum outage ratio of all units.}
\end{figure}

\subsection{Descriptive analysis of extreme weather effect}

Now we study the impact of the accumulation of weather effects by analytically evaluating the relation with the outage ratio for the cumulative weather effect in each geographical unit, where the discount rates are learned from the data. 
Here we select three representative weather variables to demonstrate such kind of relations, including composite radar reflectivity (REFC), derived radar reflectivity (REFD), and wind speed (WIND). These features describe the intensity of precipitation and wind, which are among the most relevant factors in weather-induced outages.
We calculate the accumulations of these features for the last three days at every time point during three extreme weather incidents.
Strikingly, the outage ratio stays close to zero until the accumulation of the weather variable reaches a certain ``threshold'' where a sharp jump occurs (indicated by blue lines) as shown in Figure~\ref{fig:acc-weather-vs-outage}. 
Such a trait can be well captured by a sigmoid function with respect to the value of the accumulated weather effect. These ``thresholds'' have a deep connection to DTC of the system for the corresponding weather variable.

The result reveals that power grids have DTC within which the system can well withstand the external impacts without substantial outages. 
Systems with larger capacity are more resilient to weather effects (more accumulated weather effect is needed to initiate large-scale power outages). 
The prevalence of this phenomenon in three major service territories confirms the existence of infrastructural resistance. 
For example, in Figure~\ref{fig:acc-weather-vs-outage}, we can observe that the DTC for wind speed in Massachusetts, North Carolina, and South Carolina (around 10 m/s) is much smaller than that of Georgia (around 30 m/s).
This implies that the power systems in Massachusetts, North Carolina, and South Carolina are more resilient to strong wind compared to the system in Georgia. 
We can also find that there is a considerable difference in the distribution of maximum outage ratios (indicated by green shaded area) among three service territories,
and the maximum outage ratio is negatively related to its DTC.
For example, the extreme weather is devastating to nearly half of the units (296) in Georgia as the maximum outage ratio for these units reaches close to 1 ($> 0.95$), meanwhile, the other half of the units are barely affected by the natural hazardous events (the maximum outage ratio stays around 0). 
As opposed to the polarization phenomenon in Georgia, the outage ratio for the majority of the units in North Carolina and South Carolina remains remarkably low throughout the weather events.
We note that the analyzed outages in North Carolina and South Carolina are most likely due to the damaged distribution network infrastructures, which are spread out and have lower mutual impacts. In other words, the transmission network infrastructures in these territories perform better in terms of withstanding extreme weather events compared with those in Georgia. This could be partly due to that the transmission line structures in North Carolina and South Carolina are required to withstand a gust wind with a speed of 45 m/s or greater as being installed increasingly closer to the coast per the National Electric Safety Code.

\subsection{Resilience quantification}

The resilience can be specifically described from two aspects: infrastructural resistance and operational recoverability, which correspond to the responses of the system facing external impacts and restoration tasks, respectively \citep{Jufri2019}, as shown in Figure~\ref{fig:resilience-trapezoid} (a). 
Infrastructural resistance refers to the inherent ability of the power grid to anticipate, resist, and absorb the effects of hazardous events, and determines a system's DTC.
Operational recoverability indicates the system's recovery capability from severe existing damages induced by extreme weather through a series of operational measures adopted by system operators, utility companies, and other support entities. 
To be specific, infrastructural resistance in our model can be described by the relationship between the accumulation of weather effect and the power system functionality (number of remaining customers with power supply), and 
operational recoverability can be regarded as the speed of a unit recovering from power outages, described by the discount rate of outage occurrence rate.

\begin{figure}[!t]
\centering
\FIGURE{\includegraphics[width=.9\linewidth]{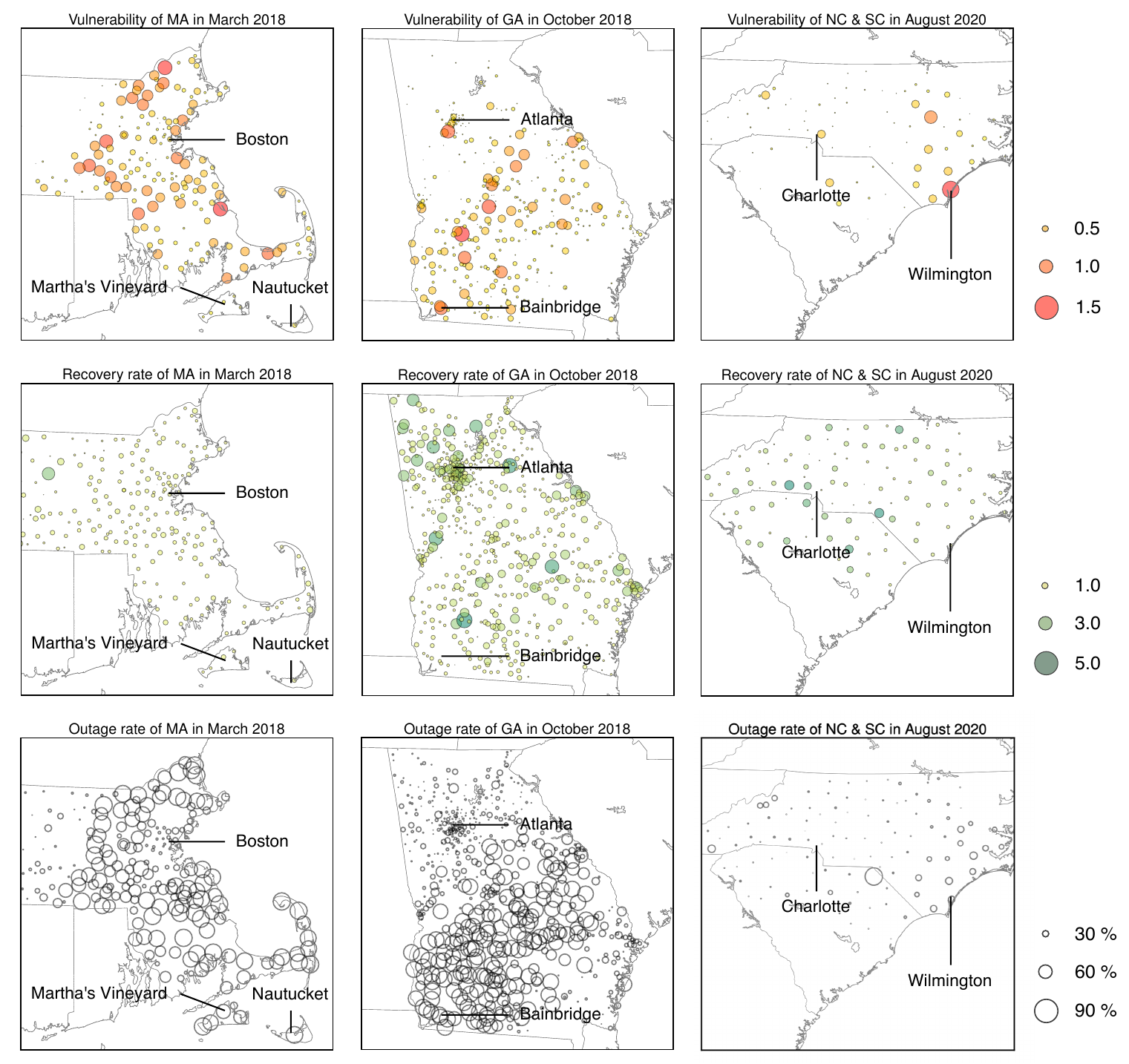}}
{Planning vulnerability and operational recoverability. \label{fig:model-coef}}
{Bubble sizes in each row from top to bottom represent planning vulnerability ($\gamma$), recovery rate ($\beta$), and outage ratio of units during studied events, respectively. 
The value of planning vulnerability represents the average number of outages that occurred in the given unit induced by a unit of cumulative weather effect. 
The number of outages in each unit decays exponentially, where the exponential rate is specified by the recovery rate. }
\end{figure}

\begin{figure}[!t]
\centering
\FIGURE{\includegraphics[width=.8\linewidth]{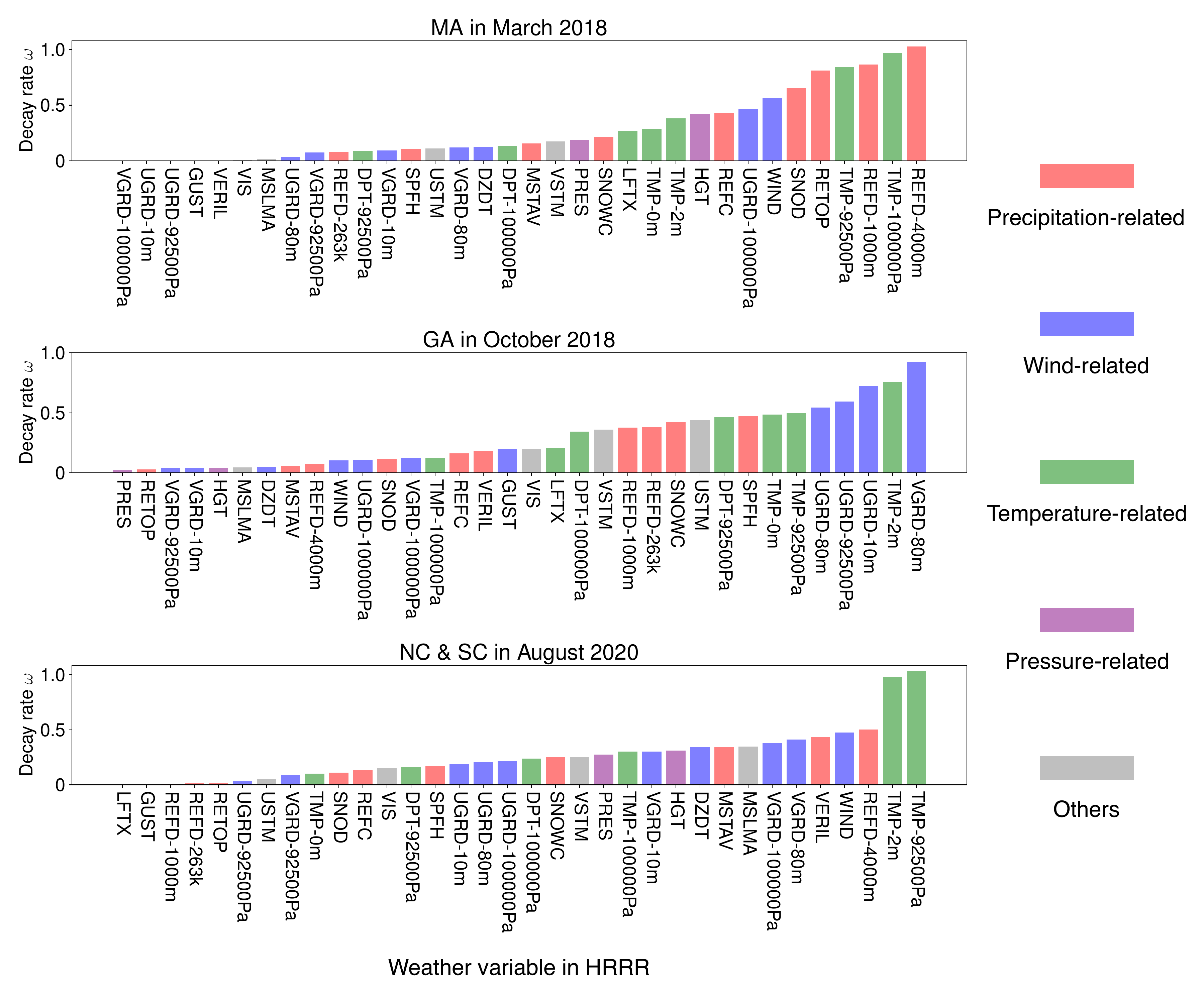}}
{Discount rate in accumulating weather effect. \label{fig:sorted-omega}}
{Bars represent the learned discount rate $\omega$ for each weather variable in the corresponding service territories, where colors indicate their categories. The weather variables have been sorted by their values. }
\end{figure}

\paragraph{Infrastructural resistance}

Our model suggests that there are three factors affecting power systems' infrastructural resistance: \emph{planning vulnerability}, \emph{maintenance sufficiency}, and \emph{criticality} (Figure~\ref{fig:resilience-trapezoid} (b)):

First, planning vulnerability determines the long-term risks of exposure to extreme weather hazards.
In particular, power infrastructural planning, including site selection of power facilities, transmission or distribution route planning, and model selection of power devices, needs to consider the likelihood of natural threats (e.g., high winds, flood, landslide, icing, and vegetation growth). 
For example, power lines going through zones with a high density of tall trees could put the system at a higher risk of outages caused by fallen tree limbs. 
In our model, we introduce a set of trainable parameters $\{\gamma_i\}_{i\in\{1,\dots,K\}}$ to each unit representing their overall planning vulnerabilities to extreme weather, where the coefficient value represents the number of outages occurred in this unit induced by a unit of cumulative weather effect. 
A larger coefficient indicates the corresponding geographical unit is more vulnerable to extreme weather and more likely to suffer a massive blackout when the extreme weather effect is excessively accumulated (Figure~\ref{fig:model-coef}).
It can be observed in the top row of Figure~\ref{fig:model-coef} that the vulnerability indices for metropolitan areas (e.g., Boston, Atlanta, and Charlotte) are relatively low, which is likely because the vegetation in well-developed areas is less dense and therefore, easier to be maintained and less likely to cause faults (e.g., a tree falling on a power line). 

Second, the power industry regularly makes inspections and maintenance for electrical components and facilities to sustain the health of the infrastructure. 
Power grid maintenance refers not just to the maintenance of assets and equipment but also to the environment in the vicinity, e.g., vegetation inspection and trimming.
Insufficient maintenance may leave the system vulnerable to the impact of extreme weather and exacerbate existing planning vulnerabilities \citep{Ji2016}. 
This factor can be captured by the discount rates $\{\omega_{m}\}_{m \in \{1, \dots, M\}}$ defined in the exponential kernel function; the smaller discount rate a unit has, the less maintenance has been completed, and the faster weather effect will be accumulated (Figure~\ref{fig:sorted-omega}). 

\begin{figure}[!t]
\centering
\FIGURE{\includegraphics[width=.9\linewidth]{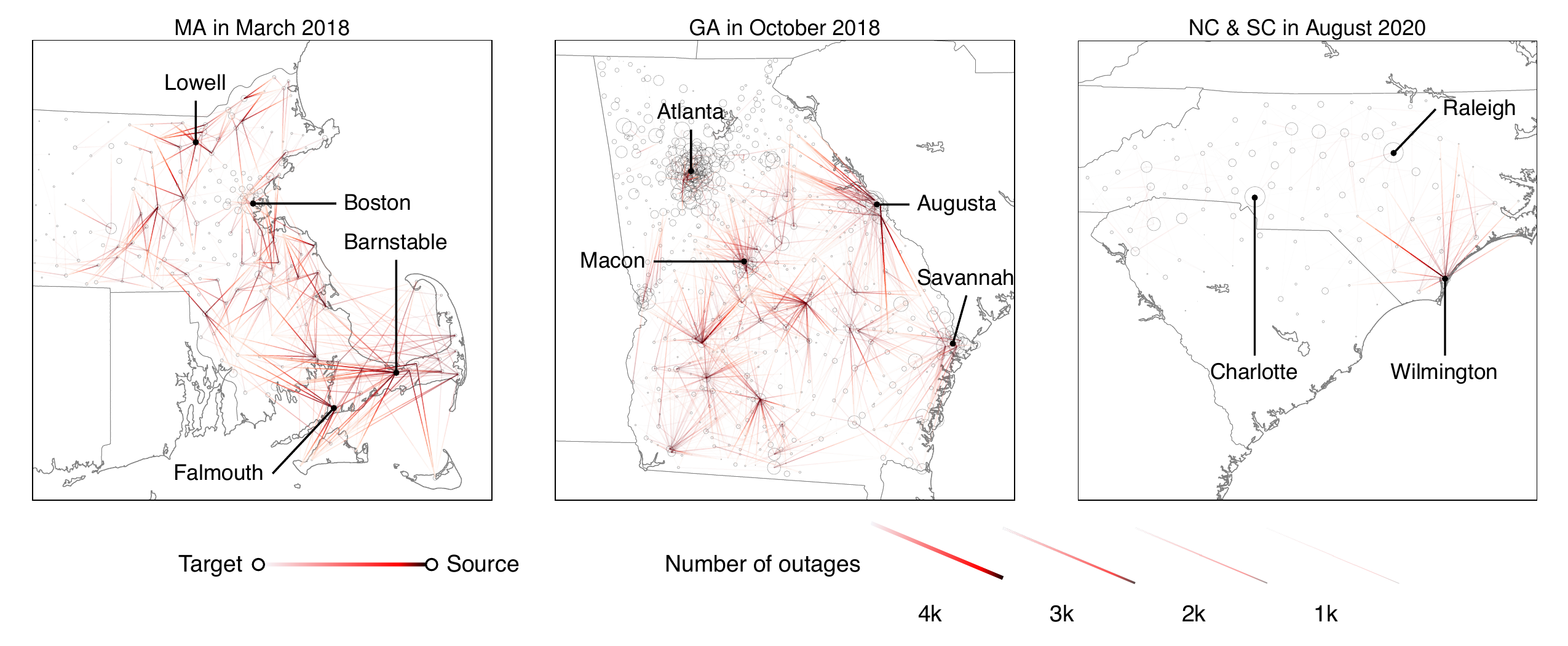}}
{Power outages propagation map. \label{fig:power-network}}
{The map shows the spatial propagation of power outages among geographical units during extreme weather. An edge between two units indicates the power outages that occurred in one unit (light red) resulted in the other (dark red). Edge width and color depth represent the number of customer power outages at the target attributed to the source. The size of the dots represents the total number of customers of the corresponding geographical unit.}
\end{figure}

Last but not least, some of the components or facilities hold critical status in the resilience of a power system.
Modern power grids are interdependent networked systems \citep{Buldyrev2010, Hines2017, Kurant2006, Laprie2007, Panzieri2008, Rinaldi2001}, and failure of some critical portions may not only affect the customers' power supply within their own service territories but also degrade the entire system's resistance ability and lead to failure of other dependent nodes in the same networks. 
This is a common mechanism of cascading failures and blackouts. 
Note that we consider each unit in the power network as a node, and model such connectivity between nodes using a directed graph \citep{Thulasiraman2011}, where directed edges between nodes represent the direction of power outages' propagation and the weight of each edge indicates the average number of outages in the target unit resulted by occurrence of every outage in the source unit. 
As we assumed the increase in the number of power outages at a unit can lead to a proportional increase (determined by their edge value $\alpha_{ij},~i,j\in\{1,\dots,K\}$) in the number of power outages at the unit it connects to, the criticality of a unit can be precisely evaluated by the total number of customer power outages in other directly connected units resulted by its failure.
Our results show disruptions in only a few critical units can lead to large-scale subsequent customer power outages in three service territories (see the units with major outgoing links in Figure~\ref{fig:power-network}).
This indicates that these units are critical in the evolution of power outages: a major portion of the power outages can be directly or indirectly attributed to these units. Such a characteristic is important because it shows some bottlenecks in the resilience of the grid, and people need to concentrate resources on enhancing these critical units.
In addition, we can observe in Figure~\ref{fig:power-network} that the most sources of the outage propagation are the mid-size urban areas (e.g., Barnstable in MA, Macon in GA, and Wilmington in NC) instead of the metropolitan cities (e.g., Boston, Atlanta, and Charlotte) or the rural areas.
As a result, we can conclude that an area that is not a dominant load center (i.e., metropolitan areas), and interconnects multiple transmission facilities or hosts a large generation capacity is more likely to be a power outage propagation source.

\paragraph{Recoverability}

We next discuss the operational recoverability of power systems. 
Note that we assumed the recovery process in each unit reduces the number of outages at an exponential rate, which is captured by another set of parameters $\{\beta_j\}_{j\in\{1,\dots,K\}}$; the parameters may be different in different units.
On a higher level, 92\% of customer power outages recovered within 6 hours, and all customers were back to normal within one day in daily operations, shown in Figure~\ref{fig:recovery-period}. 
However, recoveries from damages induced by extreme weather lasted 4, 4, and 2 days for the three service territories, and units in different regions may also carry out their restoration activities differently. 
The differences in the restoration process may be incurred by various factors, including the infrastructure damage severity, restoration plan efficacy, resources (human, funding, and supplies), and the coordination of restoration progress \citep{National2017}.
For example, six months after Hurricane Maria hit Puerto Rico, more than 10 percent of the island’s residents were still without power \citep{Baez2018}; 
Similarly, it took the U.S. Virgin Islands more than four months to fully restore power after being hit by Hurricanes Irma and Maria. 
Meanwhile, the recovery in the mainland of the US has been much quicker --- in part because the resources are more abundant. 
In particular, after Irma hit the Southeast US, knocking out power to 7.8 million customers, 60,000 utility workers from across the country deployed and restored power to 97 percent of the population in about a week \citep{NPR2017}. 

According to the fitted model, the recovery rates for urban areas are usually higher than those for rural areas because it is easier to find repair personnel, identify fault locations, and access and repair a fault that occurs in an urban area. 
Note that a rural area generally lacks of resources and includes terrains that are challenging for repair personnel to access and manage (e.g., mountains, rivers, forests). This explains why the recovery rates for the Atlanta areas are relatively higher. As a result of lower venerability and higher recoverability, the outage rates for metropolitan areas (e.g., Boston, Atlanta, and Charlotte) should be relatively low, which are verified by the outage rate maps on the bottom row of Figure~\ref{fig:model-coef}.
Combined with the published transmission and generation information, we provide more policy insights derived from our numerical study in Appendix~\ref{append:policy-insights}.


\subsection{Resilience enhancement analysis}

A large body of previous efforts \citep{AbiSamra2013, Jufri2019, Korkali2017, Mahzarnia2020, Mureddu2016, National2017, Executive2013, Shea2018} have assessed a variety of techniques that can be employed before an event occurs in an effort to enhance system resilience.
These include but are not limited to
(1) Improving system architectures to further reduce the criticality of individual units in the power network, 
(2) Enhancing the health and reliability of the individual units,
(3) Making a restoration plan ahead of events to speed up help to those who may need it. 
(4) asset health monitoring and preventive- and reliability-centered maintenance.
However, examining and evaluating the effectiveness of resilience enhancement strategies could be technically intractable. 
There are a number of reasons:
first, due to the low-frequency nature of extreme weather events, there is a lack of weather data and corresponding power outage records for such evaluation;
Second, the uniqueness of each extreme weather event and its ripple effects hinders the reproducibility of large-scale outages in resilience study;
Last but not least, accurate evaluation of these strategies requires a partnership across all levels of government and the private sectors, which is time- and cost-consuming to be carried out in practical terms. 

\begin{figure}[!t]
\centering
\FIGURE{
\centering
  \begin{tabular}[c]{ccc}
    \begin{subfigure}[c]{0.3\linewidth}
    \includegraphics[width=\linewidth]{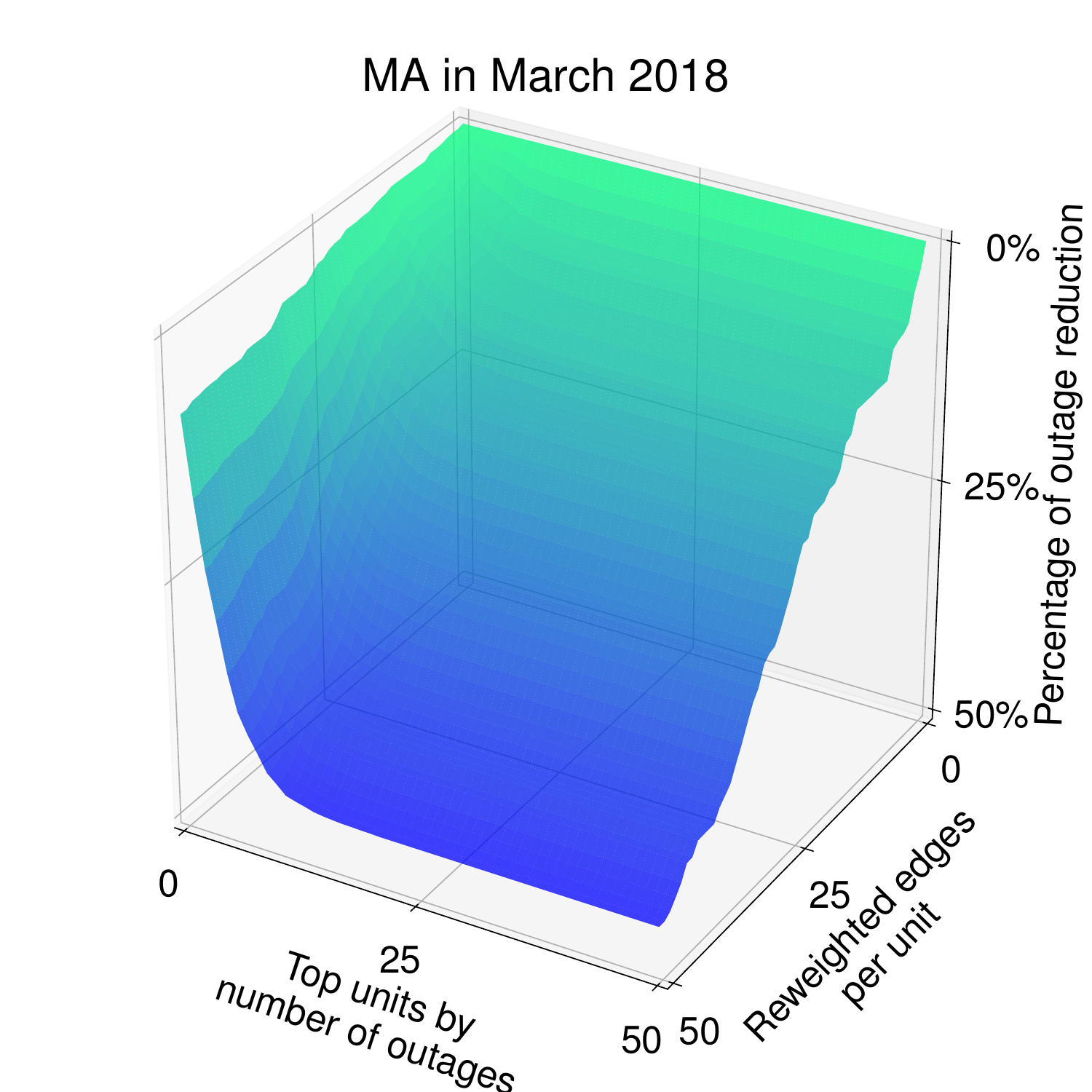}
    \end{subfigure} & 
    \begin{subfigure}[c]{0.3\linewidth}
    \includegraphics[width=\linewidth]{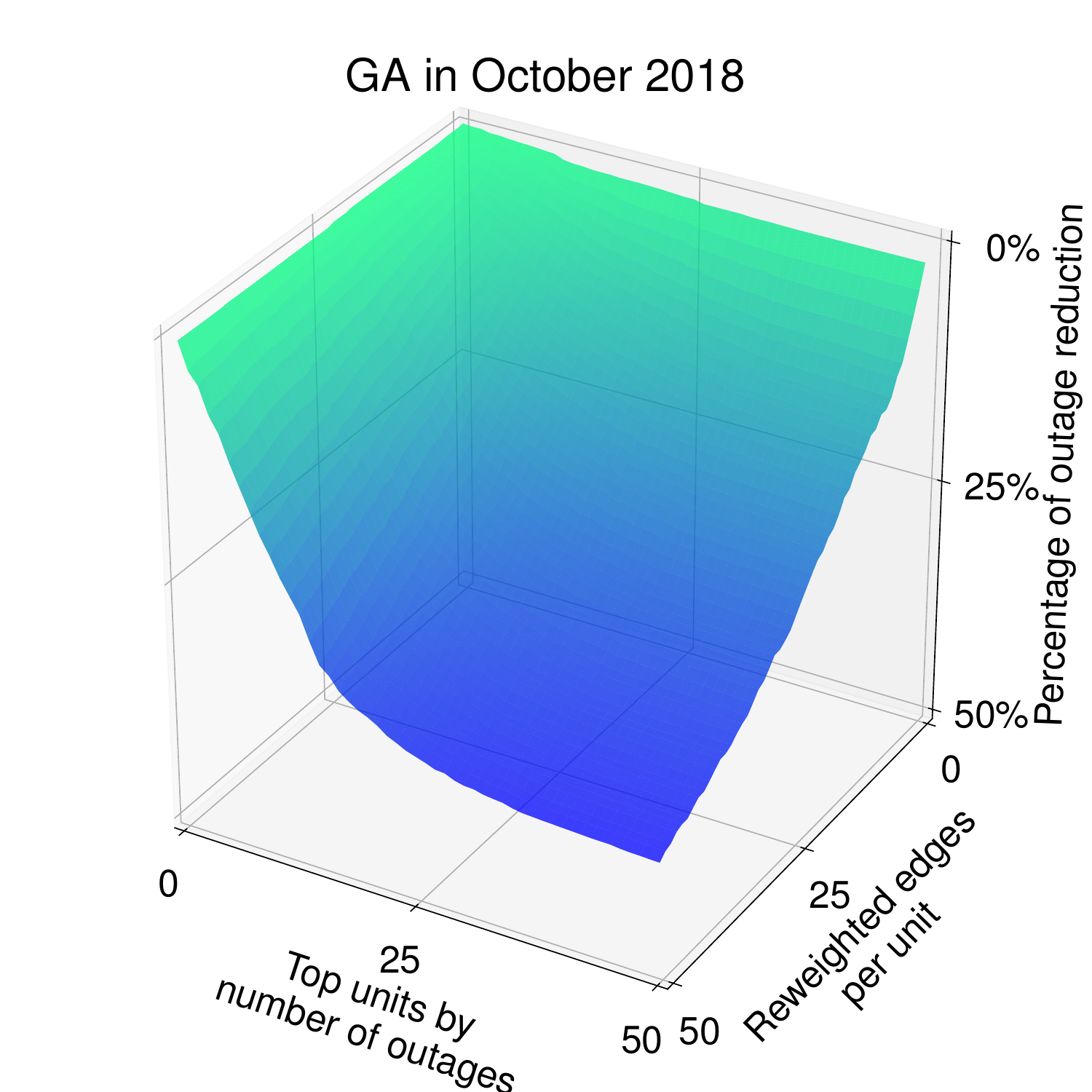}
    \end{subfigure} & 
    \begin{subfigure}[c]{0.3\linewidth}
    \includegraphics[width=\linewidth]{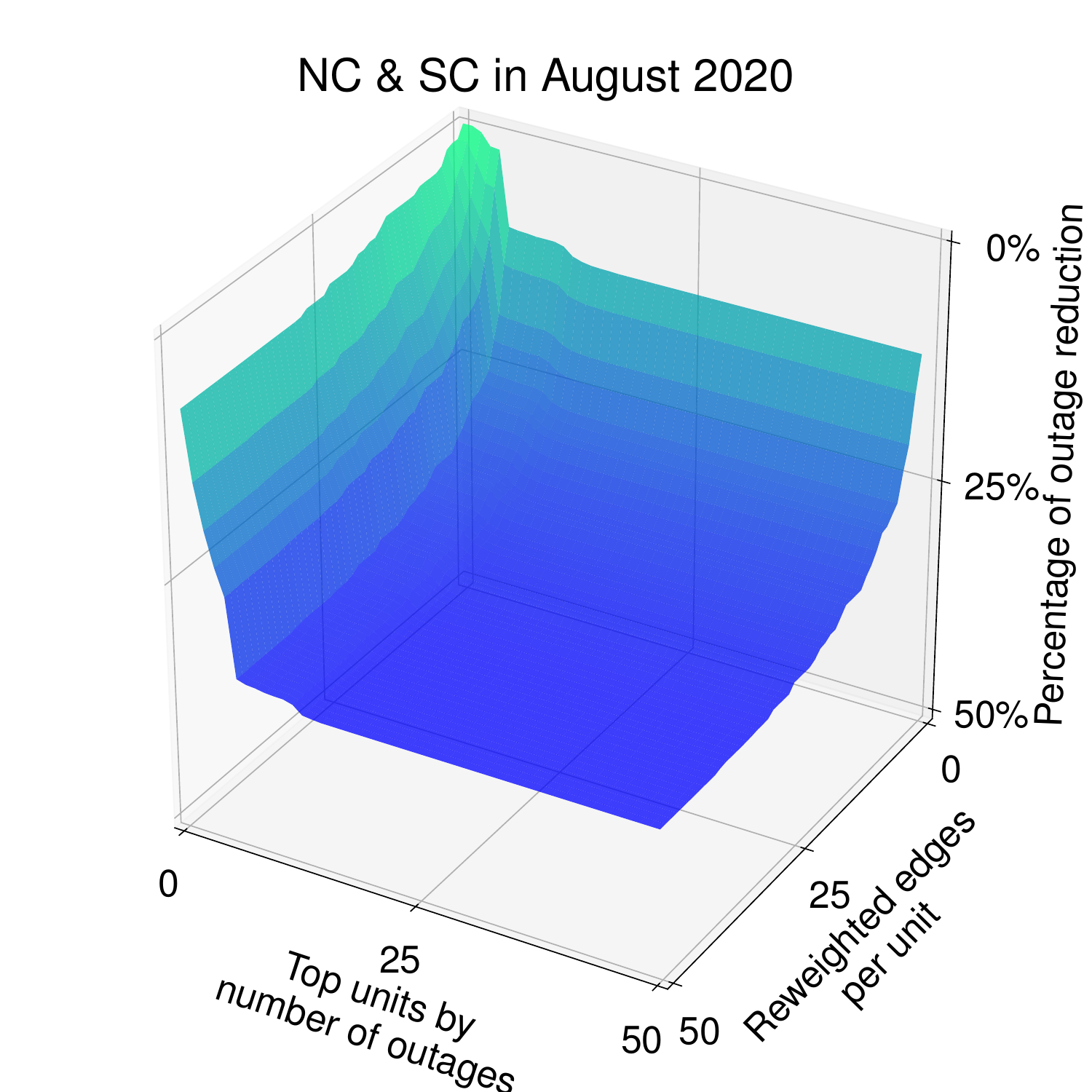}
    \end{subfigure}
    \\
    \begin{subfigure}[c]{0.3\linewidth}
    \includegraphics[width=\linewidth]{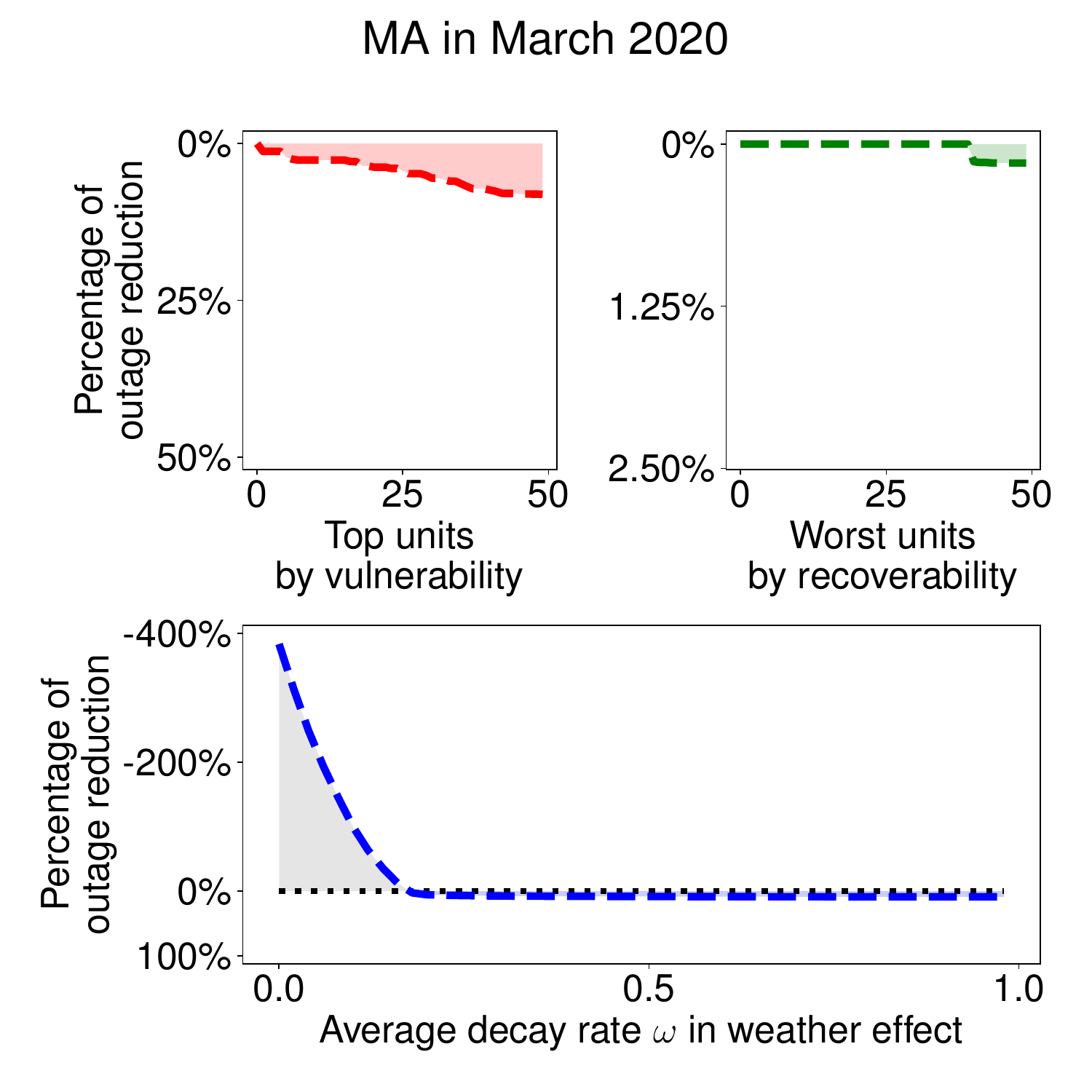}
    \end{subfigure} & 
    \begin{subfigure}[c]{0.3\linewidth}
    \includegraphics[width=\linewidth]{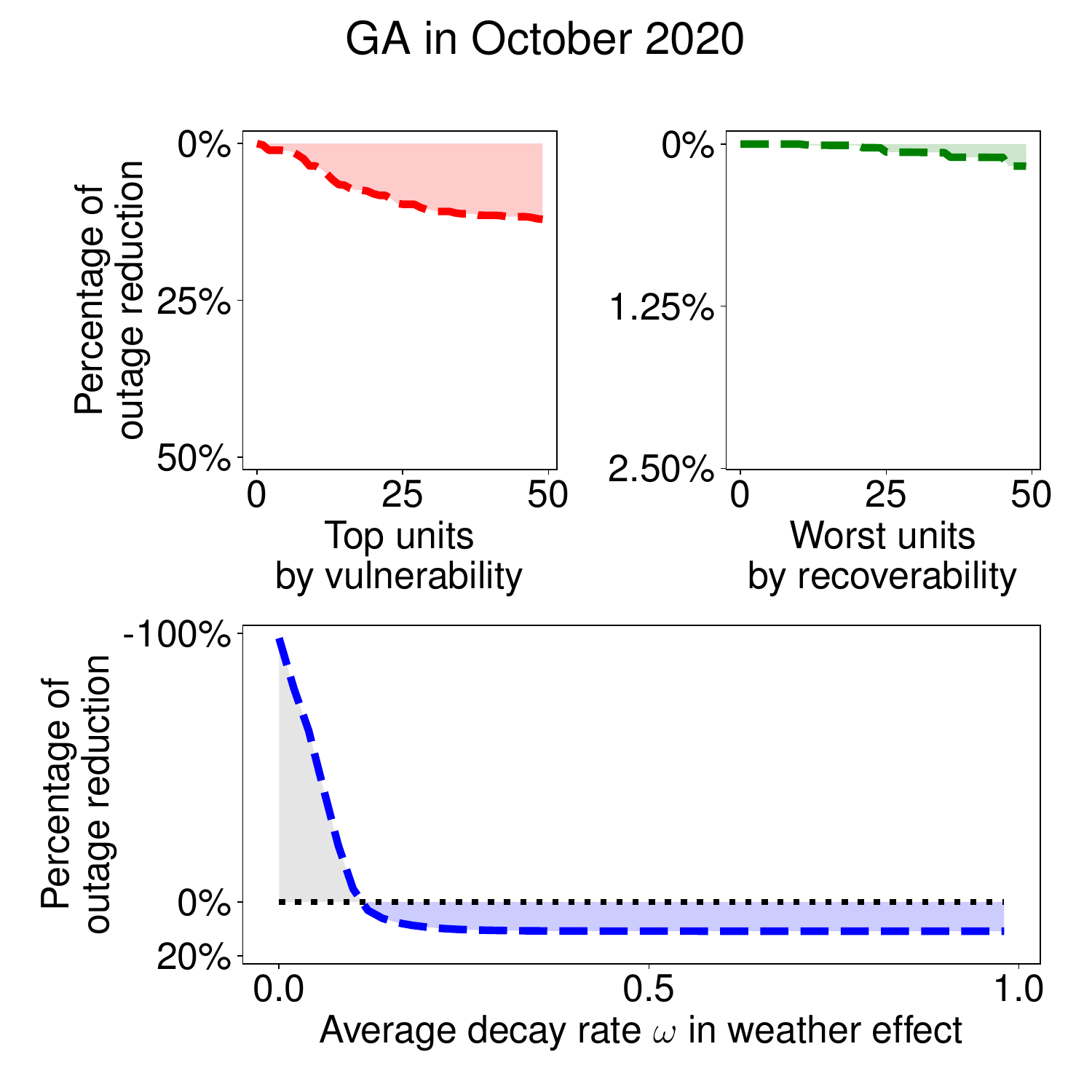}
    \end{subfigure} & 
    \begin{subfigure}[c]{0.3\linewidth}
    \includegraphics[width=\linewidth]{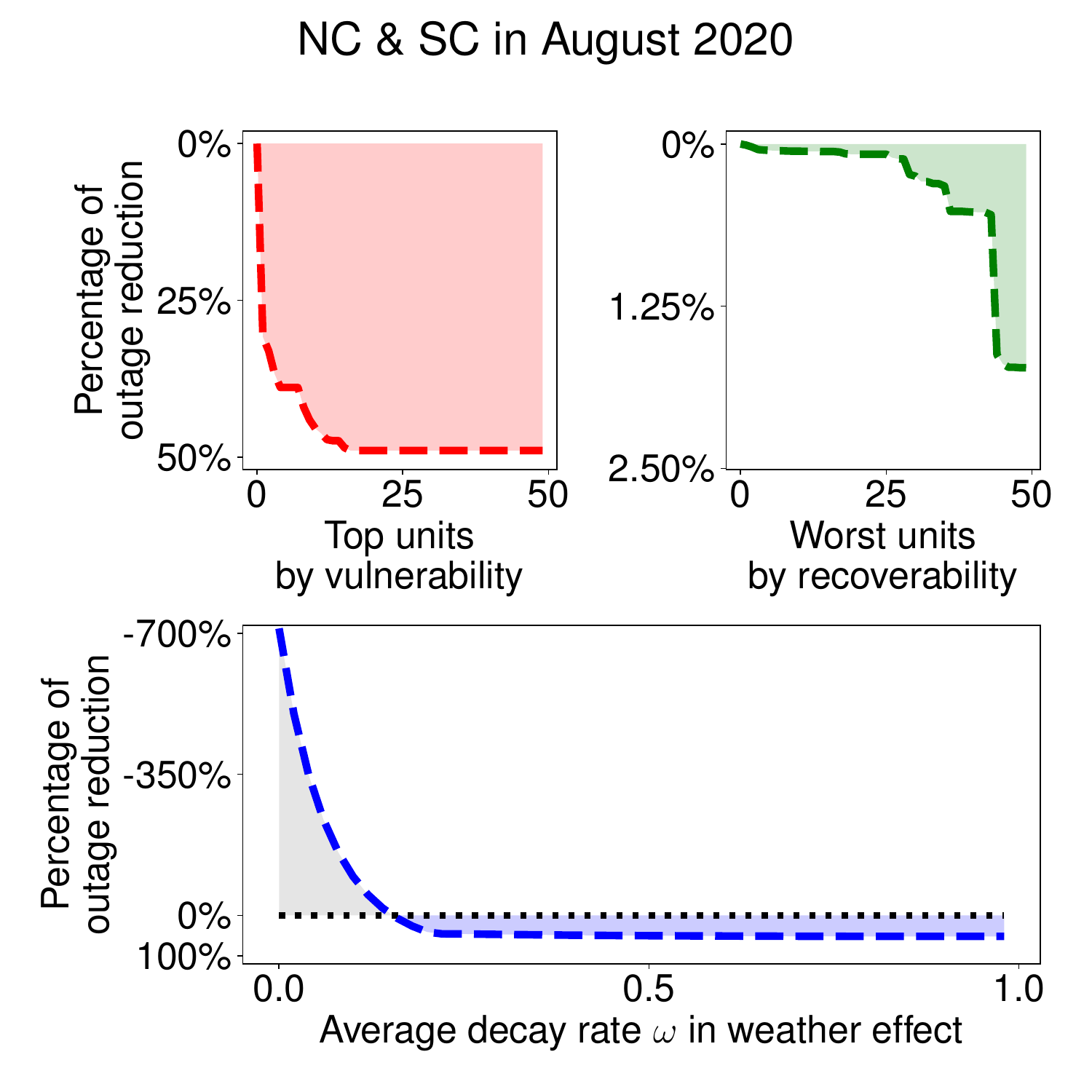}
    \end{subfigure}
  \end{tabular}}
{Resilience enhancement simulation results. \label{fig:resilience-enhancement-strategy-evaluation}}
{Top: Outage reduction by re-weighting the edges with the largest weights in the network to the average level. The left horizontal coordinate represents the number of top units by their maximum number of outages that we aim to improve during the extreme weather incidents; the right horizontal coordinate represents the number of edges per unit we need to re-weight; The vertical coordinate represents the simulated corresponding percentage of outage reduction with respect to the real number. Bottom: Outage reduction by re-weighting the units with the largest vulnerability and smallest recoverability coefficients, respectively, to their average levels. The horizontal coordinate represents the number of top units with the largest vulnerability and smallest recoverability coefficients, respectively, that we aim to improve during extreme weather incidents; The vertical coordinate represents the corresponding percentage of outage reduction with respect to the real number.}
\end{figure}

We now aim to identify and evaluate strategies in four general aspects (planning vulnerability, recoverability, criticality, and maintenance) that, if adopted, could significantly increase power grid resilience.
Even though these strategies could not be reliably and accurately assessed in the absence of detailed failure and restoration records, our models still offer an opportunity to simulate the entire process of large-scale customer power outages under different circumstances and explore potential opportunities to help alleviate the effects of extreme weather. 
To be specific, we ``improve'' certain aspects of the power grid resilience by adjusting the corresponding parameters of the fitted model, then simulate the outage process given the same weather condition, and finally calculate the outage reduction rate compared to the original number of outages (Figure~\ref{fig:resilience-enhancement-strategy-evaluation}). 
We observe that de-escalating the criticality of a small number of units with the largest amount of customer power outages can effectively mitigate the ultimate impact of extreme weather. For example, in Massachusetts, reweighting the largest 50 edges of the top 10 units by their number of customer power outages, such as Boston and Cambridge, can reduce 47.8\% customer power outages; similarly, in Georgia, the same strategy can reduce 38.4\% customer power outages.
However, the same strategy may not achieve competitive performance in North Carolina and South Carolina; instead, by improving the planning vulnerability for the unit with the largest number of customer power outages, i.e., Wilmington, the total number of outages can be drastically reduced by 32\%; it can be further improved to 49.8\% if top 12 units are considered with the same strategy. 
We also notice that improving the discount rate of accumulating weather effect ($\{\omega_m\}_{m\in\{1,\dots,M\}}$) is not rewarding compared to other strategies. However, the system will degrade remarkably if the rate is lower than the current level. The phenomenon can be explained because the maintenance has been managed at a fairly good level by the local operators.

\section{Discussion}

We first summarize the following findings using our proposed grid resilience model:
(1) Outage rates in metropolitan or economically strong areas are generally lower due to less vegetation, more underground or steel-structure-supported power lines, and adequate repair resources. In other words, the electricity infrastructures in those areas are less vulnerable to extreme weather events and more recoverable in case damage to an infrastructure occurs. 
(2) Oppositely, rural areas, especially with various terrains such as mountains, forests, rivers, and deserts, are hard to access and locate a fault, which inevitably delays an outage recovery. In addition, those economically weak areas usually lack of budgets to maintain or upgrade their electricity infrastructures, which become increasingly vulnerable to extreme weather events. As a result, those rural areas have relatively high outage rates.
(3) The direction (source to target) of an outage propagation typically follows power flow directions. In other words, an area with a large generation capacity or dense transmission network facilities (e.g., substations) is probably a hub of outage propagation. Such an area is more likely a mid-size urban area, which can be developed to host a number of transmission or generation facilities but is not a big load center that dominantly attracts power flows. 

From the quantitative analysis and simulated resilience enhancement, some planning and operational measures to prevent and mitigate weather-induced power outages can also be inspired. 
The first observation is that the power grid shows some vulnerability to major outage propagation, which significantly exacerbates the initial impact of weather events. 
If such outage evolution and propagation can be prevented or mitigated, the weather impact would be much smaller and self-contained. 
A key to stemming the outage propagation is to enable interdependency reduction when facing outage risks. The interdependency of power grids is highly related to their interconnected nature and it is neither economical nor practical to simply break the power grids into local pieces. A desired way to reduce the risks of major outages exacerbated by interdependency is to improve the operational flexibility of the power grid. A flexible power grid should embrace diversified sources with distributed locations and versatile operation schemes. For example, distributed energy resources (DER) can provide energy to local loads and thus reduce the risk of outages caused by interruption of distant electricity transmission. Energy storage can also quickly supply energy to local or nearby loads. Moreover, microgrids characterized as local power grids with diverse energy sources and flexible operation modes can easily connect to or isolate from the main power grid, which is resilient to external threats and sustainable. 

Future work could extend this approach by incorporating physical information about the power grid, such as transmission line capacities and topology, alongside geographical features like terrain and coastal proximity. 
Integrating these factors would enhance resilience quantification by accounting for spatially varying risks and enabling more realistic scenario modeling under diverse environmental conditions. 

\section*{Acknowledgements}

This work is supported by the U.S. Department of Energy Advanced Grid Modeling Program under Grant DE-OE0000875.

\bibliographystyle{informs2014} 
\bibliography{arxiv/arxiv} 


%
%
%

\newpage

\begin{APPENDICES}

\section{Data description}
\label{append:data}

This paper studies power grid resilience to extreme weather by taking advantage of a unique set of fine-grained customer-level power outage records and comprehensive weather data.
The data cover three major service territories across four states on the east coast of the United States, including Massachusetts, Georgia, North Carolina, and South Carolina, spanning more than 155,000 square miles, with an estimated population of more than 33 million people in 2019.
The extreme weather events we studied in this paper include
winter storms in Massachusetts (MA) in March 2018, Hurricane Michael in Georgia (GA) in October 2018, and Hurricane Isaias in North Carolina (NC) and South Carolina (SC) in August 2020.
The data record the number of customers without power supply and the corresponding weather conditions, which hold sufficient information to capture the spatio-temporal interactions between disruption and restoration processes.


Customer power outage data in this work were purchased from PowerOutage.us \citep{poweroutageus}. 
The data record the number of customers without power in each geographical unit at 15-minute intervals. 
These service territories are divided into hundreds of geographical units defined by township, zip code, or county. 
There are 351 units and 2,755,111 customers in Massachusetts, 665 units and 2,571,603 customers in Georgia, and 115 units and 4,305,995 customers in North Carolina and South Carolina. See a summary of the outage data in Table~\ref{tab:dataset} for more details.
The HRRR model \citep{Blaylock2015} is a National Oceanic and Atmospheric Administration (NOAA) real-time, 3 km resolution, hourly updated, cloud-resolving, convection-allowing atmospheric model, initialized by 3 km grids with 3 km radar assimilation. 
Radar data is assimilated into the HRRR model every 15 minutes over a 1-hour period, adding further detail to that provided by the hourly data assimilation from the 13 km radar-enhanced rapid refresh.
In this work, we selected 34 variables in the HRRR model (Table~\ref{tab:weather}) that characterize near-surface atmospheric activities directly linked to weather impacts on power grids. 
To match the spatial resolution with the customer power outage data, we aggregated the regional weather data in the same geographic unit by taking the average. To reduce noise in both the outage data and the weather data, we also aggregated the number of customer power outages and the value of weather variables in each unit into three-hour time slots.

\begin{table}
\TABLE
{Data sets on the customer power outages used in this study. \label{tab:dataset}}
{\resizebox{\textwidth}{!}{
\begin{tabular}{llllllllllll}
\hline\up 
 & \multicolumn{3}{c}{Basic information} &  & \multicolumn{3}{c}{Extreme weather event} &  & \multicolumn{3}{c}{Daily operation} \\ 
 \cmidrule[0.4pt]{2-4} \cmidrule[0.4pt]{6-8} \cmidrule[0.4pt]{10-12} 
 & Units & Unit type & Customers &  & Time period & Average outages & Max outages &  & Time period & Average outages & Max outages \\ \hline \up
MA & 351 & township &  2,755,111 &  & 2018-03-01 $\sim$ 2018-03-15 & 2393 & 19964 &  & 2018-03-16 $\sim$ 2018-03-31 & 49 & 1096 \\
GA & 665 & zipcode & 2,571,603 &  & 2018-10-05 $\sim$ 2018-10-20 & 359 & 7182 &  & 2018-10-21 $\sim$ 2018-11-05 & 38 & 1659 \\
NC \& SC & 115 & county & 4,305,995 &  & 2020-07-31 $\sim$ 2020-08-10 & 2198 & 92424 &  & 2020-08-11 $\sim$ 2020-08-31 & 474 & 6178
\down\\ \hline
\end{tabular}}}{}
\end{table}

\begin{table}
\TABLE
{Weather effect \label{tab:weather}}
{\begin{tabular}{lll}
\hline\up 
No. & Variable & Description \\ \hline\up 
1 & REFC & Maximum / Composite radar reflectivity {[}dB{]} \\
2 & RETOP & Echo Top {[}m{]}, 0{[}-{]} CTL="Level of cloud tops" \\
3 & VERIL & Vertically-integrated liquid {[}kg/m{]}, 0{[}-{]} RESERVED(10) (Reserved) \\
4 & VIS & Visibility {[}m{]}, 0{[}-{]} SFC="Ground or water surface" \\
5 & REFD-1000m & Derived radar reflectivity {[}dB{]}, 1000{[}m{]} HTGL="Specified height level above ground" \\
6 & REFD-4000m & Derived radar reflectivity {[}dB{]}, 4000{[}m{]} HTGL="Specified height level above ground" \\
7 & REFD-263k & Derived radar reflectivity {[}dB{]}, 263{[}K{]} TMPL="Isothermal level" \\
8 & GUST & Wind speed (gust) {[}m/s{]}, 0{[}-{]} SFC="Ground or water surface" \\
9 & TMP-92500pa & Temperature {[}C{]}, 92500{[}Pa{]} ISBL="Isobaric surface" \\
10 & DPT-92500pa & Dew point temperature {[}C{]}, 92500{[}Pa{]} ISBL="Isobaric surface" \\
11 & UGRD-92500pa & u-component of wind {[}m/s{]}, 92500{[}Pa{]} ISBL="Isobaric surface" \\
12 & VGRD-92500pa & v-component of wind {[}m/s{]}, 92500{[}Pa{]} ISBL="Isobaric surface" \\
13 & TMP-100000pa & Temperature {[}C{]}, 100000{[}Pa{]} ISBL="Isobaric surface" \\
14 & DPT-100000pa & Dew point temperature {[}C{]}, 100000{[}Pa{]} ISBL="Isobaric surface" \\
15 & UGRD-100000pa & u-component of wind {[}m/s{]}, 100000{[}Pa{]} ISBL="Isobaric surface" \\
16 & VGRD-100000pa & v-component of wind {[}m/s{]}, 100000{[}Pa{]} ISBL="Isobaric surface" \\
17 & DZDT & Verical velocity (geometric) {[}m/s{]}, 0.5-0.8{[}'sigma' value{]} SIGL="Sigma level" \\
18 & MSLMA & MSLP (MAPS System Reduction) {[}Pa{]}, 0{[}-{]} MSL="Mean sea level" \\
19 & HGT & Geopotential height {[}gpm{]}, 100000{[}Pa{]} ISBL="Isobaric surface" \\
20 & UGRD-80m & u-component of wind {[}m/s{]}, 80{[}m{]} HTGL="Specified height level above ground" \\
21 & VGRD-80m & v-component of wind {[}m/s{]}, 80{[}m{]} HTGL="Specified height level above ground" \\
22 & PRES & Pressure {[}Pa{]}, 0{[}-{]} SFC="Ground or water surface" \\
23 & TMP-0m & Temperature {[}C{]}, 0{[}-{]} SFC="Ground or water surface" \\
24 & MSTAV & Moisture Availability {[}\%{]}, 0{[}m{]} DBLL="Depth below land surface" \\
25 & SNOWC & Snow cover {[}\%{]}, 0{[}-{]} SFC="Ground or water surface" \\
26 & SNOD & Snow depth {[}m{]}, 0{[}-{]} SFC="Ground or water surface" \\
27 & TMP-2m & Temperature {[}C{]}, 2{[}m{]} HTGL="Specified height level above ground" \\
28 & SPFH & Specific humidity {[}kg/kg{]}, 2{[}m{]} HTGL="Specified height level above ground" \\
29 & UGRD-10m & u-component of wind {[}m/s{]}, 10{[}m{]} HTGL="Specified height level above ground" \\
30 & VGRD-10m & v-component of wind {[}m/s{]}, 10{[}m{]} HTGL="Specified height level above ground" \\
31 & WIND & Wind speed {[}m/s{]}, 10{[}m{]} HTGL="Specified height level above ground" \\
32 & LFTX & Surface lifted index {[}C{]}, 50000-100000{[}Pa{]} ISBL="Isobaric surface" \\
33 & USTM & U-component storm motion {[}m/s{]}, 0-6000{[}m{]} HTGL="Specified height level above ground" \\
34 & VSTM & V-component storm motion {[}m/s{]}, 0-6000{[}m{]} HTGL="Specified height level above ground"
\down\\ \hline
\end{tabular}}{}
\end{table}

\begin{figure}[!t]
\centering
\FIGURE{
\centering
  \begin{tabular}[c]{cccc}
    \begin{subfigure}[b]{.24\linewidth}
    \includegraphics[width=\linewidth]{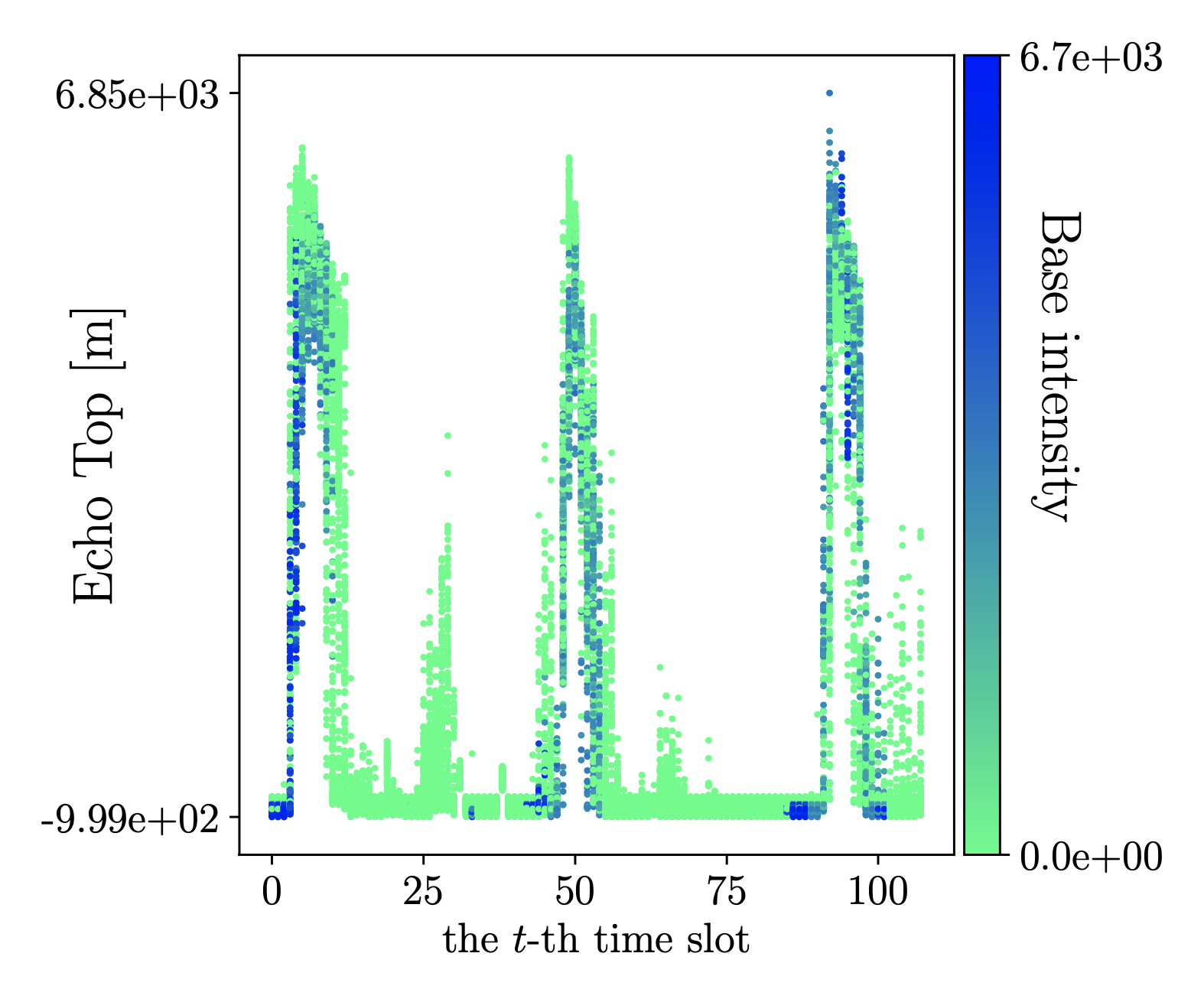}
    \end{subfigure} & 
    \begin{subfigure}[b]{.24\linewidth}
    \includegraphics[width=\linewidth]{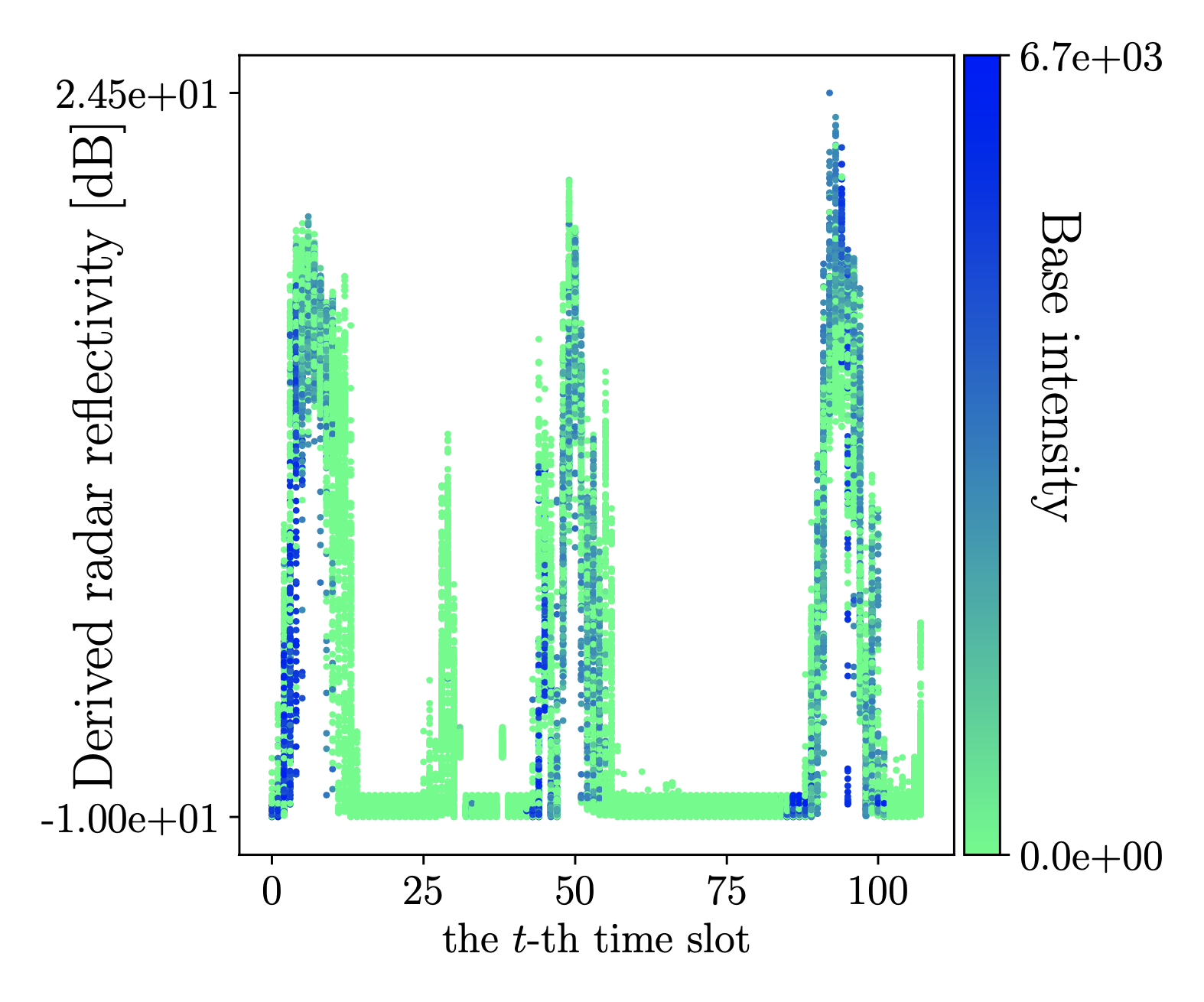}
    \end{subfigure} & 
    \begin{subfigure}[b]{.24\linewidth}
    \includegraphics[width=\linewidth]{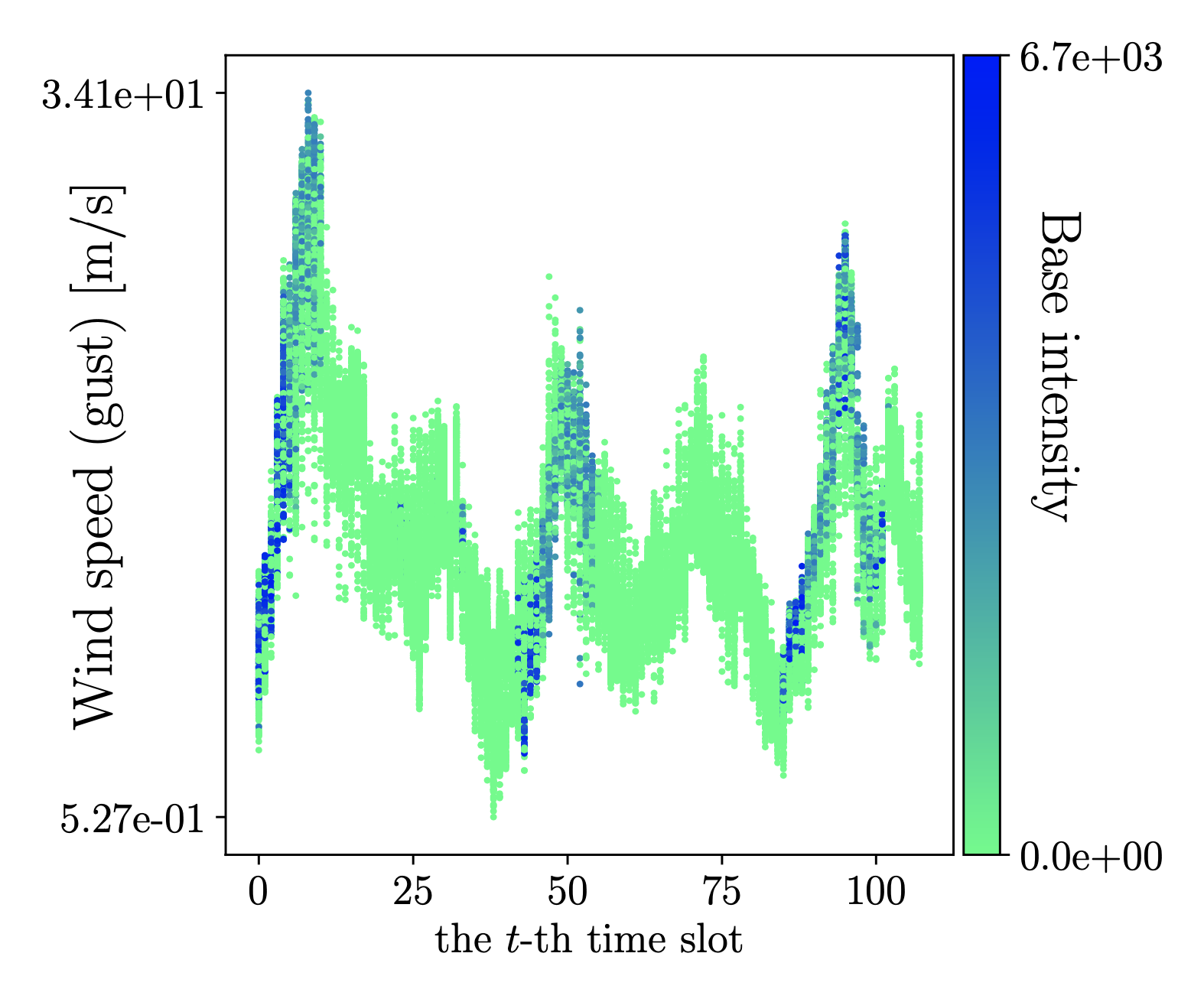}
    \end{subfigure} & 
    \begin{subfigure}[b]{.24\linewidth}
    \includegraphics[width=\linewidth]{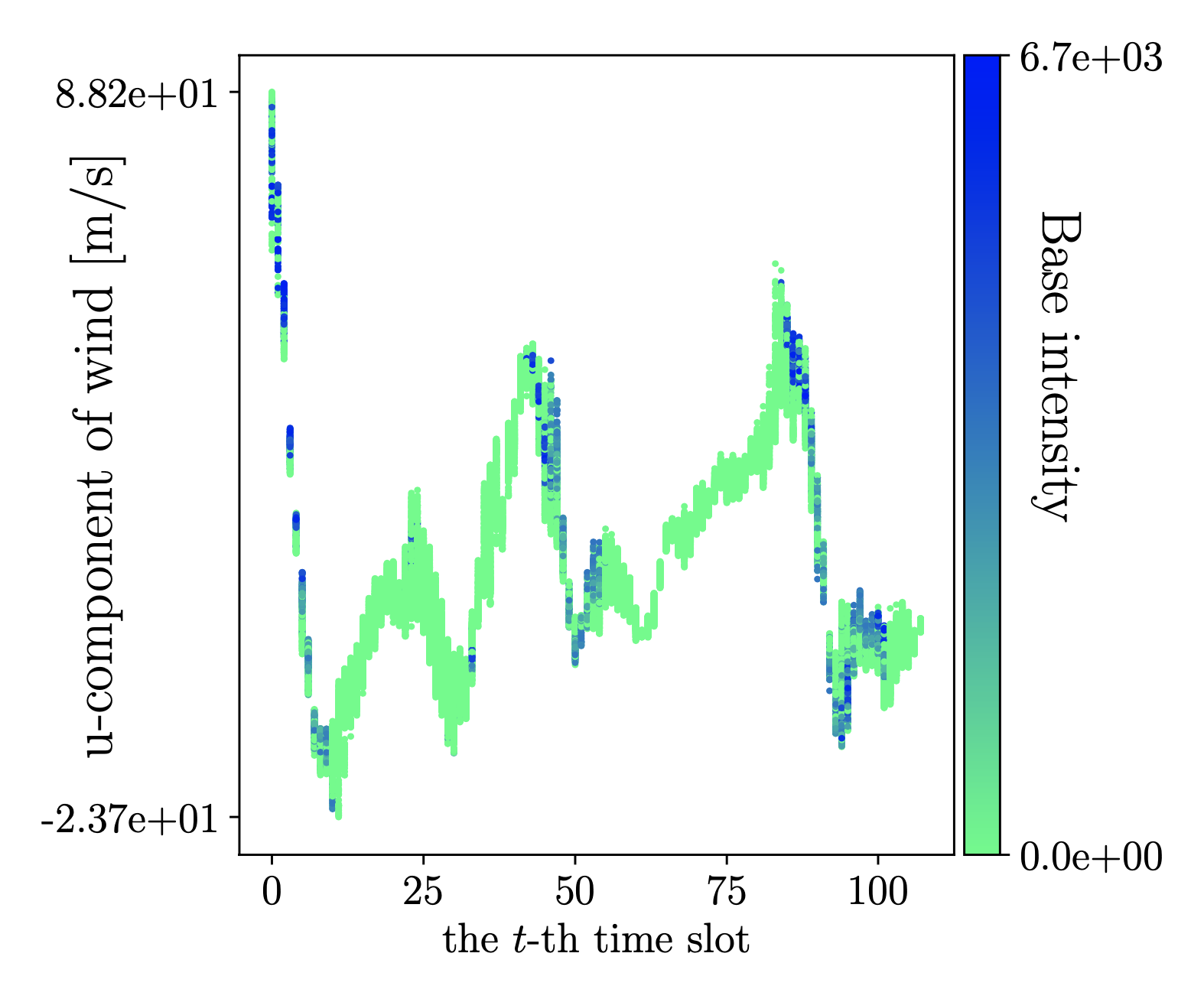}
    \end{subfigure}
    \\
    \begin{subfigure}[b]{.24\linewidth}
    \includegraphics[width=\linewidth]{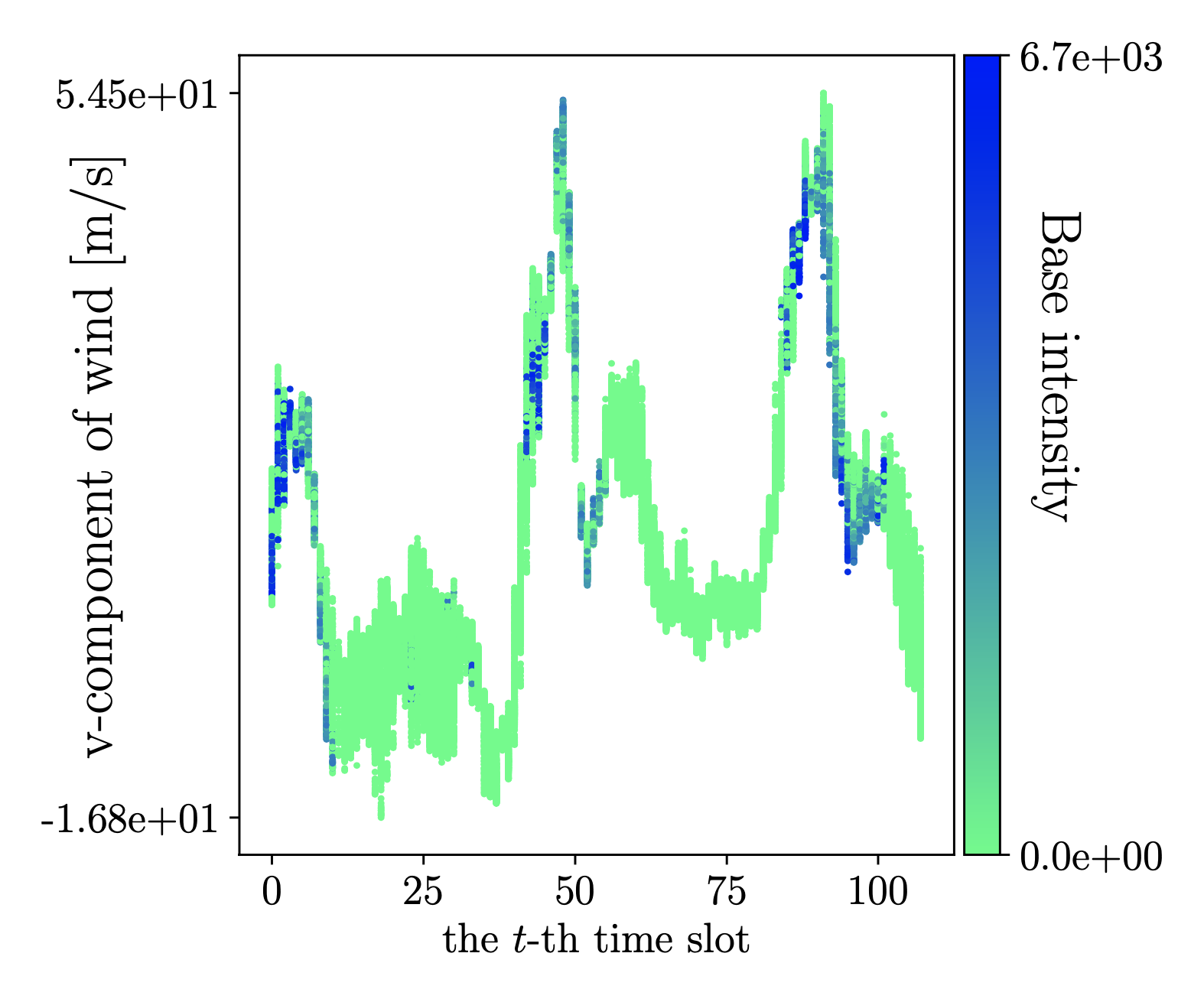}
    \end{subfigure} & 
    \begin{subfigure}[b]{.24\linewidth}
    \includegraphics[width=\linewidth]{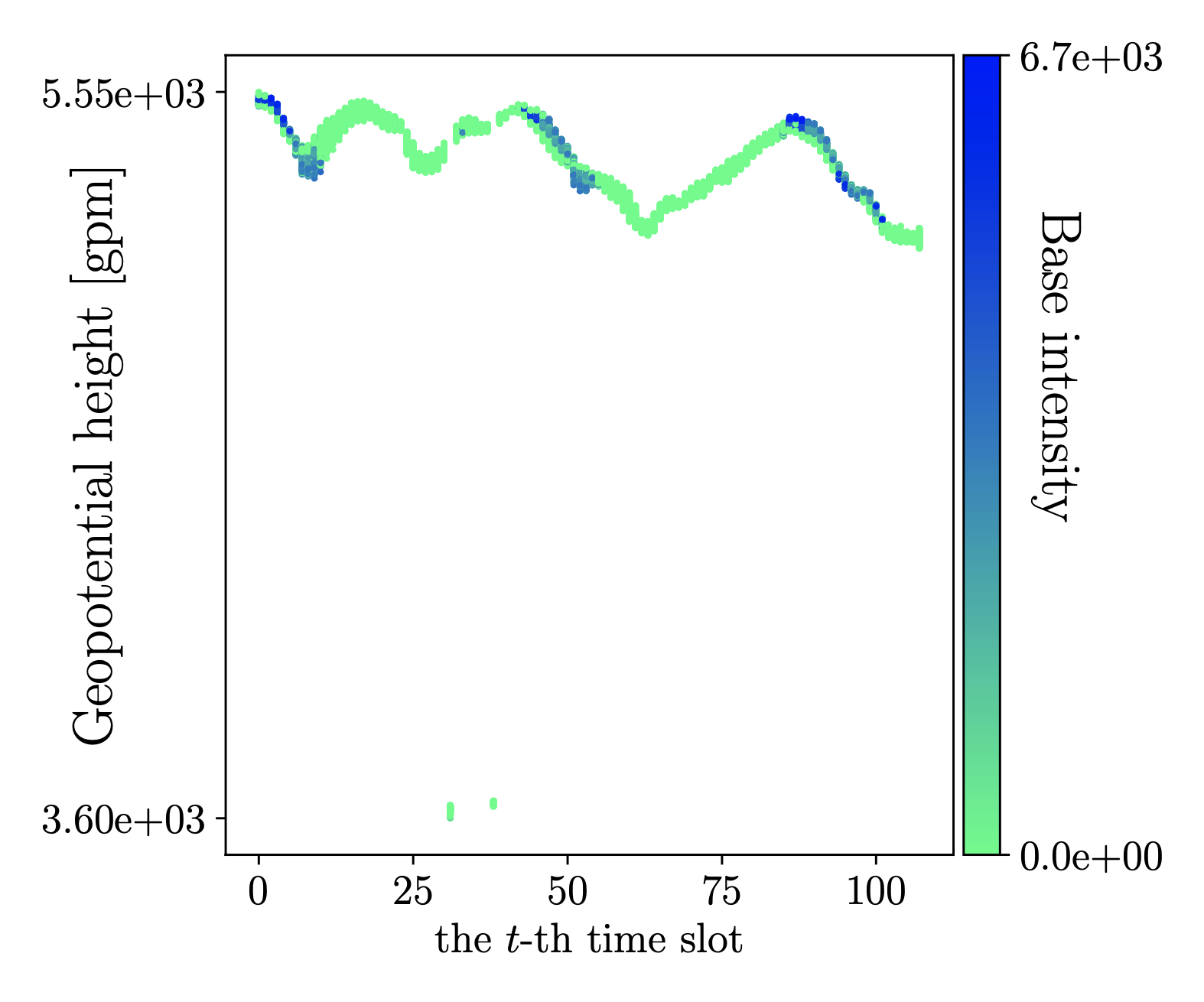}
    \end{subfigure} & 
    \begin{subfigure}[b]{.24\linewidth}
    \includegraphics[width=\linewidth]{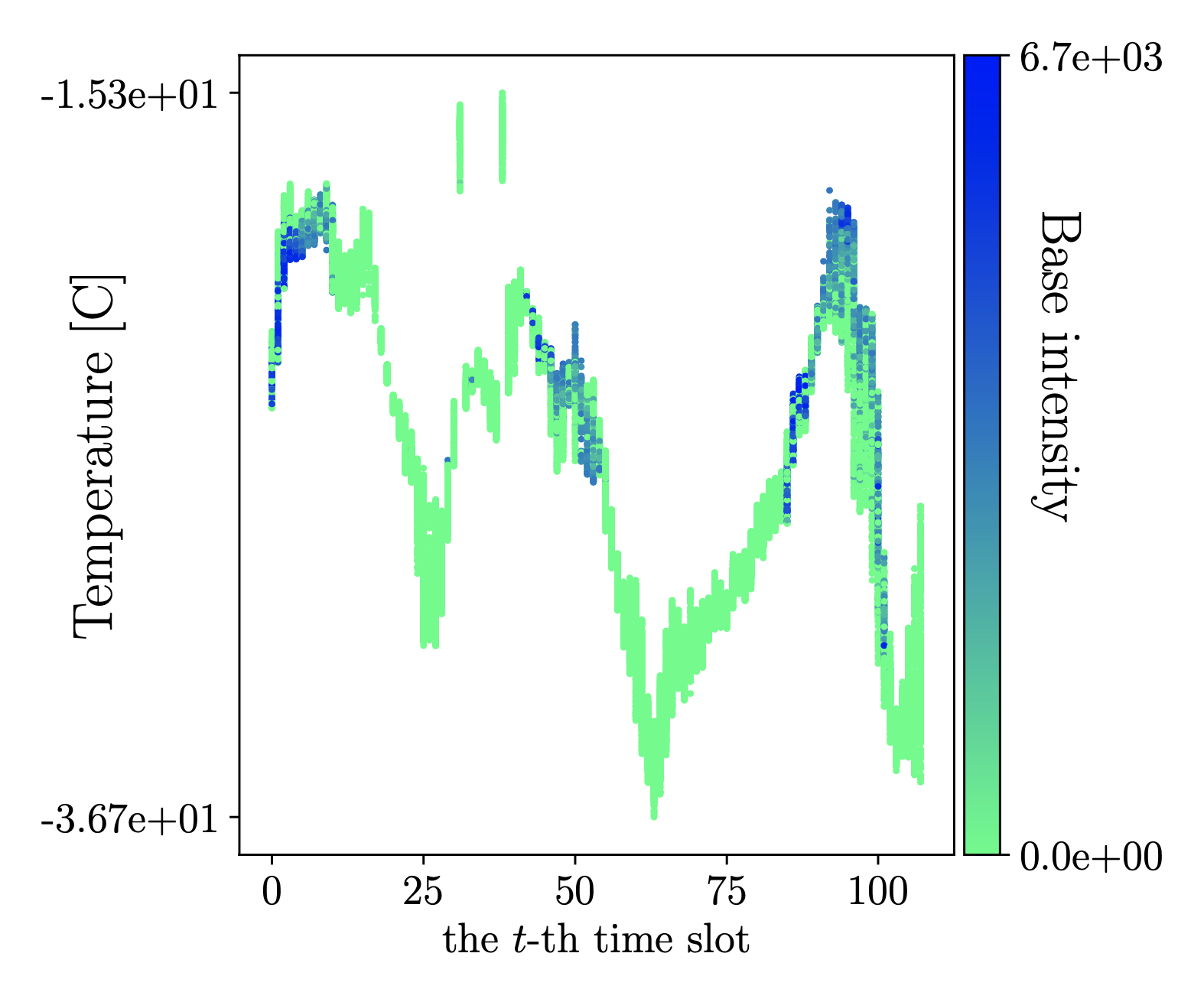}
    \end{subfigure} & 
    \begin{subfigure}[b]{.24\linewidth}
    \includegraphics[width=\linewidth]{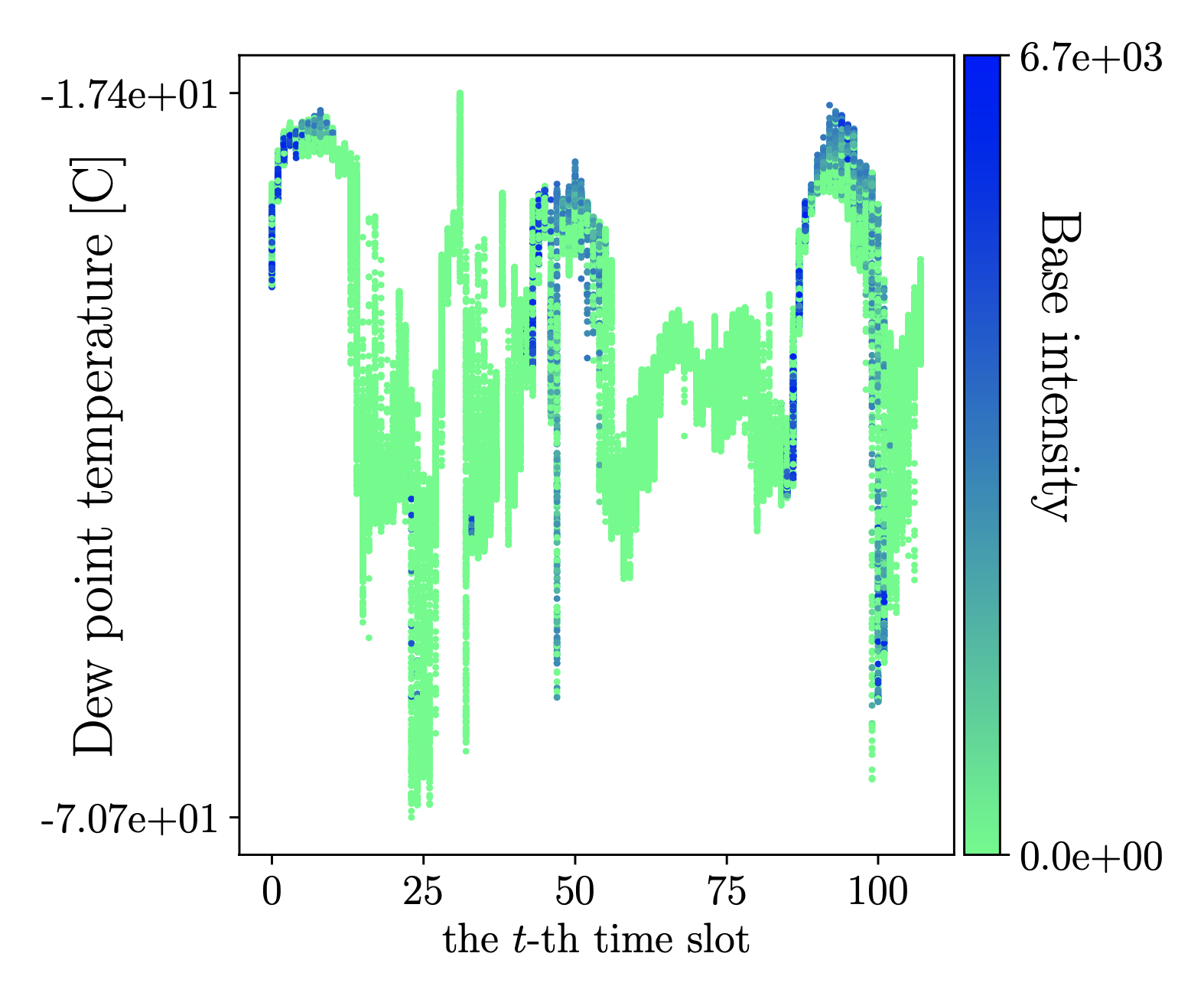}
    \end{subfigure}
  \end{tabular}}
{The temporal relationship between weather intensity and the estimated base intensity in MA. \label{fig:weather-vs-base}}
{In the sub-figures, each time slot contains 351 dots, representing the total number of townships in MA. 
The vertical position of each dot reflects the weather intensity in that city at the $t$-th time slot. 
The color depth of the dot indicates the estimated base intensity as determined by the DNN model.}
\end{figure}

We study three typical extreme weather events between 2018 and 2020 during which customers in these three service territories were significantly affected and for which the number of outages, as well as weather information for each geographical unit, are included in our data set. These three events are as follows.
\emph{March 2018 nor'easters}:
The March 2018 nor'easters included three powerful winter storms in March 2018 that caused major impacts in the northeastern, mid-Atlantic, and southeastern United States. At its peak, more than 14\% of the customers (385,744) in Massachusetts remained without power, and  outage ratios for nearly 30\% of the geographic units in the region (104) were over 50\%. 
\emph{Hurricane Michael}:
Hurricane Michael was a very powerful and destructive tropical cyclone in October 2018 that became the first Category 5 hurricane to strike the contiguous United States since Andrew in 1992. About 7.5\% of the customers (193,018) in Georgia were severely affected by the storm and lost power, and the outage rates for about 25\% of the geographic units in Georgia (166) were above 50\%. 
\emph{Hurricane Isaias.}
Hurricane Isaias was a destructive Category 1 hurricane in August 2020 that caused extensive damage across the Caribbean and the U.S. East Coast and included a large tornado outbreak. At its peak, about 13\% of the customers in North and South Carolina lost power, and the outage ratios for nearly 9\% of the geographic units in these two states were above  50\%.
For the three service territories, we determine the study periods based on power outage records and the associated weather data, including derived radar reflectivity, composite radar reflectivity, and wind speed. 
As a result, March 1 to March 16, 2018, is considered to be the period of the March 2018 nor'easters for service territories in Massachusetts, October 5 to October 20, 2018, the period of Hurricane Michael for service territories in Georgia, and August 1 to August 10, 2020, the period of Hurricane Isaias for service territories in North Carolina and South Carolina. 
To fairly compare the resilience's differences under the same condition, for each event, we also consider the rest of the days without the impact of extreme weather in the same month as daily operations (Table~\ref{tab:dataset}). 

\section{Specifics of the deep neural network used in the proposed model}
\label{append:dnn}

Our model involves approximately $25,000$ parameters, with roughly $20,000$ associated with the DNN and $10,000 \sim 25,000$ characterizing spatial correlation. Each city also has a parameter representing the decaying rate over time. 

To address concerns about parameter estimation and trustworthiness, we have implemented several validation strategies for our model: 
\begin{enumerate}
    \item \emph{Model Complexity and Parameter Estimation}: In the neural network setting, it is well-known that learning a DNN is not about identifying the ``optimal'' parameters. There can be multiple parameter combinations that are equally effective. Our focus is on finding a good function that fits the data, where the fitted parameters might differ across trials, but the function's quality remains similar in terms of generalization power \citep{pmlr-v97-du19c}.
    \item \emph{Regularization Techniques}: We use dropout in the neural network and limit graph connectivity to neighboring cities to encourage sparsity. This approach ensures that the estimated non-zero parameters range from a few hundred to a thousand, balancing model complexity and predictive performance. 
    \item \emph{Validation Against Real-World Data}: We employ extensive cross-validation to optimize hyperparameters, validate the model's performance with real-world data, and benchmark it against established models in the field. This process ensures that the model generalizes well and that the parameter estimates are robust, avoiding overfitting to the training data. 
    \item \emph{Domain Expertise and Sensible Defaults}: We leverage domain expertise to establish sensible defaults and priors for the parameters. Specifically, through discussions with domain experts, we model outage processes separately as direct impacts from weather and indirect impacts from neighboring cities. While there is no explicit knowledge of how weather affects outages at the system level, experts highlight the complex interplay between multiple weather factors and different parts of the system. This insight drives us to develop a neural network architecture that captures this complexity in a data-driven manner. These motivations ensure that our parameter estimates are based on realistic assumptions and expert knowledge, thereby enhancing the model's trustworthiness.
\end{enumerate}

Additionally, we have included some descriptive analysis in Figure~\ref{fig:weather-vs-base}. As shown, higher weather intensity does not necessarily correlate with higher base intensity (direct impact). This result highlights the strong non-linearity between the learned base intensity and the corresponding weather input, as suggested by our DNN model.

\section{Additional results}
\label{append:results}

\begin{figure}[!t]
\centering
\FIGURE{
\centering
  \begin{tabular}[c]{ccc}
    \begin{subfigure}[b]{.28\textwidth}
        \centering
        \includegraphics[width=\linewidth]{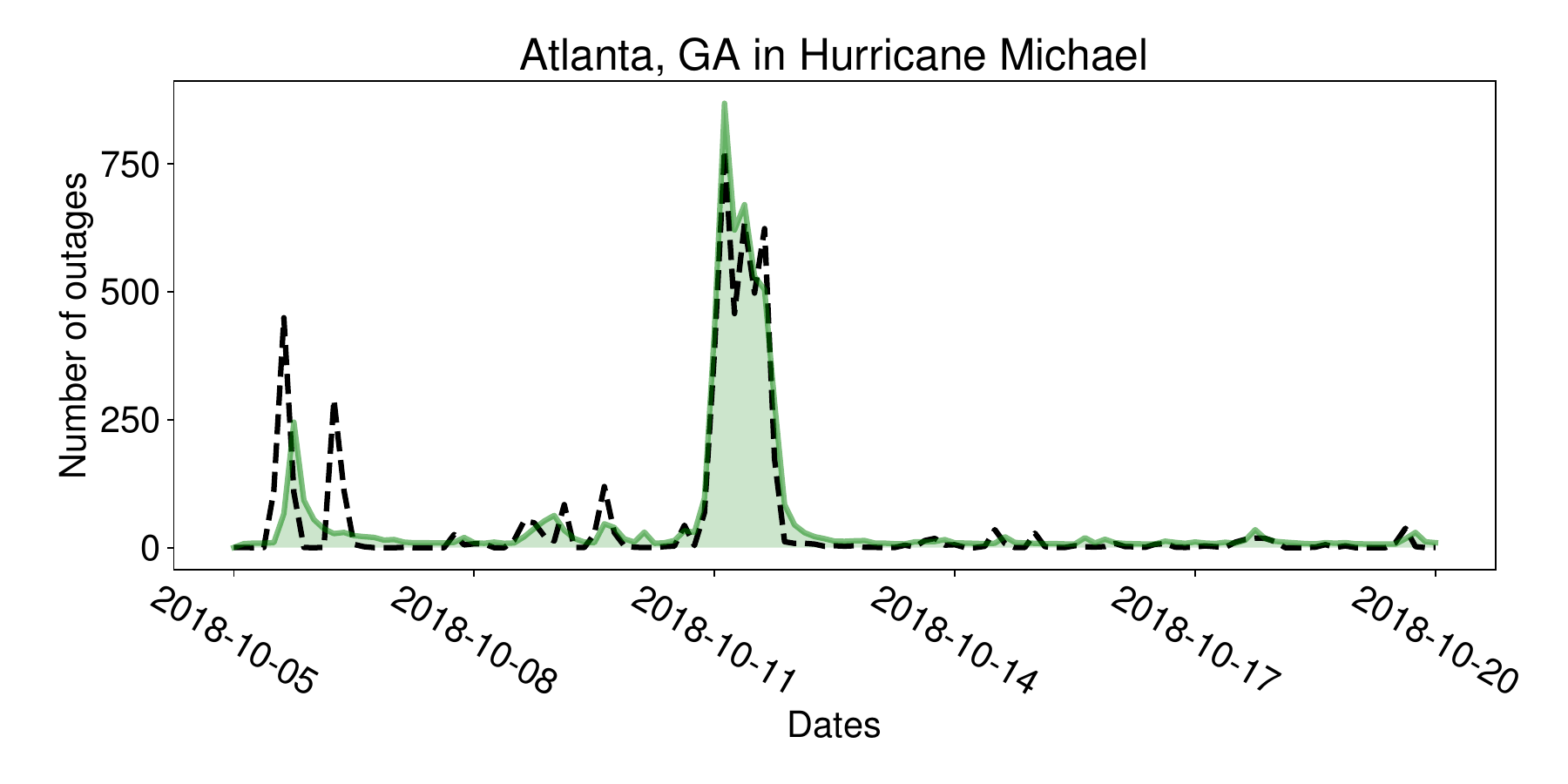}
        \vfill
        \includegraphics[width=\linewidth]{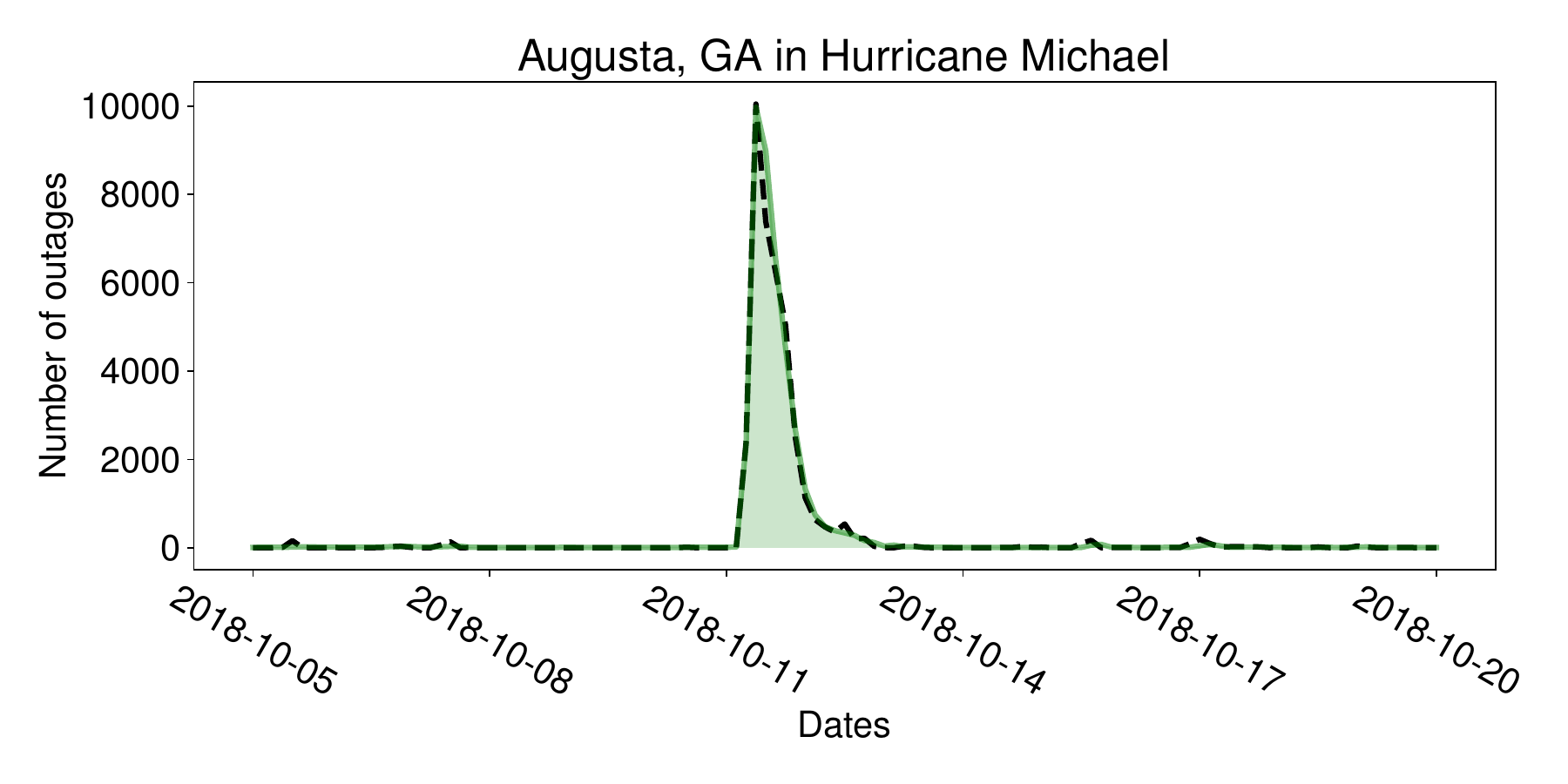}
        \vfill
        \includegraphics[width=\linewidth]{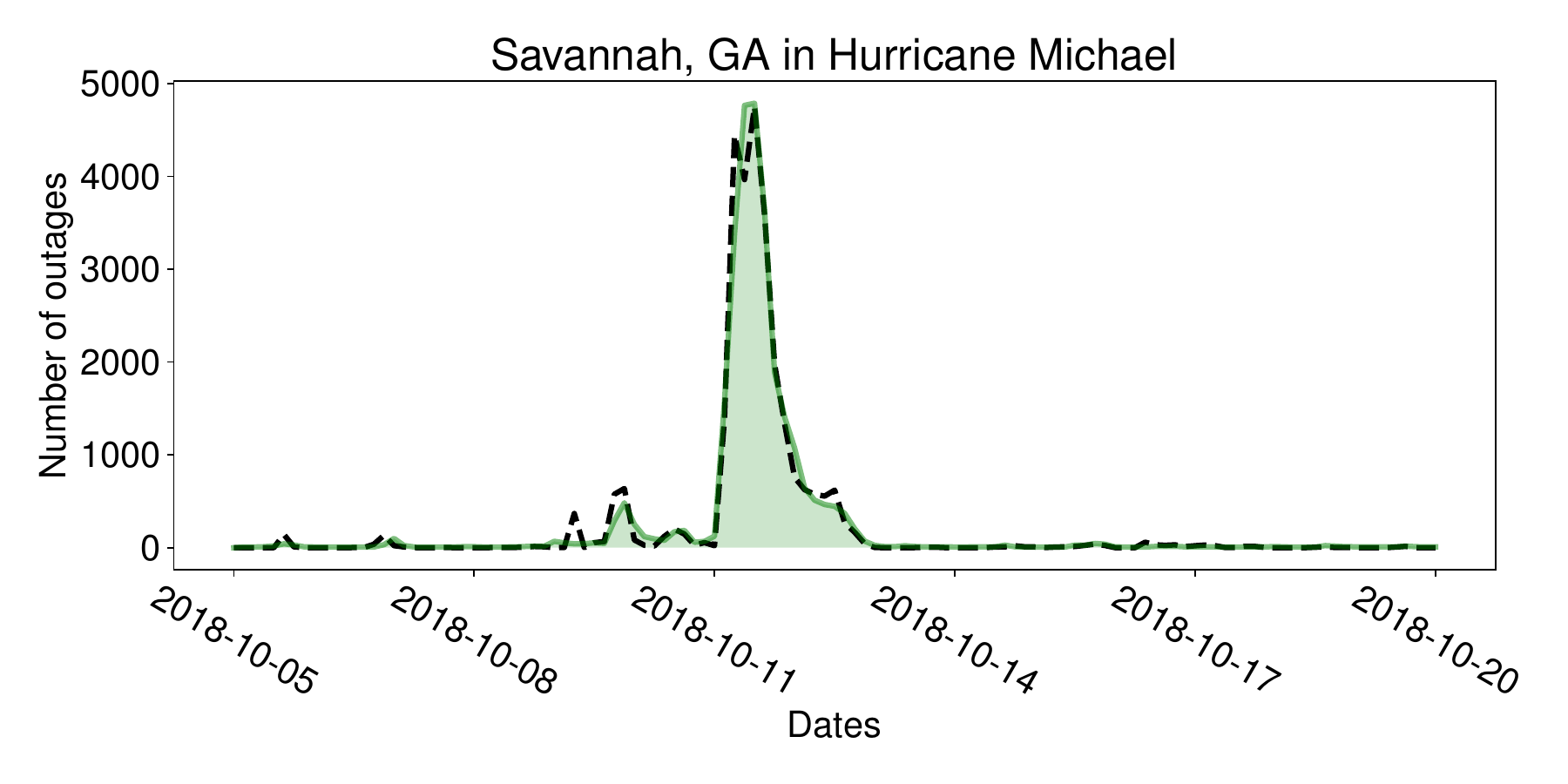}
        \vfill
        \includegraphics[width=\linewidth]{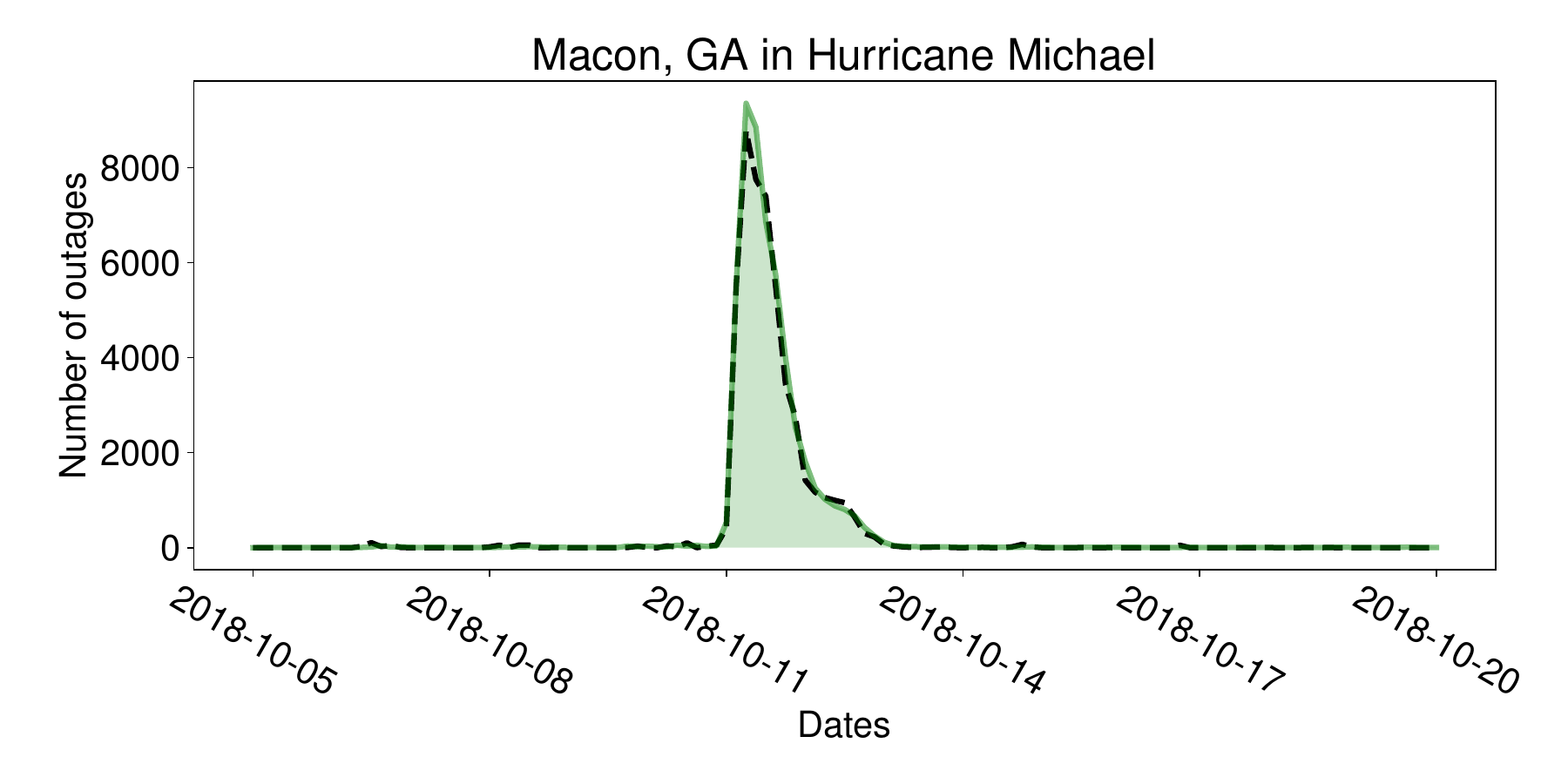}
        \vfill
        \includegraphics[width=\linewidth]{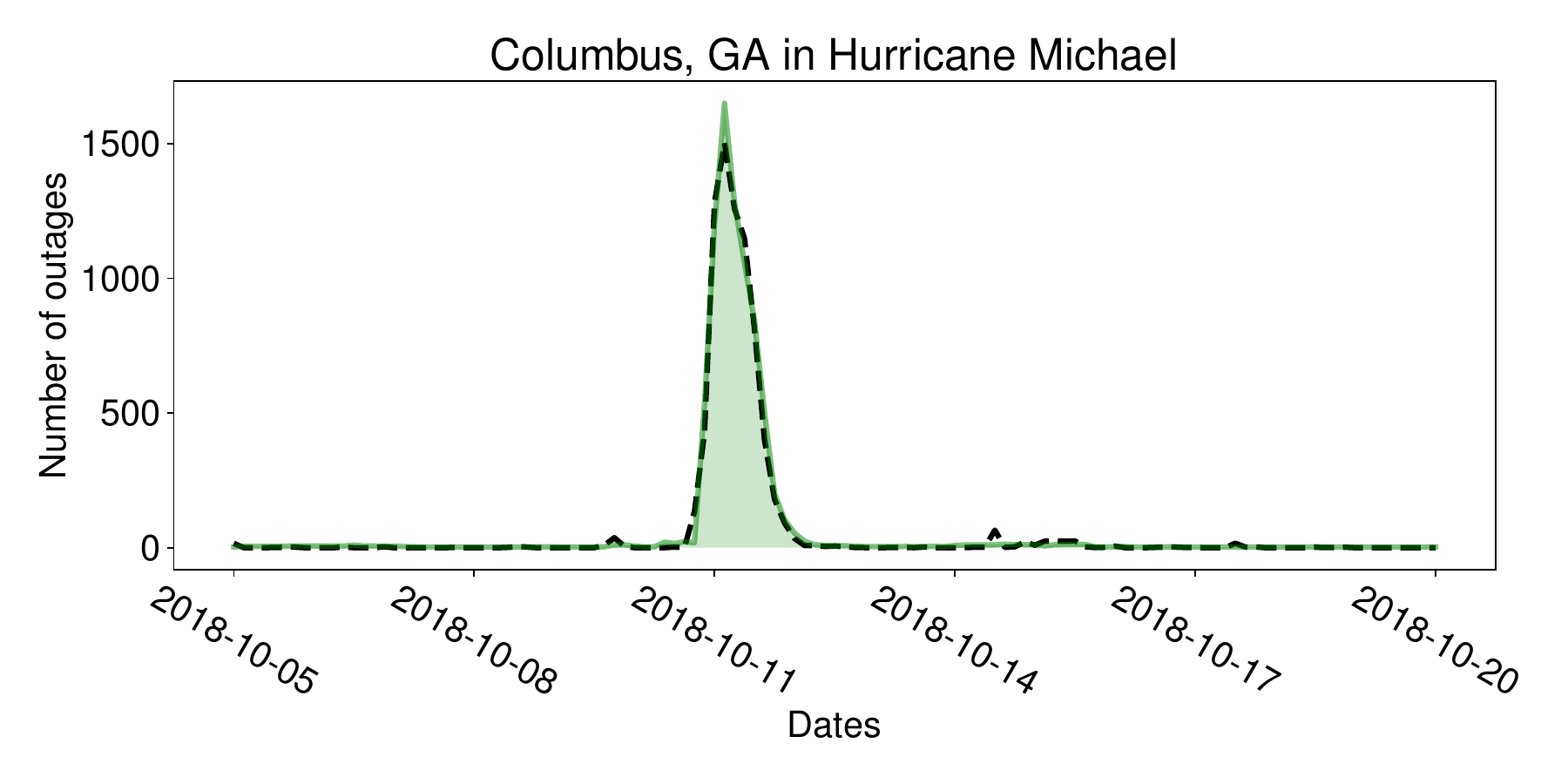}
        \caption{}
    \end{subfigure} & 
    \begin{subfigure}[b]{.28\textwidth}
        \centering
        \includegraphics[width=\linewidth]{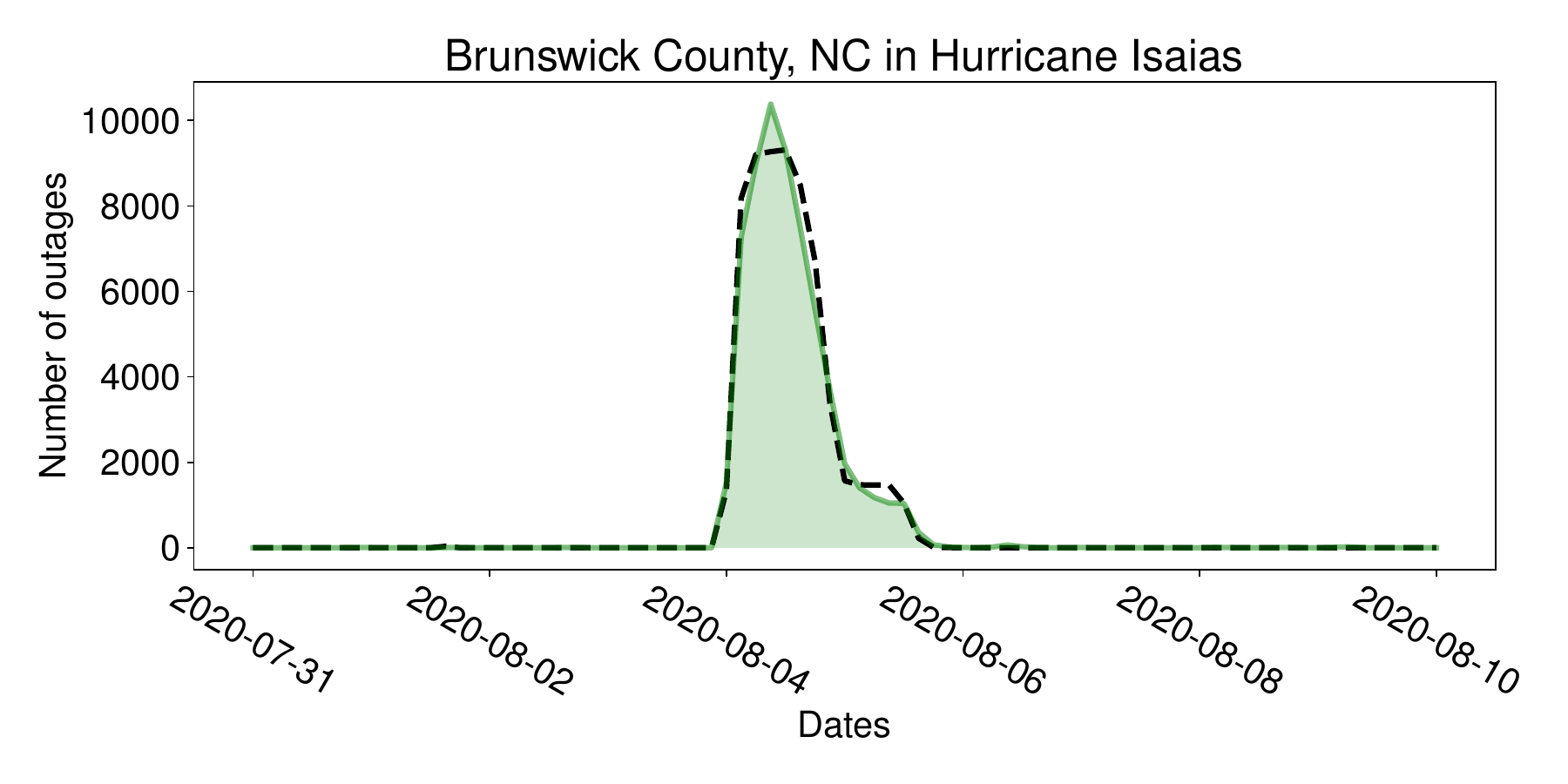}
        \vfill
        \includegraphics[width=\linewidth]{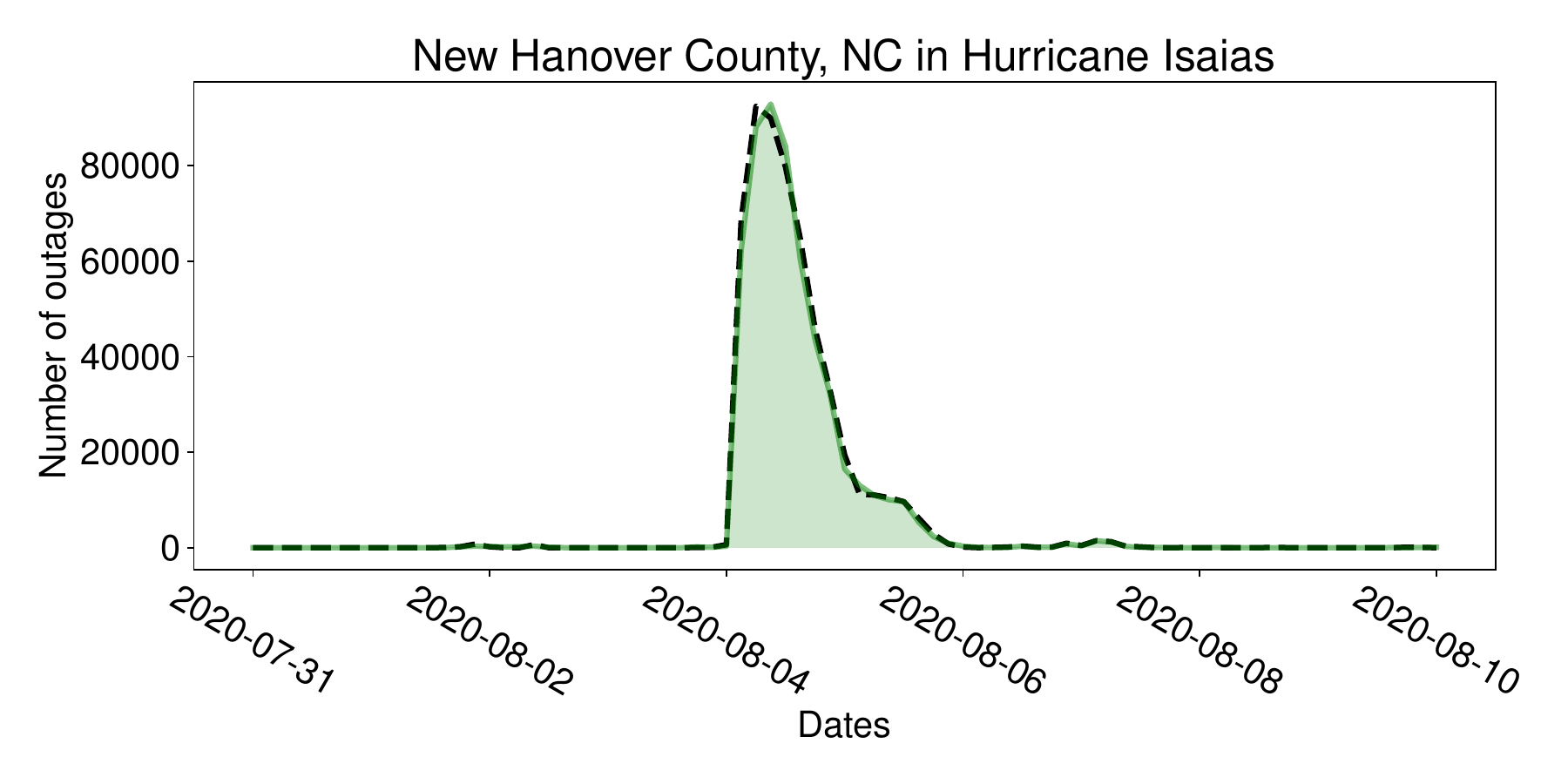}
        \vfill
        \includegraphics[width=\linewidth]{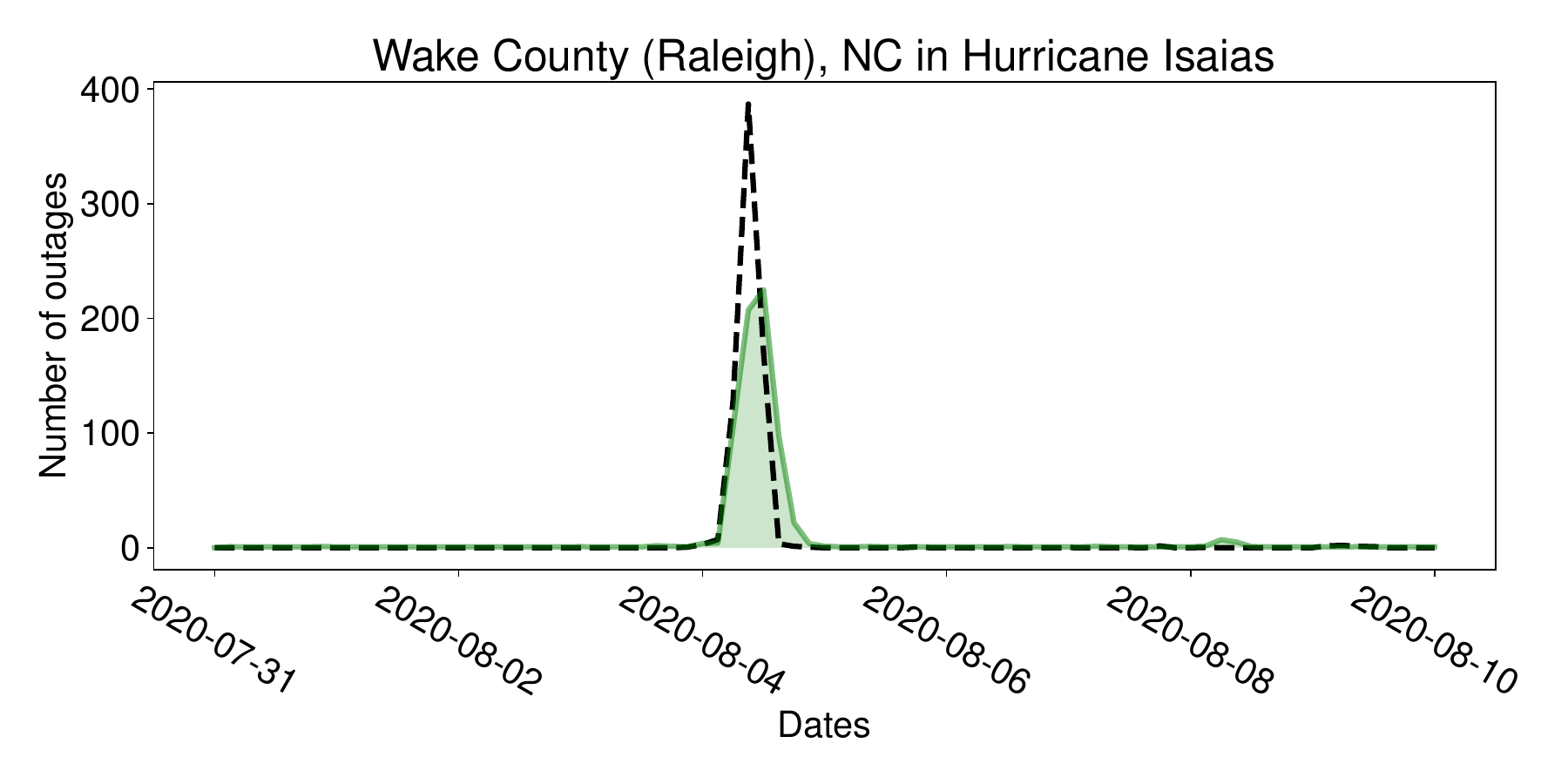}
        \vfill
        \includegraphics[width=\linewidth]{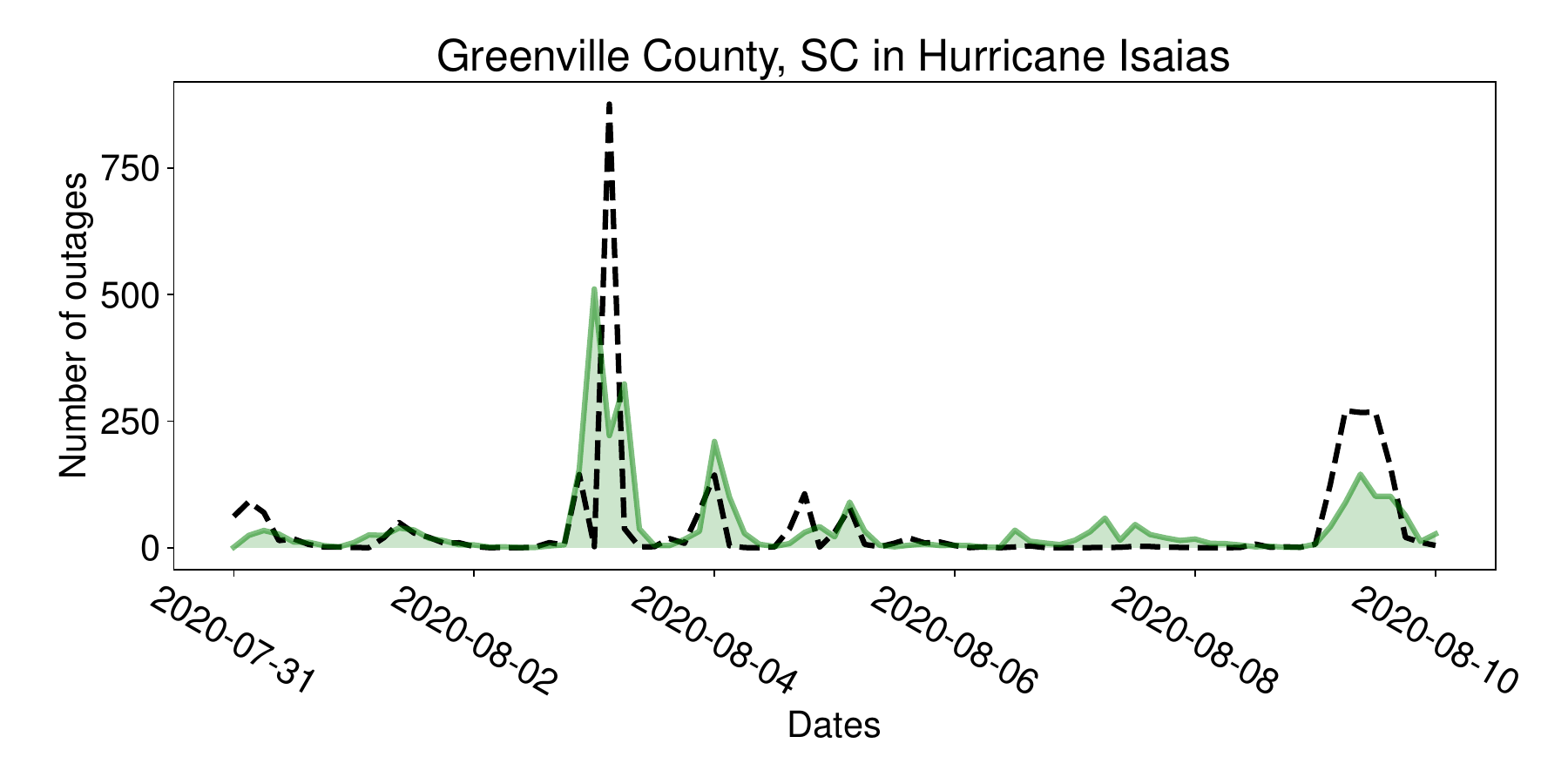}
        \vfill
        \includegraphics[width=\linewidth]{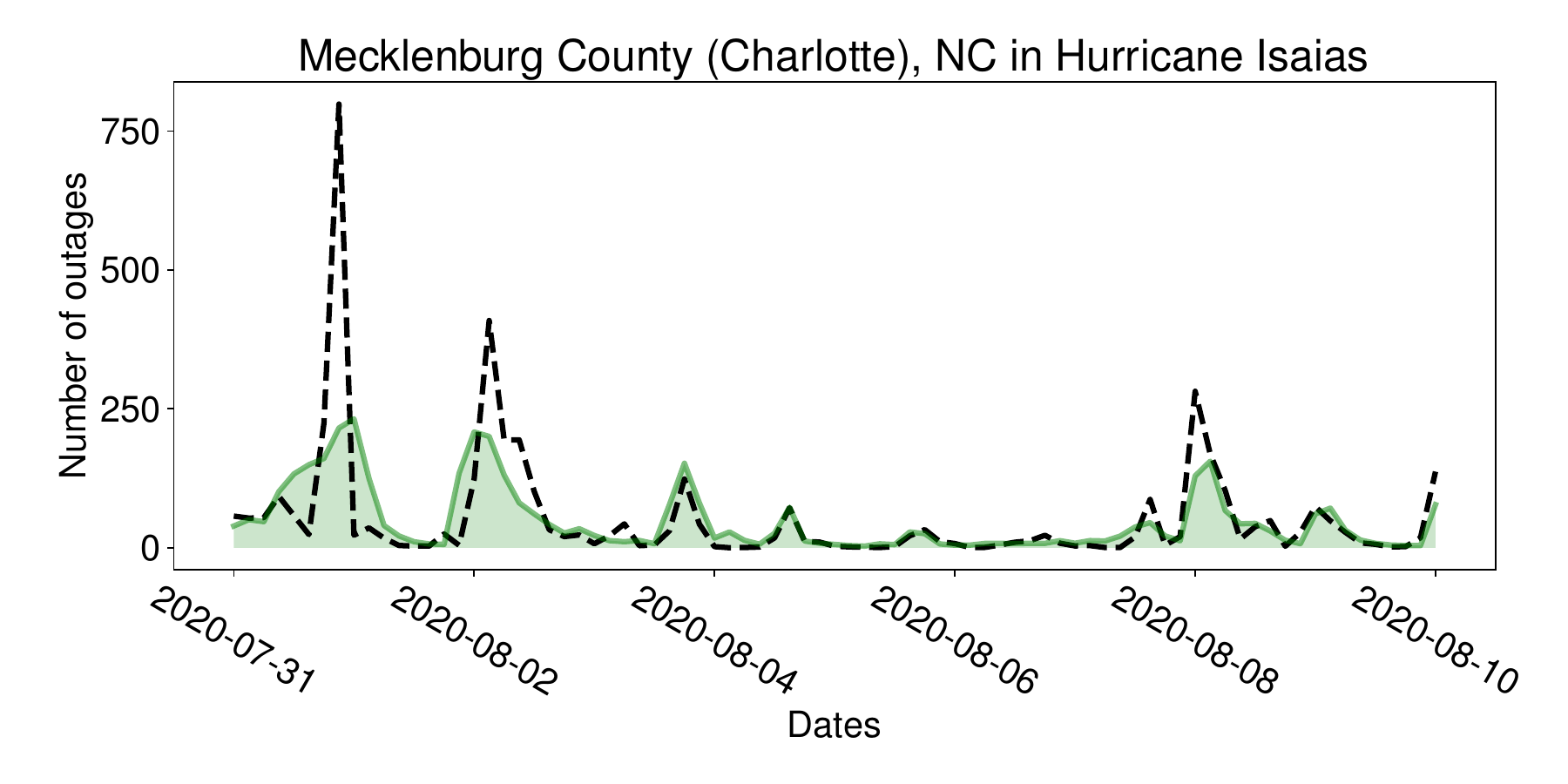}
        \caption{}
    \end{subfigure} & 
    \begin{subfigure}[b]{.28\textwidth}
        \centering
        \includegraphics[width=\linewidth]{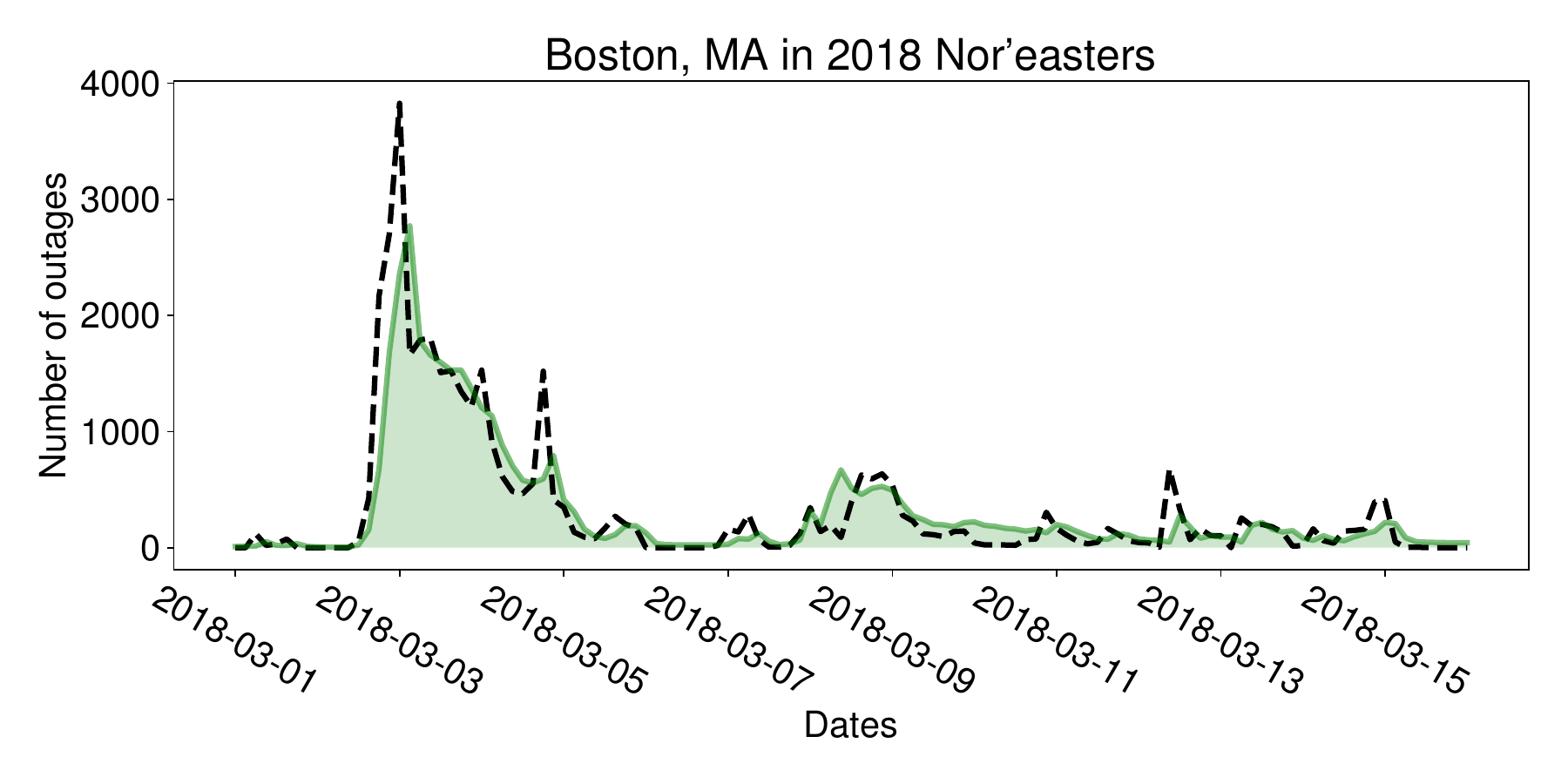}
        \vfill
        \includegraphics[width=\linewidth]{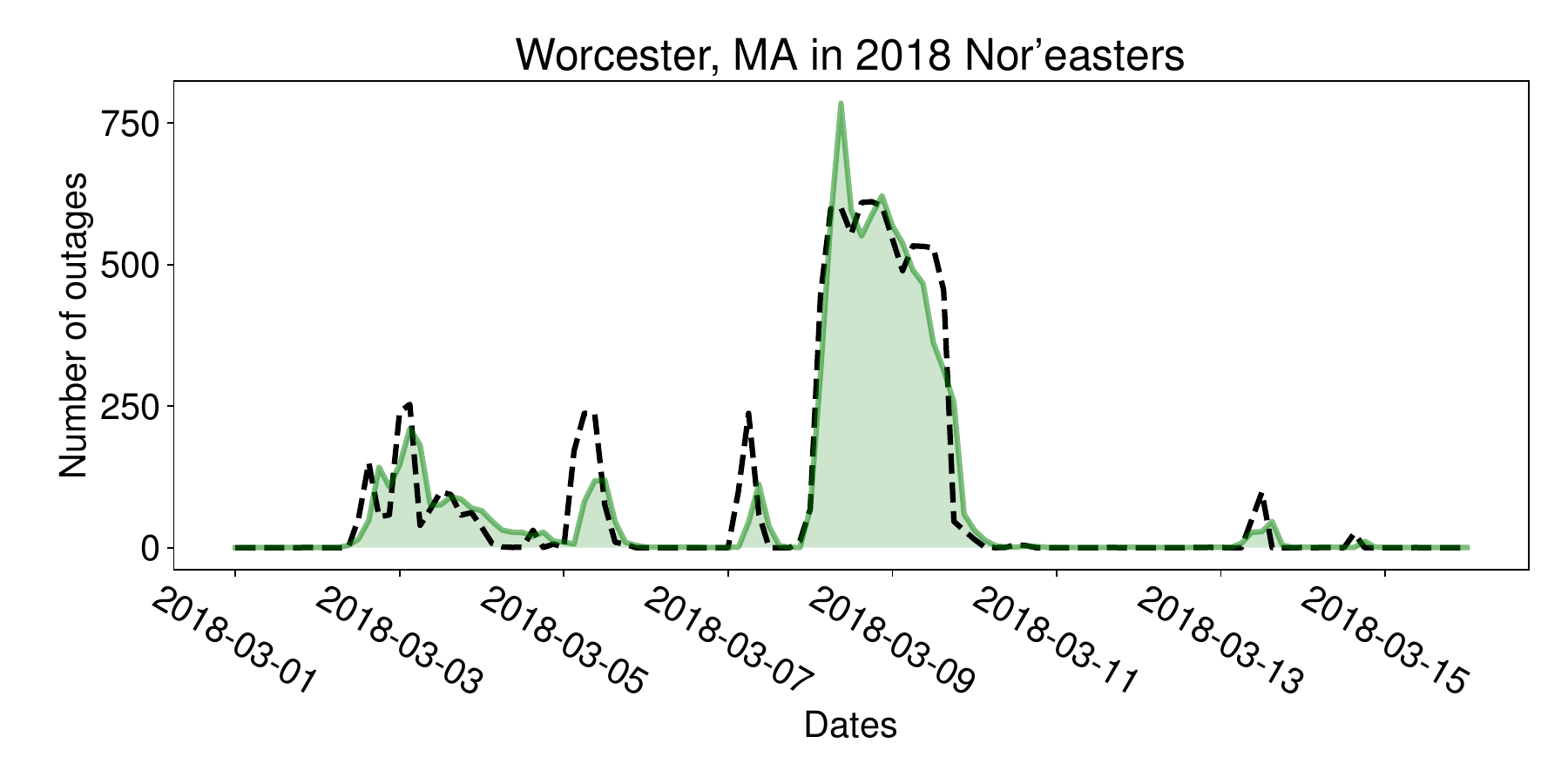}
        \vfill
        \includegraphics[width=\linewidth]{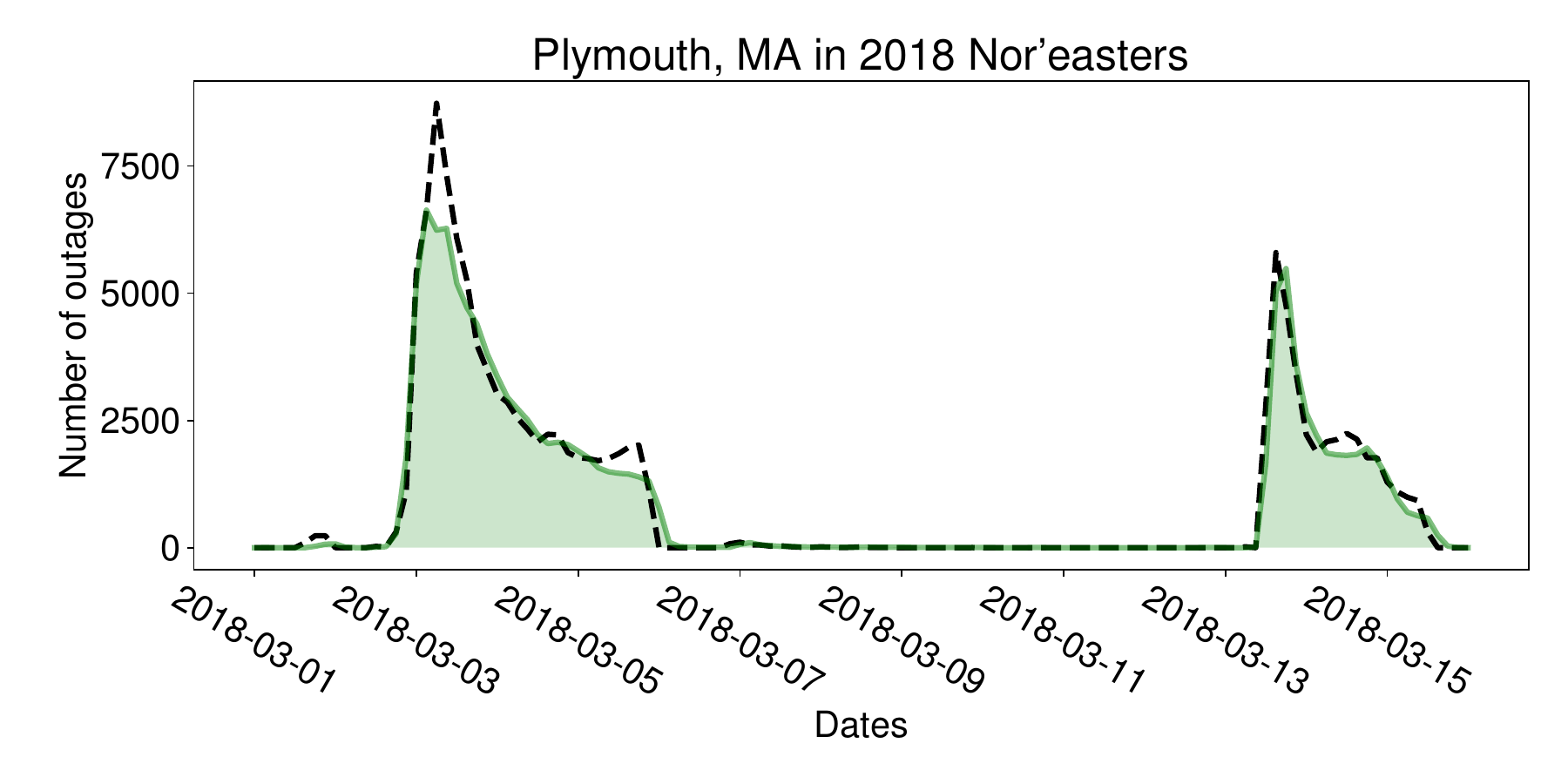}
        \vfill
        \includegraphics[width=\linewidth]{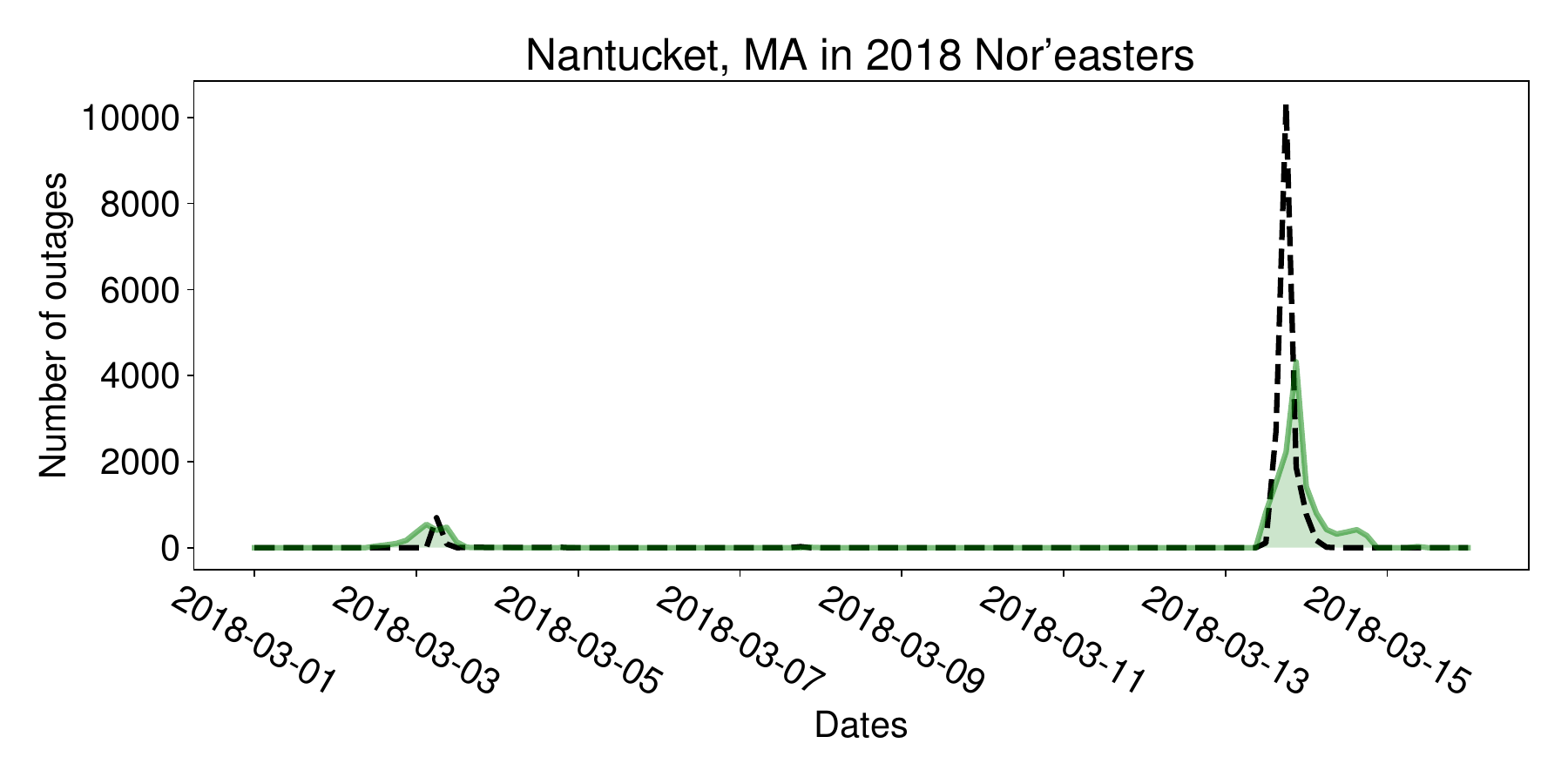}
        \vfill
        \includegraphics[width=\linewidth]{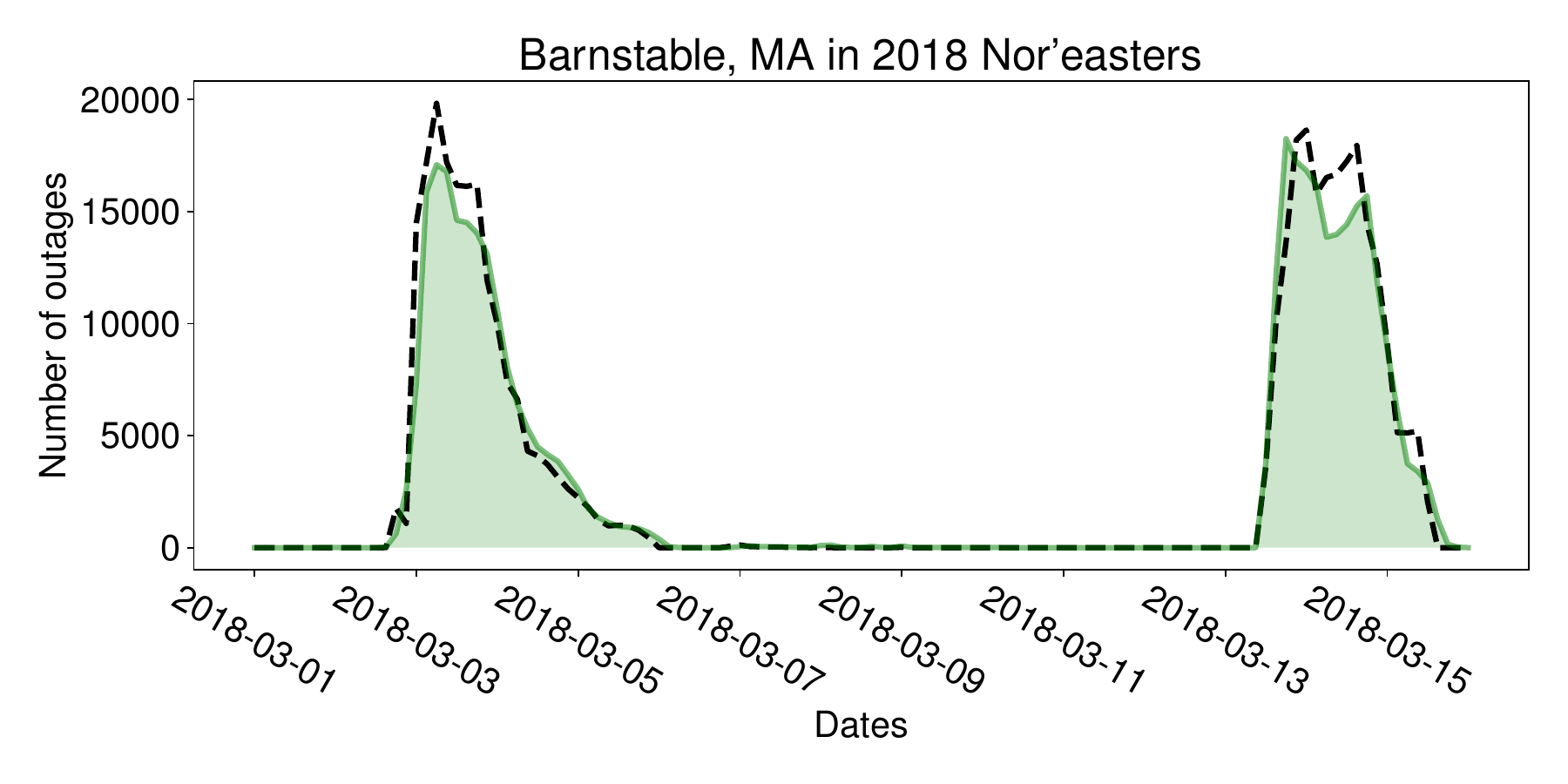}
        \caption{}
    \end{subfigure}
  \end{tabular}}
{Unit-level predictions units for three weather events. \label{fig:insample-prediction}}
{(a) Massachusetts (March 2018 nor'easters), (b) Georgia (Hurricane Michael), and (c) North Carolina and South Carolina (Hurricane Isaias). Black lines represent the real customer power outages in a certain area and green shaded regions represent the in-sample prediction for customer power outages in the same area using our model.}
\end{figure}

Figure~\ref{fig:insample-prediction} provides additional details about the unit-level in-sample prediction for three different service regions. The result demonstrate that our model is capable of accurately capturing the temporal dynamics of customer power outages at the unit level. 

\begin{figure}[!t]
\centering
\FIGURE{
\centering
  \begin{tabular}[c]{cccc}
    \begin{subfigure}[c]{0.24\linewidth}
    \includegraphics[width=\linewidth]{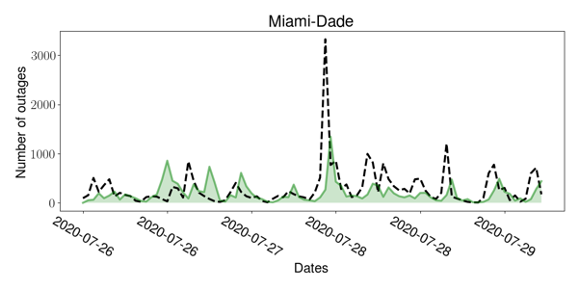}
    \end{subfigure} & 
    \begin{subfigure}[c]{0.24\linewidth}
    \includegraphics[width=\linewidth]{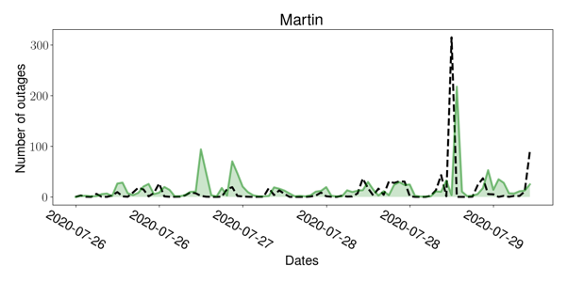}
    \end{subfigure} & 
    \begin{subfigure}[c]{0.24\linewidth}
    \includegraphics[width=\linewidth]{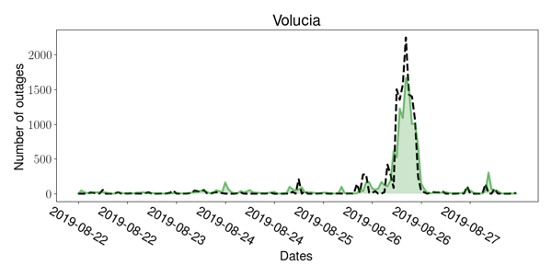}
    \end{subfigure} & 
    \begin{subfigure}[c]{0.24\linewidth}
    \includegraphics[width=\linewidth]{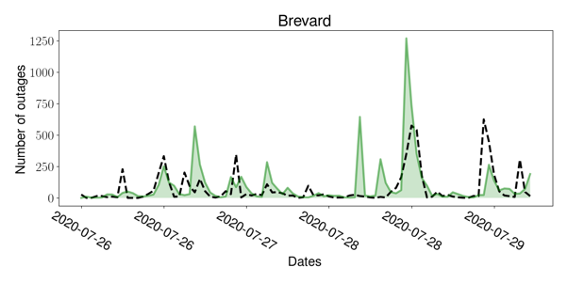}
    \end{subfigure}
    \\
    \begin{subfigure}[c]{0.24\linewidth}
    \includegraphics[width=\linewidth]{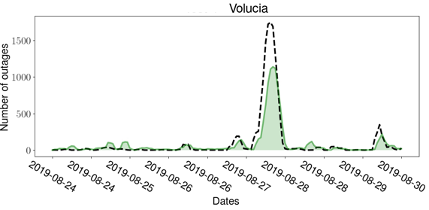}
    \end{subfigure} & 
    \begin{subfigure}[c]{0.24\linewidth}
    \includegraphics[width=\linewidth]{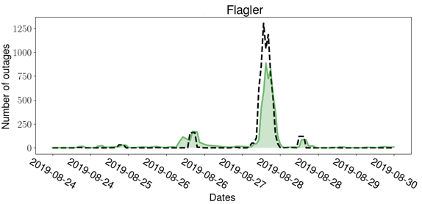}
    \end{subfigure} & 
    \begin{subfigure}[c]{0.24\linewidth}
    \includegraphics[width=\linewidth]{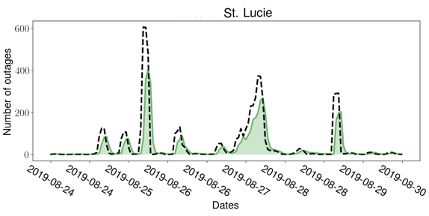}
    \end{subfigure} & 
    \begin{subfigure}[c]{0.24\linewidth}
    \includegraphics[width=\linewidth]{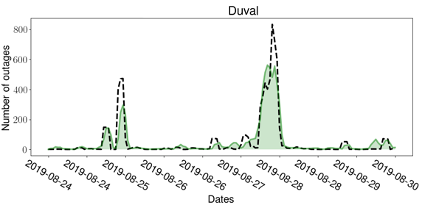}
    \end{subfigure}
    \\
    \begin{subfigure}[c]{0.24\linewidth}
    \includegraphics[width=\linewidth]{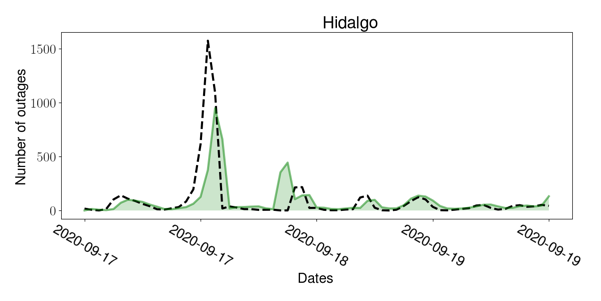}
    \end{subfigure} & 
    \begin{subfigure}[c]{0.24\linewidth}
    \includegraphics[width=\linewidth]{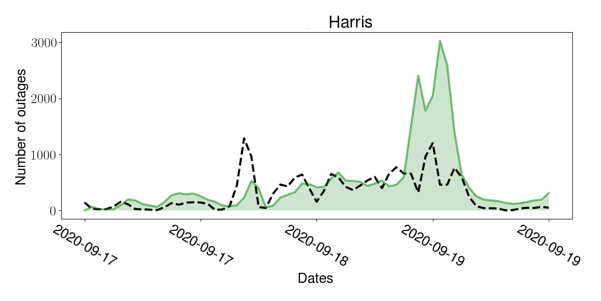}
    \end{subfigure} & 
    \begin{subfigure}[c]{0.24\linewidth}
    \includegraphics[width=\linewidth]{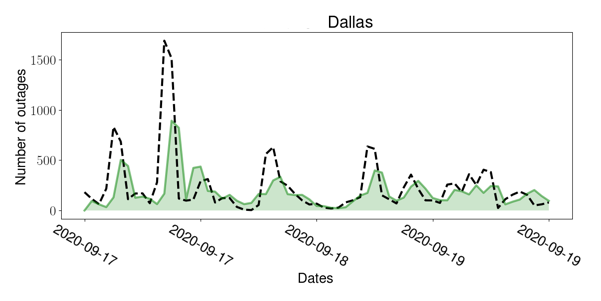}
    \end{subfigure} & 
    \begin{subfigure}[c]{0.24\linewidth}
    \includegraphics[width=\linewidth]{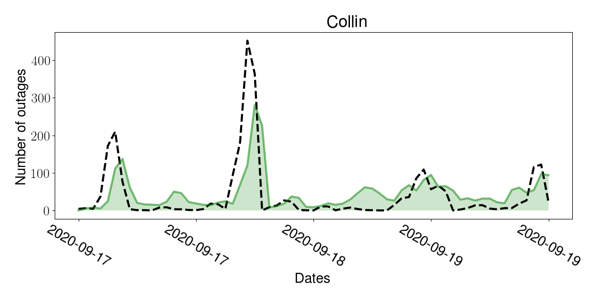}
    \end{subfigure}
    \\
    \begin{subfigure}[c]{0.24\linewidth}
    \includegraphics[width=\linewidth]{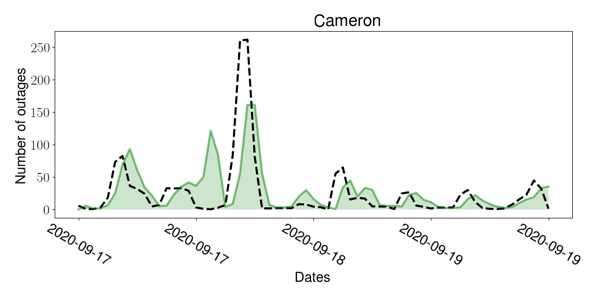}
    \end{subfigure} & 
    \begin{subfigure}[c]{0.24\linewidth}
    \includegraphics[width=\linewidth]{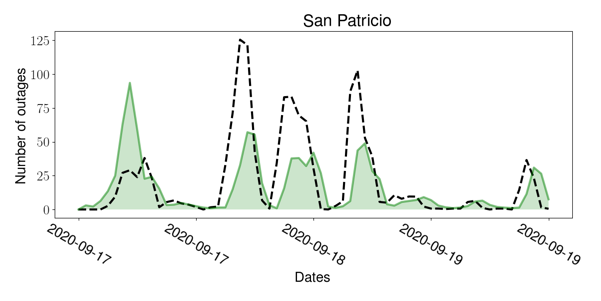}
    \end{subfigure} & 
    \begin{subfigure}[c]{0.24\linewidth}
    \includegraphics[width=\linewidth]{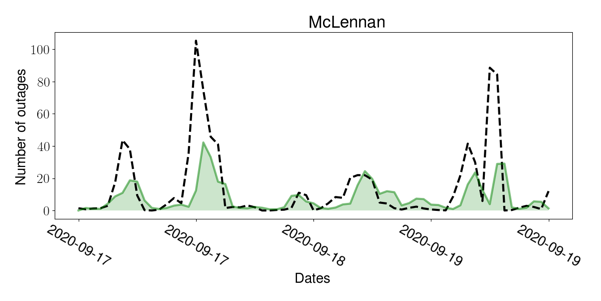}
    \end{subfigure} & 
    \begin{subfigure}[c]{0.24\linewidth}
    \includegraphics[width=\linewidth]{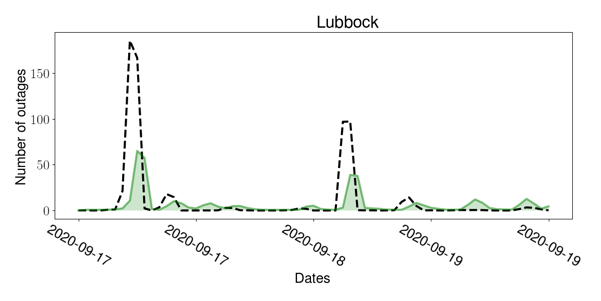}
    \end{subfigure}
  \end{tabular}}
{Three unit-level out-of-sample prediction results. \label{fig:outofsample-prediction}}
{First row: Training set: Hurricane Dorian (08-24-2019 to 09-07-2019). Testing set: Hurricane Isaias (07-30-2020 to 08-04-2020). 
Second row: Training set: Tropical Storm Alberto (05-25-2018 to 06-02-2018), Tropical Storm Gordon (09-03-2018 to 09-11-2018), Hurricane Michael (10-07-2018 to 10-17-2018). Testing set: Hurricane Dorian (08-24-2019 to 09-11-2019).
Third and fourth rows: Training set: Tropical Storm Imelda (09-17-2019 to 09-20-2019), Hurricane Hanna (07-23-2020 to 07-27-2020), Hurricane Laura (08-20-2020 to 08-30-2020). Testing set: Tropical Storm Beta (09-17-2020 to 09-25-2020).
Black lines represent the real customer power outages, and green shaded regions represent the three-hour-ahead prediction for customer power outages in the same area using our model.}
\end{figure}

In addition to the in-sample estimation, we assess the model’s predictive power by performing the one-step-ahead (out-of-sample) prediction, shown in Figure~\ref{fig:outofsample-prediction}. 
The prediction procedure withholds the future data from the model estimation and then uses the fitted model to make predictions for the (hold-out) data at the next time slot.
The result confirms that our model attains a good predictive performance on real data. 
We also apply the methodology outlined above to daily operations and achieve similar predictive accuracy, indicating that grid resilience captured by our model is inherent to the power distribution infrastructure and operations.

The inclusion of out-of-sample prediction serves several purposes:
($i$) Out-of-sample prediction is a crucial step in validating the model's robustness and generalizability. By testing the model on data that was not used during the training phase, we can assess its ability to generalize beyond the specific instances it was trained on. This is important to ensure that the model is not overfitting to the historical data and simply explaining the noise and 
can accurately reflect the underlying patterns and dynamics.
($ii$) The out-of-sample performance provides insights into how well the model captures the dynamics of extreme weather events and their impact on power grid resilience. By analyzing the discrepancies between predicted and actual out-of-sample events, we can refine our understanding of the factors that influence grid resilience and improve our strategies for mitigating adverse effects.

\begin{figure}[!t]
\centering
\FIGURE{\begin{subfigure}[b]{.48\linewidth}
\includegraphics[width=\linewidth]{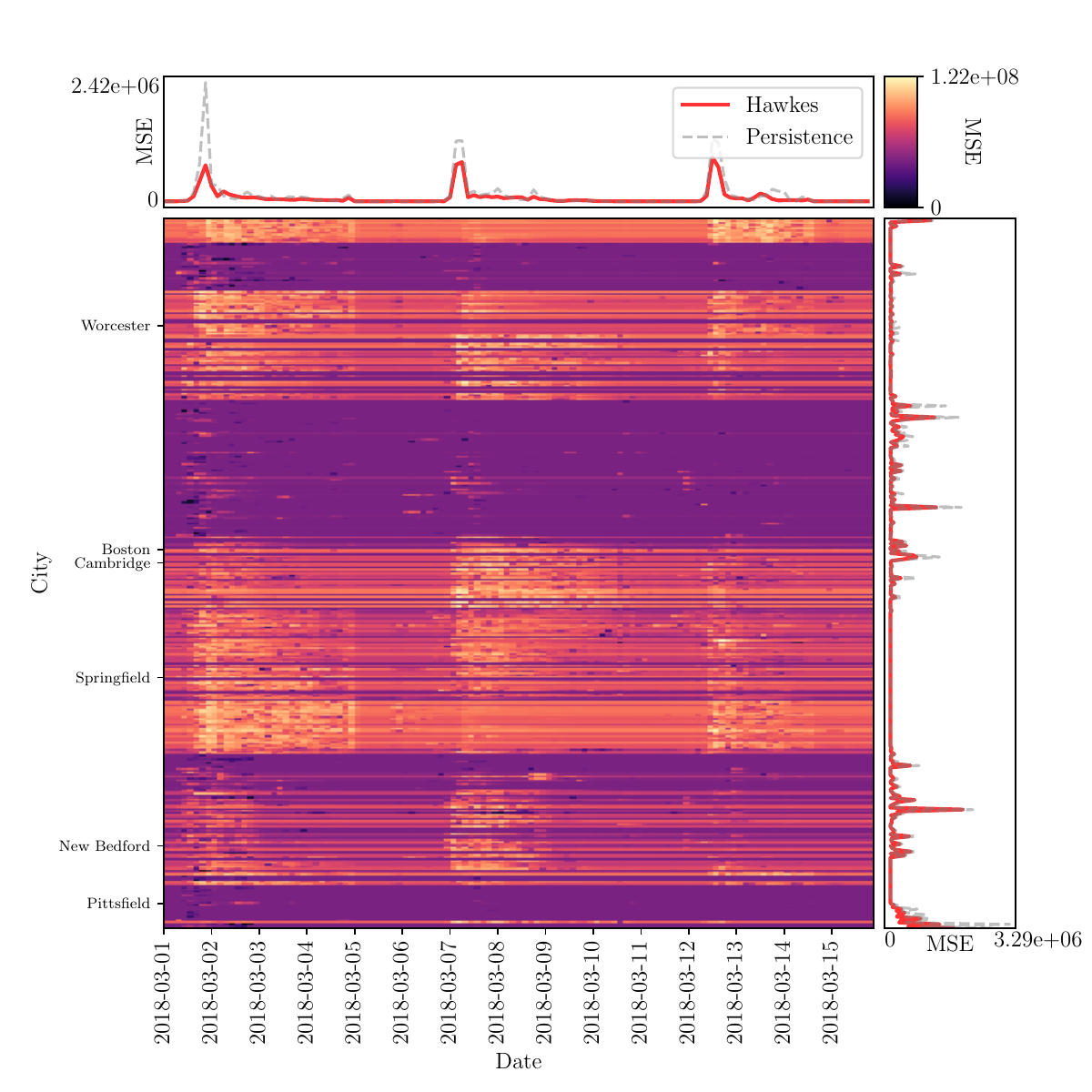}
\caption{Vanilla Hawkes}
\end{subfigure}
\begin{subfigure}[b]{.48\linewidth}
\includegraphics[width=\linewidth]{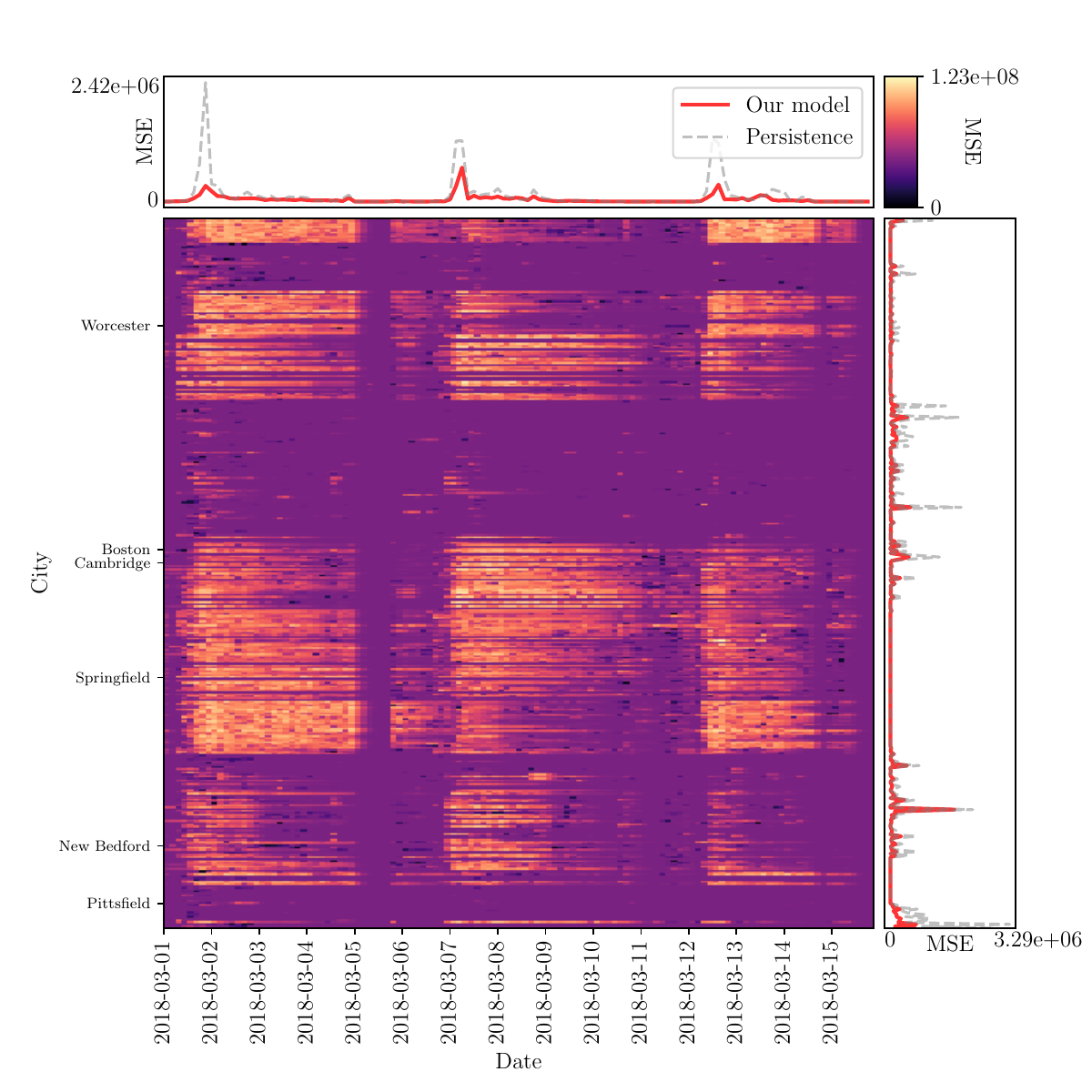}
\caption{Proposed method}
\end{subfigure}}
{Comparison of mean square error (MSE) among our proposed method, a basic Hawkes process without DNN, and a naive persistence model. \label{fig:mae}}
{Each entry in the heatmap represents the MSE for a unit at a specific time. Brighter colors indicate higher MSE values. The vertical and horizontal line charts display MSE across units and over time, respectively.}
\end{figure}

To further validate our model's performance, we assess it against two baseline methods. The first baseline, referred to as ``Hawkes,'' uses a basic Hawkes process with a constant base intensity, omitting the DNN. The second baseline is a naive persistence model that predicts the next time step using the last observation as a sanity check. We present the quantitative evaluation of their performance on the MA data sets in Figure~\ref{fig:mae}. The results indicate that our model outperforms both baselines in terms of mean square error (MSE), demonstrating that the DNN significantly enhances the model's predictive power and data explainability.

\begin{figure}[!t]
\centering
\FIGURE{\includegraphics[width=.7\linewidth]{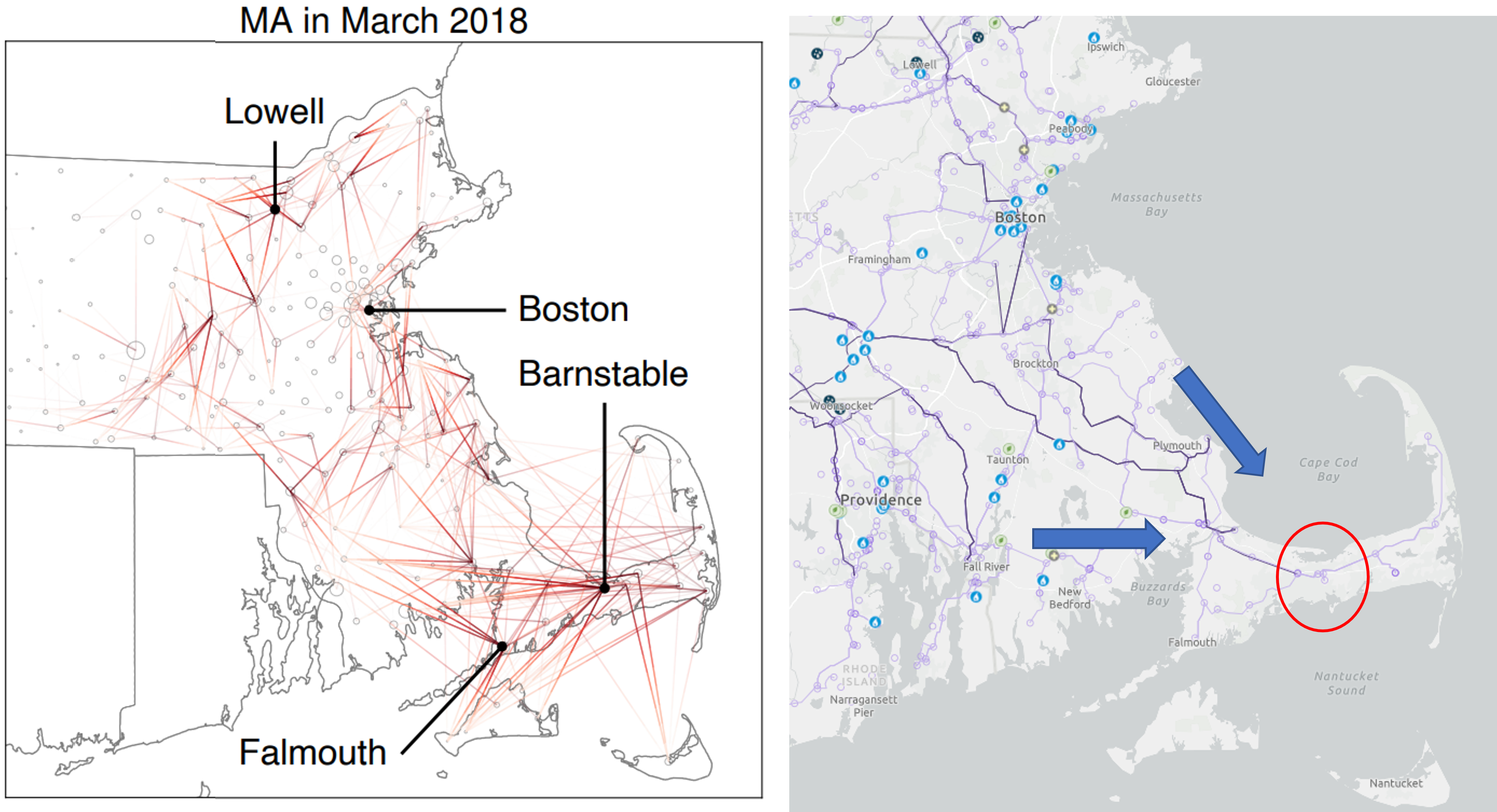}}
{Power flows influence outage interdependence. \label{fig:interdependense-power-flow}}
{This figure illustrates how power flows may influence outage interdependence relationships. Barnstable is used for this illustration, which is located as a bottleneck for power flowing from the inner MA to the area east of Barnstable. If an extreme weather event damages some power infrastructures at Barnstable disabling the power flow into its eastern area, the outages on the eastern side of Barnstable are interdependent on the outage occurred within Barnstable, as shown in the figure. Note that the outage interdependence from Barnstable to its western side is likely related to the same weather events which moved from the coast to the inner MA.}
\end{figure}

\begin{figure}[!t]
\centering
\FIGURE{\includegraphics[width=.7\linewidth]{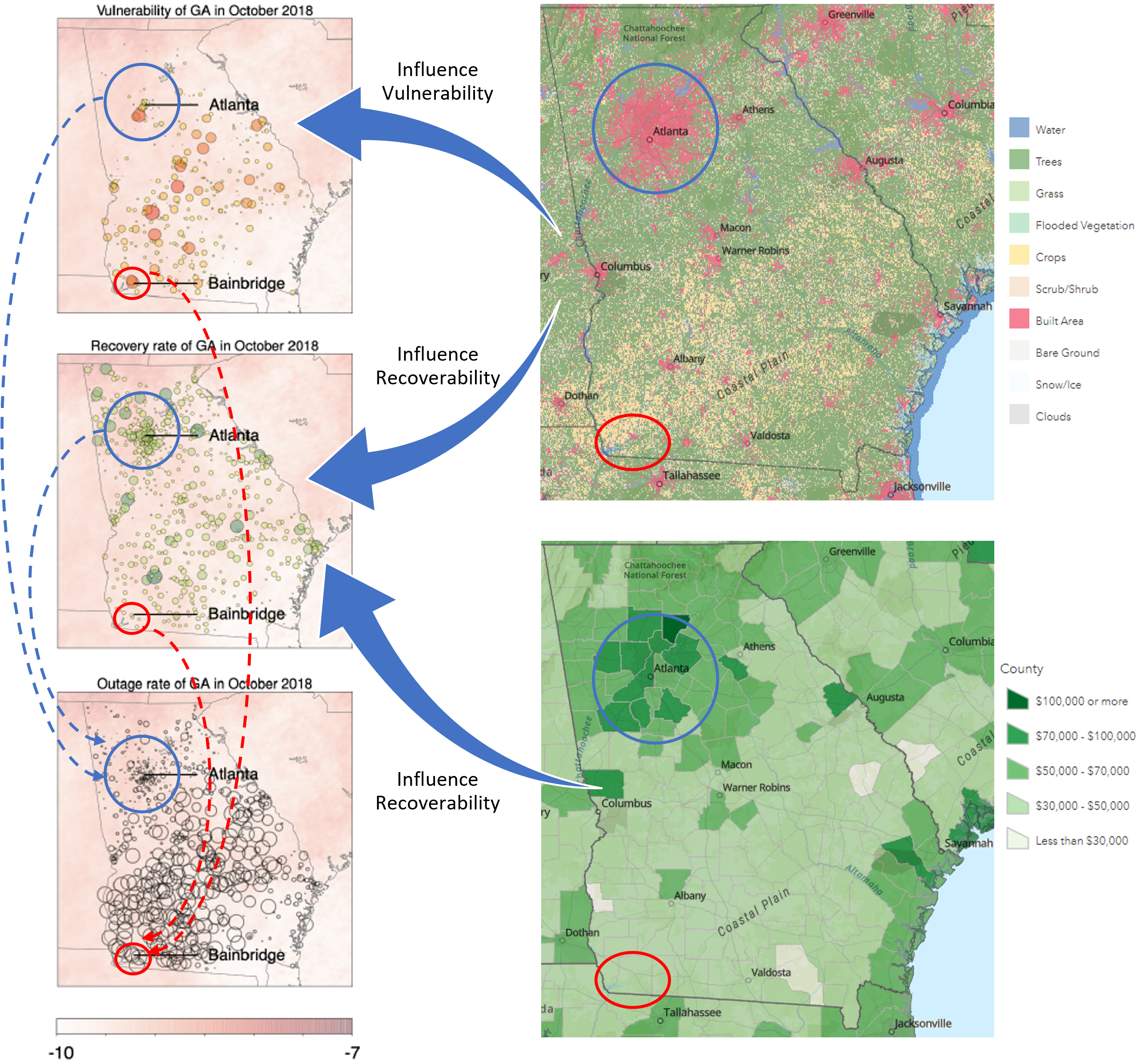}}
{Contributing social factors to outages. \label{fig:social-factors}}
{This figure shows how social development factors, including land cover and household income information at different regions of Georgia, influence the vulnerability and recoverability to extreme weather events, which further influence the outage rates at those areas. Two areas are selected as examples, i.e., Atlanta (highlighted in the blue circles) and Bainbridge (highlighted in the red circles). Due to well developed territories, highly dense population, and relatively rich household incomes of the residents, Atlanta area is benefited by lower vulnerability and higher recoverability compared with those of Bainbridge area. As a result, the outage rates at Atlanta area are relatively low compared with those for Bainbridge area.}
\end{figure}

\section{More policy insights derived from our study}
\label{append:policy-insights}

In this section, we provide more policy insights derived from our data-driven results, combined with the published transmission and generation information: 
\begin{enumerate}
    \item 
    \emph{Outage Propagation}: The phenomenon of outage interdependence, as illustrated in Figure~\ref{fig:interdependense-power-flow}, often results from damage to transmission networks, which carry large power transfers (100 MVA+ per line). The loss of a transmission line can overload nearby lines, leading to load shedding in distant areas. Mid-size urban areas (e.g., Barnstable, MA; Macon, GA; Wilmington, NC), rather than metropolitan or rural areas, are more likely to be the source of outage propagation. These areas serve as important power interconnection points or have large generation capacities. Enhancing resilience in such critical areas can prevent major outages.
    \item 
    \emph{Vulnerability and Recovery}: Vulnerability indices are lower in metropolitan areas (e.g., Boston, Atlanta, Charlotte) due to better vegetation management, reducing the likelihood of outages. Urban areas also recover faster due to easier access for repairs, while rural areas face more challenges like difficult terrain. Consequently, outage rates in metropolitan areas are lower, as confirmed by our study in Georgia shown in Figure~\ref{fig:social-factors}.
\end{enumerate}

\end{APPENDICES}



\end{document}